\documentclass[11pt]{article}
\usepackage[utf8]{inputenc}
\usepackage{verbatim}
\usepackage{amsfonts}
\usepackage{amssymb,amsmath,amsthm}
\usepackage[linesnumbered,algoruled,boxed,lined]{algorithm2e}
\usepackage{array, booktabs, makecell}
\usepackage[paper=a4paper,margin=1.2in]{geometry}
\usepackage{graphicx}
\usepackage{appendix}
\usepackage{authblk}
\usepackage{caption}
\usepackage{xcolor}
\usepackage{lipsum}
\usepackage{multicol}
\usepackage{multirow}
\usepackage{subcaption}
\usepackage{natbib}
\usepackage{makecell}
\usepackage{adjustbox}
\usepackage{eurosym}
\usepackage{enumitem}
\usepackage{bm}
\usepackage{colortbl}
\usepackage{titlesec}
\usepackage{pifont}
\newcommand{\R}{\mathbb{R}}

\newcommand{\Pro}{\mathbb{P}}

\definecolor{cadmiumgreen}{RGB}{34,139,34}
\definecolor{carnelian}{rgb}{0.7, 0.11, 0.11}
\definecolor{Gray}{gray}{0.9}
\newcolumntype{g}{>{\columncolor{Gray}}c}
\newcolumntype{h}{>{\columncolor{Gray}}l}
\def\arrvline{\hfil\kern\arraycolsep\vline\kern-\arraycolsep\hfilneg}
\usepackage{subfiles}
\usepackage[breaklinks]{hyperref}
\usepackage{breakcites}

\graphicspath{{./Images/}}

\title{Identifying and Responding to Outlier Demand in Revenue Management}

\author[a,*]{\small Nicola RENNIE}
\author[b]{\small Catherine CLEOPHAS}
\author[c]{\small Adam M. SYKULSKI}
\author[d,e]{\small Florian DOST}

\affil[a]{\footnotesize STOR-i Centre for Doctoral Training, Lancaster University, LA1 4YW, UK. (\href{mailto:n.rennie@lancaster.ac.uk}{n.rennie@lancaster.ac.uk})}
\affil[b]{\footnotesize Institute for Business, Christian-Albrechts-University Kiel, Kiel, Germany. (\href{mailto:cleophas@bwl.uni-kiel.de}{cleophas@bwl.uni-kiel.de})}
\affil[c]{\footnotesize Dept. of Mathematics and Statistics, Lancaster University, LA1 4YF, UK. (\href{mailto:a.sykulski@lancaster.ac.uk}{a.sykulski@lancaster.ac.uk})}
\affil[d]{\footnotesize Management Science and Marketing Division, Alliance Manchester Business School, M15 6PB, UK.}
\affil[e]{\footnotesize Institute of Business and Economics, Brandenburg University of Technology, 03046 Cottbus, Germany. (\href{mailto:florian.dost@b-tu.de}{florian.dost@b-tu.de})}
\affil[*]{Corresponding Author: Nicola Rennie (\href{mailto:n.rennie@lancaster.ac.uk}{n.rennie@lancaster.ac.uk})}

\date{\vspace{-8ex}}

\begin{document}

\maketitle

\begin{abstract}
Revenue management strongly relies on accurate forecasts. Thus, when extraordinary events cause outlier demand, revenue management systems need to recognise this and adapt both forecast and controls. Many passenger transport service providers, such as railways and airlines, control the sale of tickets through revenue management. State-of-the-art systems in these industries rely on analyst expertise to identify outlier demand both online (within the booking horizon) and offline (in hindsight). So far, little research focuses on automating and evaluating the detection of outlier demand in this context. To remedy this, we propose a novel approach, which detects outliers using functional data analysis in combination with time series extrapolation. We evaluate the approach in a simulation framework, which generates outliers by varying the demand model. The results show that functional outlier detection yields better detection rates than  alternative approaches for both online and offline analyses. Depending on the category of outliers, extrapolation further increases online detection performance. We also apply the procedure to a set of empirical data to demonstrate its practical implications. By evaluating the full feedback-driven system of forecast and optimisation, we generate insight on the asymmetric effects of positive and negative demand outliers. We show that identifying instances of outlier demand and adjusting the forecast in a timely fashion substantially increases revenue compared to what is earned when ignoring outliers. 

\textbf{Keywords:} Revenue management; Simulation; Forecasting; Outlier detection; Functional data analysis.
\end{abstract}

\newgeometry{margin=1.5in}

\section{Introduction} \label{sec:introduction}
In the last 40 years, \emph{revenue management (RM)} has become an indispensable business practice, particularly for transport service providers such as airlines and railways \citep{weatherford2016history}. RM solves an optimisation problem, where firms decide on offers for perishable products, usually with the objective of maximising revenue. This optimisation assumes a fixed capacity, low marginal cost, and a given \emph{demand forecast}. In that regard, \citet{Belobaba2002} highlight that inaccurate demand forecasts can significantly diminish the achieved revenue. \citet{banerjee2019passenger} point out that detailed demand forecasts also support in further planning steps, such as network resource and fuel planning.

\citet{Cleophas2017} list several causes for \emph{forecast inaccuracies}: On the one hand, the unavoidable variance of day-to-day demand prohibits perfectly accurate forecasts. On the other hand, any flaw in the forecast model, including both the predictive time series component and the customer choice model naturally causes model-based forecast errors. Finally, sudden shifts in the market may cause short-term, temporal \emph{outliers}. For example, when the system does not account for special events such as a sports championship or a trade fair, these will cause observed demand to systematically deviate from predictions. 

We focus on such \emph{demand outliers} in the domain of revenue management for passenger transport, specifically railways and airlines. In this domain, RM via capacity controls optimises \emph{booking limits}, which specify the number of units that can be sold per fare class and time in a fixed \emph{booking horizon}. Accordingly, sold units are also termed \emph{bookings}. The distribution of bookings over intervals of the booking horizon constitutes a \emph{booking pattern}. Booking patterns may be aggregated across fare classes and are reported either for single resources, such as flight legs, or for complementary combinations of resources, such as network itineraries. Here, we focus on aggregated booking patterns as reported for single resources, such as a single flight or a railway connection.

Common RM demand forecasting techniques estimate demand from historical booking patterns and booking limits \citep{weatherford2016history}. Accordingly, we let outlier detection rely on the same data. We follow the definition by \citet{Hawkins1980} and define an outlier as `an observation which deviates so much from the other observations as to arouse suspicions that it was generated by a different mechanism.' Detection can either apply \textit{online}, within the booking horizon and considering partial booking patterns, or \textit{offline}, after a booking horizon, when the complete pattern can be analysed. Demand outliers affect revenue management systems in two ways: (i) in \emph{foresight}, the flawed forecast results in non-optimal capacity allocations; and (ii) in \emph{hindsight}, the outlier can contaminate the data underlying future forecasts. Accordingly, online detection can improve foresight, whereas offline detection can improve hindsight. To detect outliers, functional data analysis, where each booking pattern is treated as an observation of a function over time, is a natural place to turn to. Functional approaches can detect outliers in both magnitude and shape of an observed booking pattern. In other words, it can detect outliers that deviate across the entire booking horizon and those that deviate in only part of the booking horizon. Effective detection in online and offline settings has to be capable of identifying both types of outliers.

By investigating practical RM implementations in the airline and railway industry, we find that the current process relies on analysts, who manually examine booking patterns. When analysts perceive demand outliers, they attempt to compensate by adjusting the reported data, the forecast, or booking limits. The decision of whether an adjustment is necessary and in what form depends on the analysts' intuition. As noted by \citet{Cleophas2017} and \citet{banerjee2019passenger}, little existing work systematically measures the effect of such interventions. There is even less consideration of providing systematic analytics support for the related decisions. However, research on human decision making in general, and judgemental forecasting in particular, clearly demonstrates fallibility and bias \citep{OConnor1993, lawrence2000field, lawrence2006}. This motivates the need for automated alerts to highlight outliers and thereby support analysts.

To our knowledge, we are the first to propose an automated methodology for outlier detection in the RM domain. Specifically, this paper makes the following contributions: (i) it proposes a novel outlier detection approach, combining functional data analysis and time series extrapolation, which improves overall detection performance; (ii) it provides a simulation-based framework for generating regular and outlier booking patterns  and evaluating their effect throughout the RM process; (iii) it demonstrates the asymmetric effects of outliers on RM performance; (iv) it quantifies the benefits from successful online or offline outlier detection for RM; (v) it demonstrates the use of such outlier detection in an application to empirical railway booking data.

\newpage
\section{RM Forecasts and Forecast Evaluation}
\label{sec:RMForecasting}

The importance of accurate forecasts as input to revenue optimisation is well-documented in the literature. Authors are largely concerned with forecasting customer demand (\citet{Pereira2016}, \citet{Weatherford2002}, \citet{Talluri2004}), although forecasting cancellations and no-shows has also been explored (\citet{Morales2010}). \citet{Belobaba2002} confirm previous findings that inaccurate demand estimates can significantly impact revenue. Under the use of optimisation heuristics such as Expected Marginal Seat Revenue (EMSRb) (\citet{Belobaba1989}), under- or over-forecasting can even be beneficial. As described by \citet{Mukhopadhyay2007}, most RM systems require forecasts of the \textit{actual} demand, rather than the \textit{observed} demand. The actual demand consists of both observed demand and customer requests that were denied due to restrictive booking limits. Actual demand is difficult to observe in practice, and so must be estimated. To this end, \citet{Weatherford2002} survey various techniques. 

When allowing for inaccurate demand forecasts, much RM research focuses on rendering the optimisation component more robust or forecast-independent, as detailed in the contributions reviewed in \citet{gonsch2017survey}. In another review, \citet{Cleophas2017} point out that there is little research into the effects of manually adjusted forecasts in RM. \citet{Mukhopadhyay2007} propose a method for measuring the performance of adjusted and unadjusted forecasts. They find that if analysts can reliably improve demand forecasts on critical flights, significantly more revenue can be generated. \citet{Zeni2003} describe a study at US Airways, which aimed to isolate and estimate the value of analyst interactions. According to that study, around 3\% of the additional revenue generated within the duration of the study could be attributed to analyst input. 

Given that experiments in a live RM system carry significant risks, the use of simulation for evaluation is common. Additionally, simulation studies enable {\em a priori} knowledge about the true demand generation process, which can never be known in a real-world setting. \citet{Frank2008} discuss the use of simulation for RM and provide guidelines; in a related effort, \citet{Kimms2007} consider demand modelling for RM simulations. The paper at hand follows these contributions in establishing a simulation-based framework to generate outlier observations. \citet{Doreswamy2015} employ simulation as a tool to analyse the effects of different RM techniques for different airlines, when switching from leg-based controls to network controls. \citet{Cleophas2009} focus on an approach to evaluating the quality of RM forecasts both in terms of revenue and common forecast error measurements.  Another example of using simulation to evaluate the performance of forecast components is given in \citet{bartke2018benchmarking}. \citet{Temath2010} used a simulation-based approach to evaluate the robustness of a network-based revenue opportunity model when input data is flawed. In the broader context of demand forecasting, \citet{petropoulos2014horses} evaluate fitting time series forecasts for particular patterns of demand evaluation by manipulating these patterns in a simulation framework.

\section{Existing Work on Outlier Detection}
\label{sec:background}

\begin{figure}[htbp]
    \centering
        \begin{subfigure}[htbp]{1\textwidth}
            \centering
            \includegraphics[width=1\textwidth]{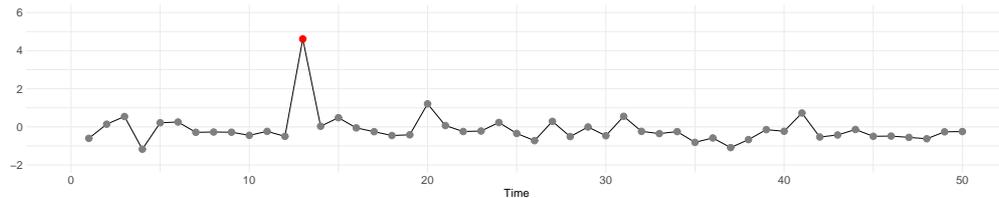}
            \caption{Outlier within a given time series}  
			\label{fig:within}
        \end{subfigure}
        \quad
        \begin{subfigure}[htbp]{1\textwidth}  
            \centering 
            \includegraphics[width=1\textwidth]{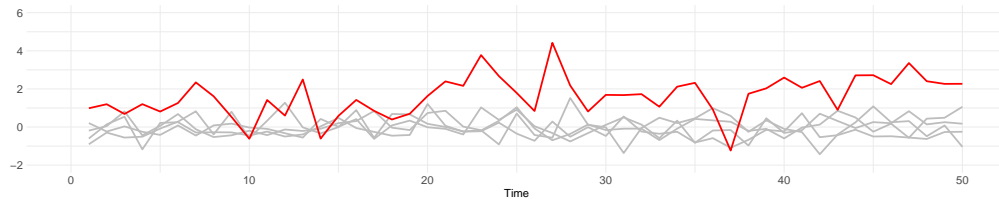}
            \caption{Outlying time series within a collection of series}
			\label{fig:asts}
        \end{subfigure}
        \quad
        \caption{Different types of outliers in time series data}
        \label{fig:typesofoutliers}
\end{figure}

To assess the existing methodological contributions to outlier detection, we distinguish between identifying outlying observations within a time series (Figure \ref{fig:within}), and identifying an entire outlying time series (in our case, booking pattern) (Figure \ref{fig:asts}). In this paper, we aim for the latter. 

Literature on handling outliers in the RM process is scarce, though there is some discussion in \citet{Weatherford2003}: the authors consider removing outliers caused by atypical events, such as holidays and special conventions, to improve future forecasting. However, they propose only to remove observations outside of the mean \(\pm\) 3 standard deviations and do not seek to identify outliers online within the booking horizon. 

Beyond RM, a wealth of literature studies outliers (also referred to as anomalies) in time series, as reviewed by \citet{Chandola2009} and \citet{Pimentel2014}. For example, \citet{Hubert2015} survey various functional outlier detection techniques for time series data, and apply their methods to multiple real data sets. \citet{barrow2018impact} consider the effect of functional outliers for call centre workload management and recommend an artificial neural network to model them as part of the forecast rather than identifying them. \citet{Hyndman2019} propose a sliding window approach for detecting outlying time series within a set of (nonstationary) time series, based on the use of extreme value theory for outlier detection. The authors also distinguish identifying outliers within a time series, and identifying a outlying series from a set. The remainder of this paper distinguishes three classes of approaches to outlier detection: (i) univariate, (ii) multivariate, or (iii) functional. Further technical details of all outlier detection methods described here are available in Appendix A.1.

\paragraph{Univariate Approaches} \mbox{}\\ 
Univariate outlier detection techniques identify anomalous observations of a single variable, and so can be applied independently at different time points in a time series, e.g., to the cumulative number of bookings per interval in a booking horizon.
\begin{itemize}[leftmargin=*]
\item \textbf{Nonparametric Percentiles}: This class of approaches uses lower and upper percentiles of the observed empirical distribution at each time point as limits for what constitutes a regular observation as opposed to an outlier. This type of percentile-based approach is discussed by \citet{Barnett1995}. It can be used as a basic way to estimate statistics in a more robust manner, by \textit{trimming} or \textit{winsorising} the data (see \citet{Dixon1974}). The downside of this approach is that a fixed percentage of the data will always be classified as outliers, even when there are fewer or more actual outliers in the data. 
\item \textbf{Tolerance Intervals}: Statistical tolerance intervals contain at least a specified proportion of observations with a specific confidence level \citep{Hahn1981}. They require two parameters: the coverage proportion, \(\beta\), and confidence level, \(1-\alpha\). For booking patterns, at each interval of the booking horizon, these approaches define a tolerance interval for the cumulative number of bookings by that time. If the number of observed bookings lies outside of this tolerance interval, the pattern is deemed an outlier. Nonparametric tolerance intervals do not assume an underlying distribution, and instead are based on the order statistics of the data \citep{Wilks1941}. Parametric tolerance intervals assume an underlying distribution \citep{Hahn1981}. The choice of distribution is not arbitrary, and a bad choice of distribution will perform poorly. \citet{Liang2018} choose to fit a Normal distribution to hotel booking data to detect anomalous observations.
\item \textbf{Robust Z-Score}: The \(Z\)-score measures where an observation lies in relation to the mean and standard deviation of the overall data \citep{Iglewicz1993}. The robust z-score uses the median and the median absolute deviation to provide a similar measurement. As such, an observation with a robust z-score above some threshold is classified as an outlier. This score-based method assumes that the observations in a given booking interval are approximately normally distributed based on two justifications: (i) A large proportion of univariate outlier detection methods rely on distributional assumptions (often normality). (ii) Although the discrete, non-negative integer nature of booking data suggests the use of a Poisson distribution, in the presence of trend or seasonal adjustments, the data may no longer have these properties. 
\end{itemize}

\paragraph{Multivariate Approaches} \mbox{}\\ 
Univariate outlier detection approaches ignore the dependence both within and between time series. We next turn to multivariate approaches as potential methods for capturing within (but not between) time series dependence. In this setting, a time series of length $\tau$, that is, a booking pattern observed over $\tau$ intervals, is considered as a point in a $\tau$-dimensional space. This lets the multivariate approaches compute the distance between any two booking patterns, but ignores the time ordering of observations.
\begin{itemize}[leftmargin=*]
\item \textbf{Distance}: 
Each booking pattern (observed over $\tau$ intervals) can be characterised by its $\tau$-dimensional distance to every other booking pattern. Aggregating these distances transforms the problem into a univariate outlier detection problem, based on the mean distances.Depending on the length of the booking pattern, issues relating to sparsity due to high dimensionality may arise. As discussed by \citet{Aggarwal2001}, some distance metrics perform better than others in a high dimensional space. However, in relation to distance metrics, high dimensionality often refers to at least hundreds of dimensions. The number of booking intervals in RM applications is often fewer than this, ranging from 20 to 50 in examples known to the authors. Therefore, we consider the classical Euclidean and Manhattan distance metrics in our comparative evaluation.
\item \textbf{K-Means Clustering}: \(K\)-means clustering splits the observed booking patterns into $K$ groups by iteratively minimising the ($\tau$-dimensional) distance between each booking pattern and the centre of its assigned cluster (see e.g. \citet{MacQueen1967}). This approach uses a distance threshold to identify booking patterns as outliers based on their distance the centre of their cluster \citep{DebDay2017}. As in the distance-based approaches, the choice of distance metric is highly relevant for clustering. Once more, this paper compares Euclidean and Manhattan distance metrics. The approach requires as its input parameter a given $K$ to indicate the number of clusters. Information on the methodology used to determine $K$ is available in Appendix A.1, including a comparison of performance under different choices of $K$, and the distribution of genuine outliers across such clusters. \end{itemize} 

\paragraph{Functional approaches} \mbox{}\\ 
There are two main issues with the use of multivariate outlier detection approaches in this application: (i) the effects of high-dimensionality on distance metrics when considering a large number of booking intervals, and (ii) the lack of accounting for the consecutive, time-ordered, nature of the observations. For such issues, functional data analysis is an intuitive place to turn. Functional data analysis addresses both issues by (i) treating booking patterns as functions observed $\tau$ times rather than points in a $\tau$-dimensional space, and (ii) explicitly taking into account the time-ordering of the observations. We provide further analysis of the importance of time-ordering in Appendix C.4.

The functional analysis setting, as discussed by \citet{Febrero2008}, lacks a rigorous definition of an outlier. We use the same definition as \citet{Febrero2008}: `a curve is an outlier if it has been generated by a stochastic process with a different distribution than the rest of curves, which are assumed to be identically distributed'. We view this as a more specific version of the definition by \citet{Hawkins1980}. 

A \textit{depth} function attributes a sensible ordering to observations, such that observations near the centre should have higher depth and those far from the centre should have lower depth. In the functional data setting, this idea provides an ordering to a set of smooth functions observed over discrete time-intervals, with the most central curve trajectory having highest depth. Functional depth not only accounts for the magnitude of the observations, but also for the variability in amplitude i.e. the shape of the curve \citep{Febrero2008}. Given this definition of functional depth, the degree of abnormality of a curve can be characterised by its functional depth, if its depth is particularly low \citet{Hubert2015}. Depth-based approaches for detecting outlying curves are discussed in detail by \citet{Hubert2012}. In this paper, we focus on the multivariate halfspace depth described by \citet{Claeskens2014}. We state and explain the mathematical definition of the multivariate halfspace depth in Appendix A.1.

\section{Proposed Methodology: Functional Outlier Detection With Extrapolation} \label{sec:functional}
To improve foresight, RM systems need to identify demand outliers online and as early as possible in the booking horizon. This enables the RM system to update controls for the remainder of the horizon. We term this problem {\em online outlier detection}. When tasked with online detection at time \(t_{\tau}\) in the booking horizon, all approaches discussed in the previous section exclusively consider the first \(\tau\) observation intervals.

In the online setting, only a partial booking pattern is available for analysis. Therefore, we propose to supplement the outlier detection by extrapolating the expected bookings from the current time \(t_{\tau}\) up to the end of the booking horizon, \(t_T\). We solve the resulting missing data problem by extrapolating from the bookings observed so far. In the computational study, we compare simple exponential smoothing (SES, \citet{Chatfield1975}), autoregressive integrated moving average models (ARIMA, \citet{BoxJenkins}), and integrated generalised autoregressive conditional heteroskedasticity models (IGARCH, \citet{Tsay2002}). Appendix A.2 provides a detailed list of univariate forecasting methods that can be used for extrapolation.

Algorithm \ref{alg:method} outlines the procedure on a set of \(N\) booking patterns observed until time \(t_{\tau}\): Given an entire booking horizon of length \(t_T\) with \(t_1, \hdots, t_{\tau}, \hdots, t_T\), then \(\bm{y}_n(t_{\tau})\) is a time series describing the bookings for pattern \(n\) up to time \(t_{\tau}\): \(\bm{y}_n(t_{\tau}) = \left(y_{n}(t_1), y_{n}(t_2), \hdots, y_{n}(t_{\tau}) \right)\).

\begin{algorithm}[ht] 
\SetAlgoLined
 At time \(t_{\tau}\) forecast the accumulation of bookings at each time \(\tau +1, \hdots, T\), \(\hat{y}_{n}(t_{\tau+1}), \hdots, \hat{y}_{n}(t_{T})\) for each booking pattern \(n\) \;
 Calculate \(\mathcal{D}_n(\bm{\hat{y}}_n(t_{\tau}))\), the functional depth of the observed and extrapolated booking pattern \(\bm{\hat{y}}_n(t_{\tau}) = \left(y_{n}(t_1), y_{n}(t_2), \hdots, y_{n}(t_{\tau}), \hat{y}_{n}(t_{\tau+1}), \hdots, \hat{y}_{n}(t_{T}) \right)\), for each booking pattern \(n\) at time \(t_{\tau}\)\;
 Calculate a threshold, \(C\), for the functional depth\;
 \hspace{1cm} \shortstack[l]{Bootstrap the original booking patterns, with probability proportional to \\ their functional depths;} \\
 \hspace{1cm} Smooth the bootstrap samples\;
 \hspace{1cm} Let $C^b$ be the $1^{st}$ percentile of the depths of the $b^{th}$ bootstrapped sample\;
 \hspace{1cm} Set $C$ as the median value of the $C^b$\;
 \If{\(\mathcal{D}_n(\bm{\hat{y}}_n(t_{\tau})) \leq C\)}{
 Define booking pattern \(n\) as an outlier. Delete booking pattern \(n\) from the sample of \(N\) patterns.
 }
 \While{\(\exists \mbox{ } n \mbox{ s.t. } \mathcal{D}_n(\bm{\hat{y}}_n(t_{\tau})) \leq C\)}{
 Recalculate functional depths on the new sample, and remove further outliers.
 }
 \caption{Using extrapolation to improve functional outlier detection}
 \label{alg:method}
\end{algorithm}

\begin{figure}[!ht]
        \centering
        \begin{subfigure}[b]{0.47\textwidth}
            \centering
            \includegraphics[width=0.8\textwidth,height=0.8\textwidth]{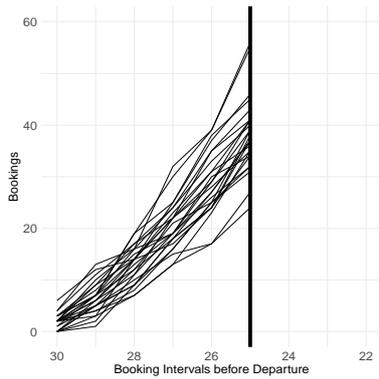}
            \caption{Observed booking patterns up to \\ \(t_{\tau} = 25\)}  
					\label{fig:ex1}
        \end{subfigure} \hspace{0.3cm}
        \begin{subfigure}[b]{0.47\textwidth}  
            \centering 
            \includegraphics[width=0.8\textwidth,height=0.8\textwidth]{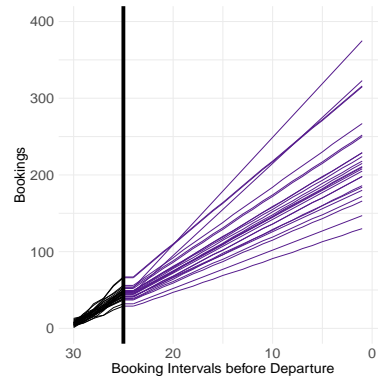}
            \caption{ARIMA extrapolation of booking patterns (purple)}
					\label{fig:ex2}
        \end{subfigure}
        \quad
        \begin{subfigure}[b]{0.47\textwidth}
            \centering
            \includegraphics[width=0.8\textwidth,height=0.8\textwidth]{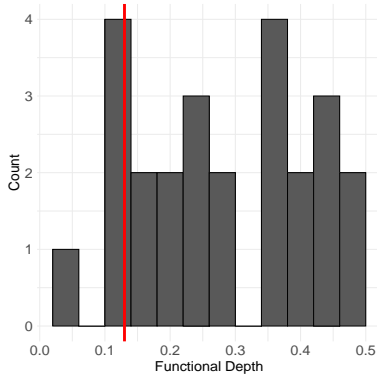}
            \caption{Histogram of functional depths with threshold (red)}
						\label{fig:ex3}
        \end{subfigure} \hspace{0.3cm}
        \begin{subfigure}[b]{0.47\textwidth}  
            \centering 
            \includegraphics[width=0.8\textwidth,height=0.8\textwidth]{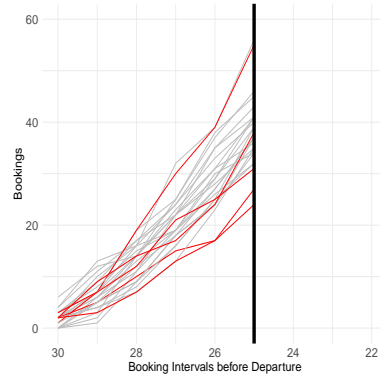}
            \caption{Observed booking patterns with detected outliers (red)}
						\label{fig:ex4}
        \end{subfigure}
        \quad
        \caption{Example: functional halfspace depth with ARIMA extrapolation outlier detection}
        \label{fig:example}
\end{figure}

Figure \ref{fig:example} demonstrates the algorithmic approach; in the extensive simulation analysis, we apply it to a variety of booking patterns and outliers. Figure \ref{fig:ex1} shows 25 booking patterns that have been observed during the first five of thirty intervals of the booking horizon. The extrapolation step is shown in Figure \ref{fig:ex2}, where the purple lines depict the ARIMA extrapolation of accumulated bookings until the end of the horizon. The empirical distribution of the functional depths of the extrapolated sample are shown in Figure \ref{fig:ex3}, with the threshold shown in red (computed via the bootstrapping routine described in Algorithm \ref{alg:method}, lines 3-7). The booking patterns classified by the algorithm as outliers are highlighted red in Figure \ref{fig:ex4}. 

The input parameters relating to the calculation of the threshold includes the number of bootstrap samples (line 4), the smoothing method (line 5), and the choice of percentile (line 6). In this paper, we select parameters as per \citet{Febrero2008} as they perform well in a wide range of settings. Further details of threshold calculation are available in Appendix A.1. The proposed approach could alternatively feature any of the multivariate or functional approaches reviewed Section \ref{sec:background}.\footnote{It is not applicable for univariate outlier detection methods as, in this setting, the number of bookings at each point in time is considered independently of past or future bookings.} However, a functional approach provides more scope for extensions, such as considering seasonality and increasing the frequency of outlier detection. In addition, the approach can utilise a variety of methods for extrapolating. Note that the methodology employed for this extrapolation is independent of the forecasting methodology to predict demand for RM.

\section{Simulation-based Framework} \label{sec:simulation}
To quantify effects from demand outliers and evaluate outlier detection approaches, we simulate a basic RM system with capacity controls. Such systems are common in the transport industry, but not limited to that domain \cite[see][Chapter~2.1]{TallurivanRyzin2004}. The system implemented here is minimal and general and does not fully mirror a real-world application system. However, the booking patterns our simulation generates are comparable with those observed in real-world RM systems -- see Appendix C.10. Since the simulation renders the process of demand generation explicit, computational experiments can yield truthful detection rates. This is impossible in empirical data analysis, where the true demand and the distinction of \textit{regular} versus outlier demand is never fully certain. Thereby, simulation modelling provides an alternative to the problem of creating reproducible forecasting research, highlighted for instance, by \citet{boylan2015reproducibility}.

The simulation implements the following steps:
\begin{enumerate}
\item Parameterise a demand model to specify both regular and outlier demand.
\item Generate multiple instances of regular and outlier demand from (1) in terms of customer requests (e.g. customers that intend to book a seat on a particular railway connection) arriving across the booking horizon.
\item From the demand model of regular demand (2), compute the forecast in terms of the number of expected requests per fare class and time in the booking horizon. 
\item Compute booking limits that maximise expected revenue from bookings based on the demand forecast (3).
\item Use the booking limits (4) to transform arriving requests (2) into booking patterns over the course of multiple consecutive simulated booking horizons.
\item Analyse booking patterns (5) to identify booking horizons with outlier demand.
\item Compare knowledge of the underlying demand model (2) to identified outliers (6) to compute detection rates.
\end{enumerate}

\begin{table}[t!]
\centering
\resizebox{\textwidth}{!}{%
\begin{tabular}{cghh} 
\hline \hline
\rowcolor{white} & \textbf{Symbol} & \textbf{Definition} &	\textbf{Regular Demand Value} \\ \hline
  & \(\mathcal{I}\) 			& \shortstack[l]{The set of customer types}  & \(\left\{\mbox{1 = Business, 2 = Tourist}\right\}\)  \\ 
\rowcolor{white}  & \(\mathcal{J}\) & The set of fare classes	& \(\left\{\mbox{A, O, J, P, R, S, M}\right\}\)	\\ 
 & \(\alpha\), \(\beta\) 	& \shortstack[l]{Parameters of Gamma distribution \\ for number arrivals}	 &  \(\alpha = 240, \beta = 1\) \\
\rowcolor{white} & \(a_i\), \(b_i\) 			& Parameters of Beta distribution, \(\acute{\lambda}_i(t)\) & \(a_1 = 5, b_1 = 2, a_2 = 2, b_2 = 5\) \\
\textbf{Fixed} & \(\phi_i\)	& \shortstack[l]{Proportion of total customer arrivals \\ stemming from type \(i\)}	& \(\phi_1 = 0.5, \phi_2 = 0.5\) \\ 
\rowcolor{white} \textbf{Input} & \(p_{ij}\)					& \shortstack[l]{Probability of type \(i\) being willing-to-pay \\ at most fare class \(j\)}	&	\shortstack[l]{ \(p_{1j} = \left\{0.35,0.1,0.25,0.15,0.05,0,0\right\}\) \\ \(p_{2j} = \left\{0.05,0.1,0,0.05,0.1,0.15,0.5\right\}\)}		\\ 
 & \(r_j\)	& Average fare for fare class \(j\)	&	
\(\left\{400, 300, 280, 240, 200, 185, 175 \right\}\) \\ 
\rowcolor{white} & C 	& Capacity	& 200 \\
 & $N_S$ 	& \shortstack[l]{Number of runs of simulation used \\ to compute forecasts \(\hat{\mu}_j\) and \(\hat{\sigma^2}_j\)}	& 100 \\ \hline \hline
\rowcolor{white} \textbf{Random}  & \(D\) 						 & \shortstack[l]{Total customer arrivals \( \sim Gamma(\alpha, \beta)\)} &  \\
\textbf{Input} & \(\lambda_i(t)\) 			& \shortstack[l]{Time-dependent rate of the Poisson \\ process of type \(i\) customer arrivals} 		& 		\\ \hline \hline
\rowcolor{white}  & \(x_{ij}(t)^{(n)}\) 			& \shortstack[l]{\(n^{th}\) realisation of  Poisson process of \\ type \(i\) customers purchasing \\ fare class \(j\) at time \(t\)}				 	&	\\
\textbf{Output} & \(\hat{\mu}_j\)				    & Forecast of mean of fare class \(j\) demand &						 		\\
\rowcolor{white} & \(\hat{\sigma^2}_j\)				& Forecast of variance of fare class \(j\) demand			&	 		\\
 & \(y_j(t)^{(n)}\) & \shortstack[l]{\(n^{th}\) realisation of cumulative bookings in \\ fare class \(j\) at time \(t\)}	&	\\ \hline \hline
\end{tabular}}
\caption{Table of notation and parameter values used for simulation}
\label{tab:simnotation}
\end{table}

Table \ref{tab:simnotation} sets out the notation used in the remainder of this section to detail the demand model, demand forecasting, revenue maximisation heuristics, and booking limits. In this, we detail both the models and algorithms, and the parameter settings implemented in the computational study.

\subsection{Generating Demand in Terms of Customer Requests} 
Heterogeneous demand is a frequently stated RM precondition, assuming that customer segments differ in value and can be identified through their idiosyncratic booking behaviour. To model this parsimoniously, the simulation features two customer types but can be easily extended to feature more. We index any parameter that characterises high-value customers with index $1$ and any parameter that characterises low-value customers with index $2$. Classical RM assumes that requests from high-value customers typically arrive later in the booking horizon than those from low-value customers. High-value customers are more likely to book expensive fare classes when cheap fare classes are not offered. 

We follow \citet{Weatherford1993} in modelling requests from either customer type as arriving according to a non-homogeneous Poisson-Gamma process. Requests from customer type $1$ arrive according to a Poisson(\(\lambda_1(t)\)) distribution; those from customer type $2$ arrive according to a Poisson(\(\lambda_2(t)\)) distribution. The total number of customer arrivals \(D\) is split between the two segments, such that 
\begin{eqnarray} \label{eqn:arrivals}
\lambda_1(t)|(D=d) &=& d \times \phi_1 \frac{t^{a_1-1}(1-t)^{b_1-1}}{B(a_1,b_1)}, \\ \label{eqn:arrivals2}
\lambda_2(t)|(D=d) &=& d \times \phi_2 \frac{t^{a_2-1}(1-t)^{b_2-1}}{B(a_2,b_2)},
\end{eqnarray}
where \(D \sim \mbox{Gamma}(\alpha,\beta)\) with probability density function:
\begin{equation}
    f(d|\alpha, \beta) = \frac{\beta^{\alpha}}{\Gamma(\alpha) d^{\alpha -1}e^{\beta d}}.
\end{equation}

\begin{figure}[htbp]
        \centering
        \begin{subfigure}[b]{0.47\textwidth}
            \centering
            \includegraphics[width=0.8\textwidth,height=0.8\textwidth]{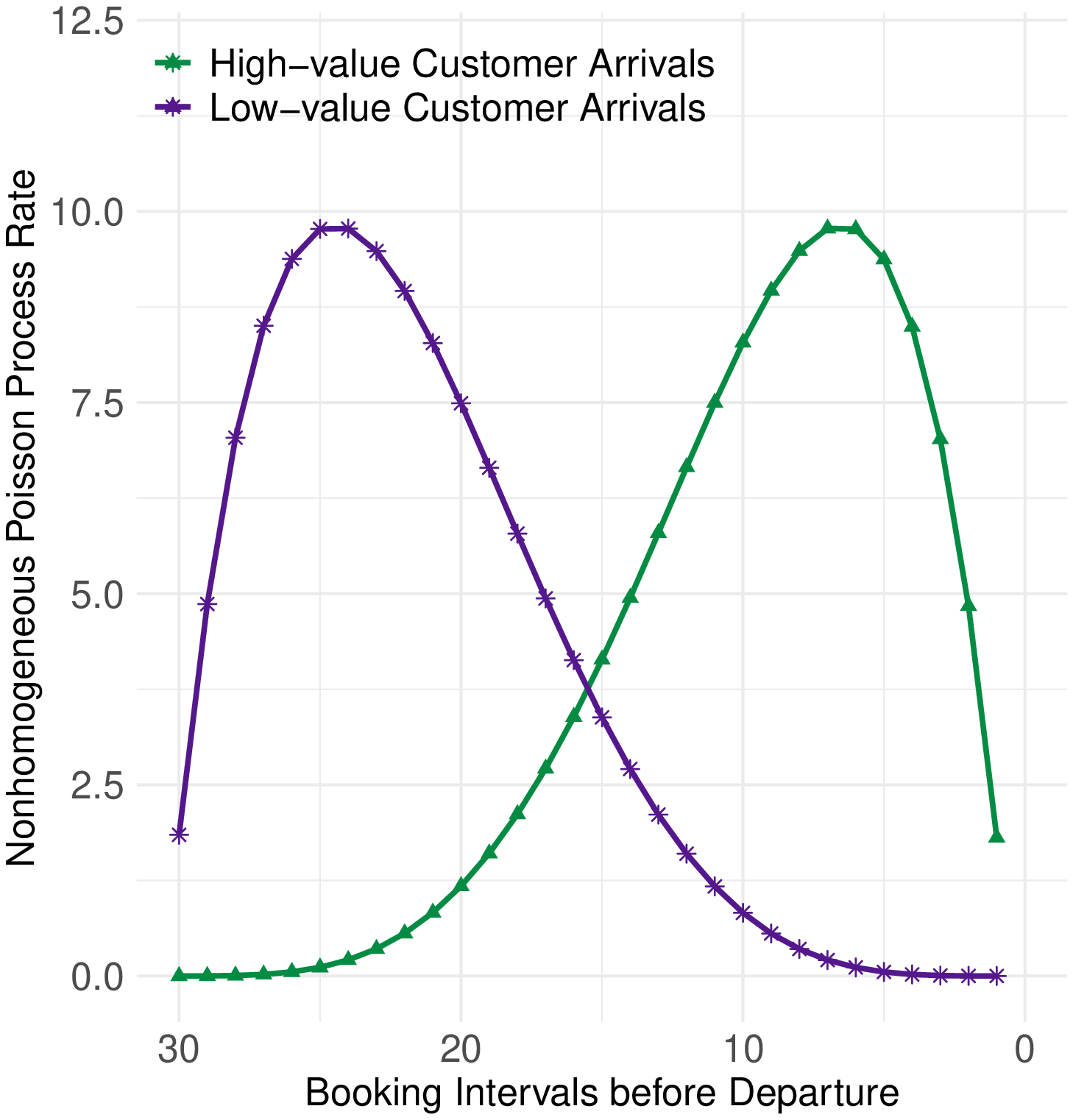}
            \caption{\(\lambda_i(t)\), Arrival rates per \\ customer type}  
					\label{fig:arrivals}
        \end{subfigure}
        \begin{subfigure}[b]{0.47\textwidth}  
            \centering 
            \includegraphics[width=0.8\textwidth,height=0.8\textwidth]{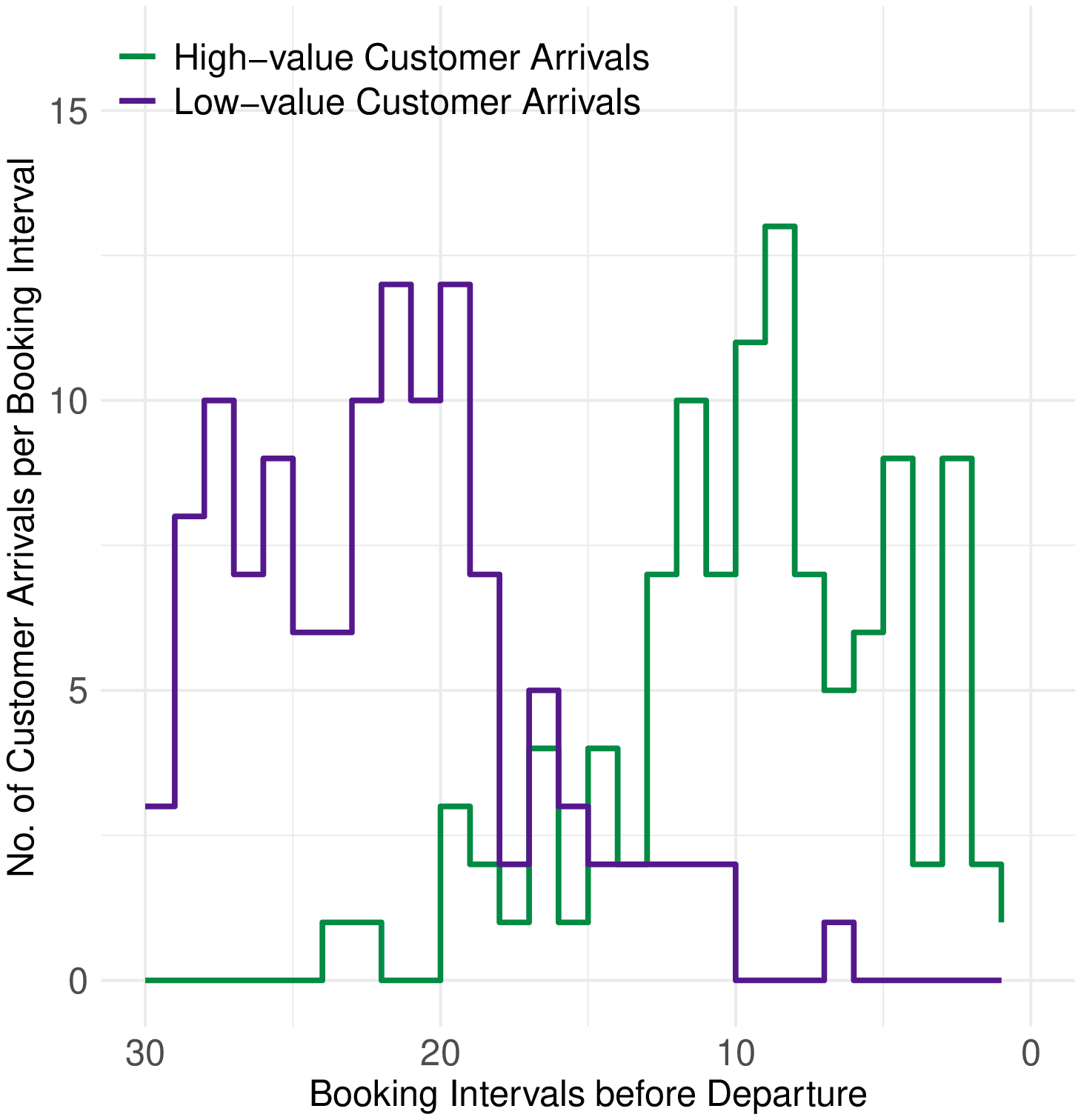}
            \caption{\(x_i(t)\), Realisation of the Poisson process with rate \(\lambda_i(t)\) per customer type}
					\label{fig:arrivals2}
        \end{subfigure}
        \quad
        \begin{subfigure}[b]{0.47\textwidth}
            \centering
            \includegraphics[width=0.8\textwidth,height=0.8\textwidth]{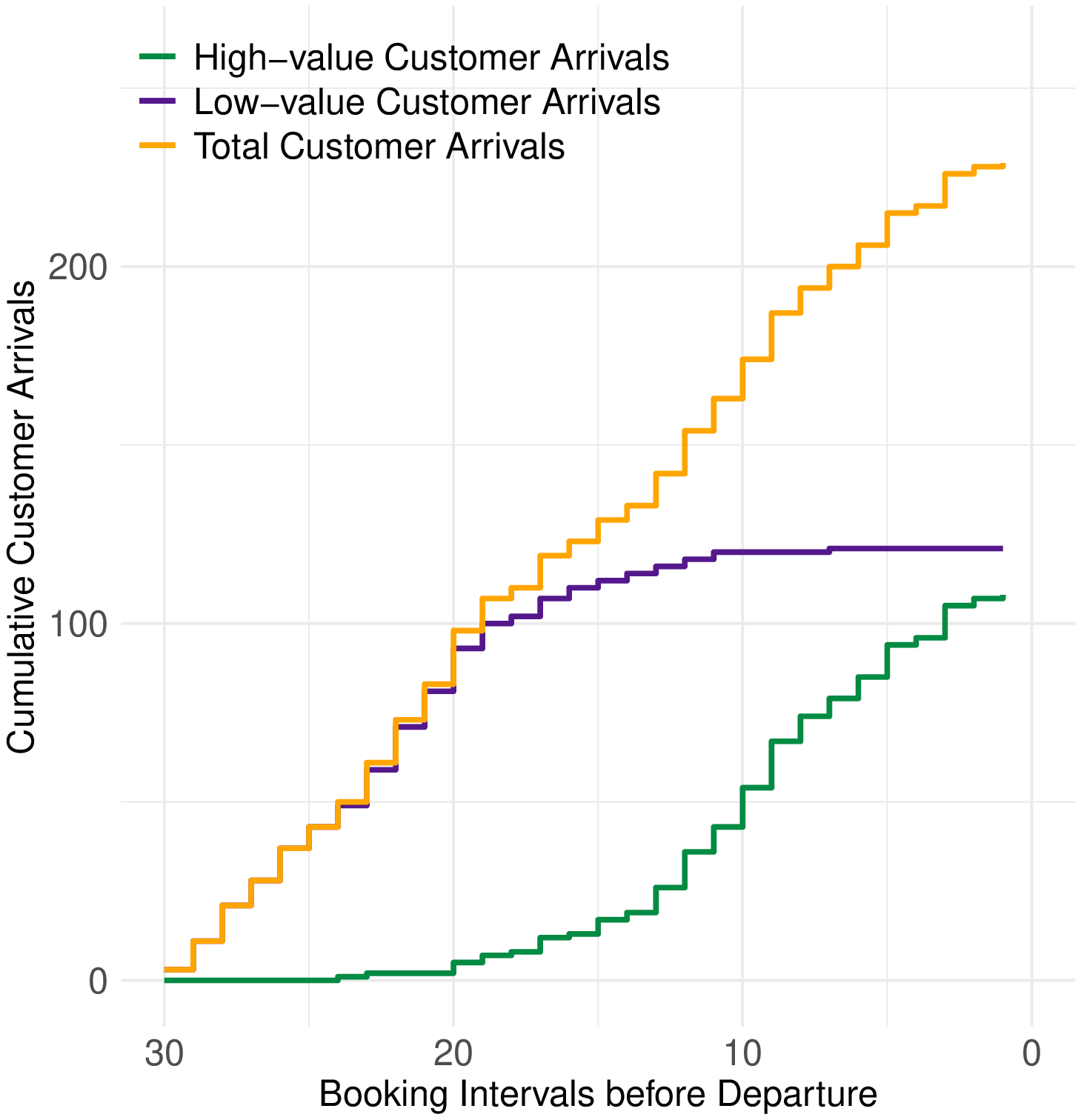}
            \caption{\(\sum_{k \leq t} x_i(k)\), Cumulative number of customer arrivals per customer type}
						\label{fig:cumarrivals}
        \end{subfigure}
        \begin{subfigure}[b]{0.47\textwidth}  
            \centering 
            \includegraphics[width=0.8\textwidth,height=0.8\textwidth]{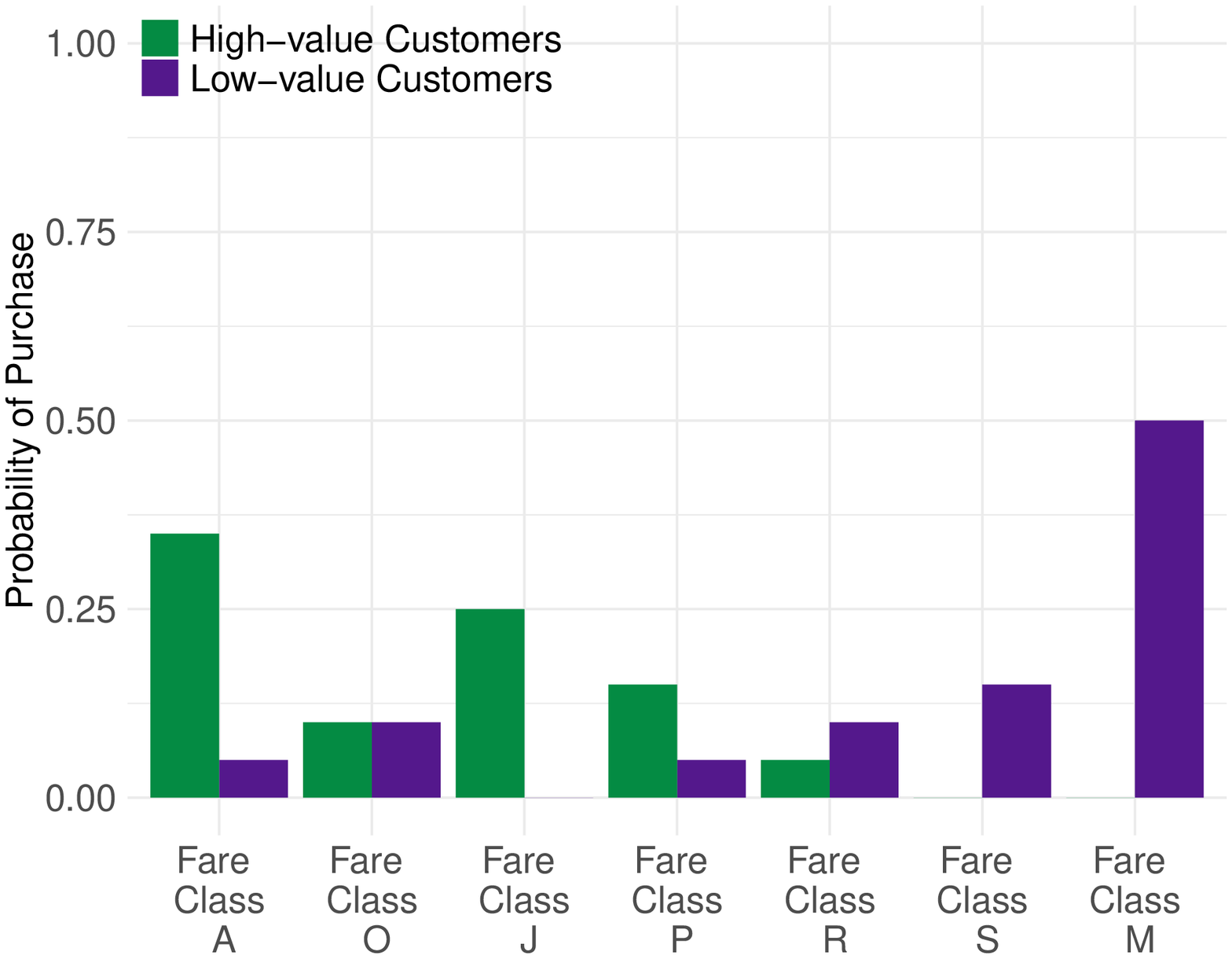}
            \caption{\(p_{ij}\), Probability of a customer of type \(i\) being willing to pay up to fare class \(j\)}	\label{fig:purchaseprobs}
        \end{subfigure}
        \quad
        \caption{Customer arrivals generated by a nonhomogeneous Poisson-Gamma process \\ \(D \sim Gamma(240, 1)\mbox{, }\phi_1 = \phi_2 = 0.5\mbox{, } a_1 = 5\mbox{, }b_1 = 2\mbox{, }a_2 = 2\mbox{, } b_2 = 5\)}
        \label{fig:nhpparrivals}
\end{figure}

The constraint \(\phi_1 + \phi_2 = 1\) ensures that all requests belong to exactly one customer type. Additionally, we set parameters \(a_1\), \(b_1\), \(a_2\) and \(b_2\) such that they follow the assumption that valuable customers are more likely to request at later stages of the booking horizon:
\begin{equation}
\frac{a_1 - 1}{a_1 + b_1 - 2} > \frac{a_2 - 1}{a_2 + b_2 - 2}
\end{equation}
Figure \ref{fig:arrivals} illustrates arrival rates \(\lambda_1(t)\) and \(\lambda_2(t)\) across the booking horizon, with Figure \ref{fig:arrivals2} showing one realisation of request arrivals in a specific horizon.

A set of fare classes, \(1, \hdots, |\mathcal{J}|\) differentiates discount levels, \(r_1 \geq r_2 \hdots \geq r_{|\mathcal{J}|}\). The simulation implements a random choice model to let customers choose from the set of currently offered classes. The model assumes all customers book the cheapest available fare class. At the same time, not all customers can afford to book any fare class. For every fare class \(k\), the probability that a customer of type \(i\) is willing to pay {\em at most} fare class \(k\) is \(p_{ik}\), as shown in Figure \ref{fig:purchaseprobs}. Each customer has a single fare class threshold, which is the most they are willing to pay, such that:
\begin{equation}
    \sum_{k=1}^{|\mathcal{J}|}p_{ik} + p_{i0} = 1
\end{equation}
where \(p_{i0}\) is the the probability of a type \(i\) customer arriving and choosing not to book based on the classes on offer. Hence, the probability of booking fare class \(j\) is: 
\begin{eqnarray}
    \Pro \left(\mbox{Book fare class }j | \mbox{No availability in classes } j + 1, \hdots, |\mathcal{J}|\right) &=& \sum_{k=1}^{j} p_k \\
    \Pro \left(\mbox{Book fare class }j | \mbox{Availability in classes } j + 1, \hdots, |\mathcal{J}|\right) &=& 0
\end{eqnarray}
where \(p_{k}\) is the weighted average of probabilities of each customer type $i$ being willing to pay up to fare class \(k\):
\begin{equation}
p_k = \sum_{i \in \mathcal{I}} \phi_i p_{ik},
\end{equation}
and \(\phi_i\) is the proportion of total customer arrivals stemming from type \(i\).

While demand arrival rates vary across the booking horizon, the simulation models arrival rates and choice probabilities as stationary between booking horizons. While, in real-world markets, demand shifts in seasonal patterns and trends, we rely on random draws from distributions with stationary parameters as when introducing and detecting outliers, the simplest case lets all \textit{regular} demand behaviour derive from the same distribution. When an approach cannot correctly detect abnormal demand when all regular demand comes from this same distribution, it is highly unlikely that it will perform better when regular demand is non-stationary. 

\subsection{Outlier Generation} \label{sec:outliergen}
We generate outlier demand by parameterising demand generation in a way that deviates from the regular setting. Combining outlier demand with booking limits (optimised based on forecasts of regular demand) creates an outlier booking pattern. Outliers can result from three approaches to adjusting the parameters in Equations \eqref{eqn:arrivals} and \eqref{eqn:arrivals2}, and the probabilities, \(p_{ij}\):
\begin{enumerate}
\item \emph{Demand-volume outliers}: Increasing or decreasing the volume of demand across the whole (or partial) booking horizon, by adjusting the parameters \(\alpha\), and \(\beta\) in the Gamma distribution for \(D\), the total demand.
\item \emph{Willingness-to-pay outliers}: Shifting the proportions of demand across fare classes, by either adjusting the choice probabilities per customer type or to the ratio of customer types, \(\phi_1, \phi_2\).
\item \emph{Arrival-time outliers}:  Shifting the arrival pattern of customer requests (from a subset of customer types) over time by adjusting parameters \(a_1, b_1, a_2, b_2\), which control the time at which requests from each customer type arrive.
\end{enumerate}

\subsection{Forecasting Demand}
Most RM approaches to capacity control rely on knowing the number of expected customer requests per offered product, potentially per set of offered products. The simulation implements heuristics that rely on the mean and the variance of expected requests per fare class (compare Section \ref{sec:HeuristicRevenueOptimisation}).

To avoid interference from arbitrary forecasting errors, we exploit knowledge of the demand model given in the simulation setting when creating the forecast: We first draw \(N_S\) sets of customer arrivals from Equations \eqref{eqn:arrivals} and \eqref{eqn:arrivals2}. Let \(x_{ij}(t)^{(n)}\) define the \(n^{th}\) realisation of type \(i\) customers who booked in fare class \(j\) at time \(t\) as drawn from the Poisson arrival process with rate \(\lambda_i(t)\), and probability \(p_{ij}\). Then, we set the forecast to be the mean demand across all customer types upon departure from \(N_S\) simulations for fare class \(j\), \(\hat{\mu_j}\):
\begin{equation}
\hat{\mu_j} = \frac{1}{N_S} \sum_{n=1}^{N_S} \left(\sum_{t \in \mathcal{T}} \sum_{i \in \mathcal{I}} x_{ij}(t)^{(n)}\right).
\end{equation}
Similarly, the simulation forecasts the variance of the demand for fare class \(j\) as:
\begin{equation}
\hat{\sigma_j}^2 = \frac{1}{N_S-1} \sum_{n=1}^{N_S} \left\{ \left[ \left( \sum_{t \in \mathcal{T}} \sum_{i \in \mathcal{I}} x_{ij}(t)^{(n)}\right) - \hat{\mu_j} \right]^2 \right\}.
\end{equation}
Here, we aggregate across the booking horizon in order to obtain forecasts for the final demand for each fare class. The resulting sum of customer requests per fare class across customer types gives the total expected demand per fare class. The mean and variance of these $N_S$ realisations are taken to be the forecasted parameters of a Normal distribution for each fare class demand. 

Note that this aggregated forecast deliberately prepares the heuristic applied for revenue optimisation in this case. Applying, for instance, a dynamic program to optimally control arriving customer requests, would require a forecast of customer arrival rates and choice probabilities. The consequence of outliers, however, would be the same, as the arrival rates and choice probabilities deviate for demand outliers.

Last but not least, the simulation forecast assumes stationarity of demand, which is correct with regard to the demand setting simulated here. Therefore, a single forecast value is predicted for all future booking periods. Naturally, in a real-world setting, this stationarity is not given, but instead trends and seasonality complicate forecasting. Future research featuring such forecast aspects  would open the path to further differentiation with regard to the effects of different types of outliers given different parameterisations of the non-stationary components.

\subsection{Heuristic Revenue Optimisation}
\label{sec:HeuristicRevenueOptimisation}
The simulation implements two well-known heuristic methods for obtaining booking controls for a single resource: EMSRb and EMSRb-MR. We pick these heuristics for their wide acceptance and pervasive use in practice. Furthermore, as opposed to, e.g., exact dynamic programming formulations, these heuristics yield the booking limits widely implemented in current practice. We expect the nature of these booking limits and their updates to be a relevant factor for the recognition and compensation of demand outliers.

\begin{itemize}[leftmargin=*]
\item \emph{EMSRb}, Expected Marginal Seat Revenue-b, was introduced by \citet{Belobaba1992}. EMSRb calculates joint protection levels for all more expensive classes relative to the next cheaper fare class, based on the mean expected demand and its variance. 
\item \emph{EMSRb-MR}: To make the EMSRb heuristic applicable when demand depends on the set of offered fare classes, e.g. when customers choose the cheapest available class, \citet{Fiig2010} introduce this variant. It applies a marginal revenue transformation to demand and fares before calculating the EMSRb protection levels based on transformed fares and predicted demand. 
\end{itemize}

Booking limits can be implemented in either a \textit{partitioned} or \textit{nested} way (\citet{Brumelle1993}, and \citet{Talluri2004}, Chapter 2). Partitioned controls assign capacity such that each unit can only be sold in one specific fare class. Conversely, nested controls let assignments overlap in a hierarchical manner i.e. units of capacity assigned to one fare class can also be sold in any more expensive fare class. Thus, nested booking limits ensure that for any offered class, all more expensive classes are also offered---as this seems an intuitive goal these booking limits are much more commonly used. Therefore, the simulation implements nested controls. 

\subsection{Evaluation of Outlier Detection}
We regard outlier detection as a binary classification problem, where the two classes are \textit{regular booking patterns} and \textit{outlier booking patterns}. By definition, for any pattern generated in the simulation, we know the true class, as we know the underlying demand model.

Several indicators evaluate the performance of binary classification outcomes, as surveyed by \citet{Tharwat2018}. Each outcome falls into one of four categories: (i) if a genuine outlier is correctly classified, it is a \textit{true positive (TP)}; (ii) if a regular observation is correctly classified, it is a \textit{true negative (TN)}; (iii) if a regular observation is wrongly classified as an outlier, it is a \textit{false positive (FP)}; and (iv) if a genuine outlier is wrongly classified as regular, it is a \textit{false negative (FN)}. 

To analyse results in this paper, we implement the \textbf{Balanced Classification Rate (BCR)} as suggested by \citet{Tharwat2018}. This indicator accounts for both the average of the true positive rate and true negative rate:
\begin{equation}
BCR = \frac{1}{2} \left(\frac{TP}{TP + FN} + \frac{TN}{TN + FP} \right).
\end{equation}  
The notions of high detection rates (fraction of genuine outliers which are correctly detected) and low false positive rates (fraction of regular observations which are incorrectly labelled as outliers) create conflicting objectives. For example, a high true positive rate does not necessarily indicate a high performing algorithm, if it is accompanied by a high false positive rate. Therefore, combining both into a single figure is useful. Nonetheless, additional results on true positive rates, false positive rates, and positive likelihood ratios \citep{Habibzadeh2019} are included in Appendices C.4 and C.5. Typically, the number of outliers is outweighed by the number of normal observations. This leads to one class being significantly larger than the other. BCR is robust to this imbalance. 

Additionally, we generated a \textbf{receiver operating characteristic (ROC)} curve by plotting the true positive rate against the false positive rate \citep{McNeil1984}. This provides an additional diagnostic for binary classifiers. The ROC curve compares the true positive to false positive ratio as the threshold (at which an outlier is classified) varies. The optimal ROC curve is that with the combination of highest true positive rate and the the lowest false positive rate i.e. with an area under the ROC curve closest to 1.

\subsection{Experimental Set-up}
We vary two main elements of the experimental set up for experimental analysis. The first is the parameter settings used to generate regular and outlier demand. The second are the settings of outlier detection.

We generate regular demand according to the parameters in Table \ref{tab:simnotation}, which results in regular total demand with a mean of 240, and standard deviation of 15.492. We benchmark detection performance on outlier demand generated in various ways. Our main focus is on analysing different magnitudes of demand-volume outliers. Our choice of parameter changes for outlier generation follows \citet{Weatherford2002}, who investigate the effects of inaccurate demand forecasts on revenue. In particular, they consider cases where forecasts are 12.5\% and 25\% higher or lower than the actual demand. We perform a similar analysis on the benefits of detecting outliers where the overall number of customers deviates from regular demand by \(\pm\) 12.5\% and \(\pm\) 25\%.  The four types of demand-volume outliers we consider are generated by varying the parameters \(\alpha\) and \(\beta\) as described in Table \ref{tab:params_mags}. This results in a change in mean of the desired magnitude and direction, but no change in variance. In addition, we consider other types of outliers, as outlined in Section \ref{sec:types}.
\begin{table}[htbp]
\centering
\begin{tabular}{r|cccc}
\hline \hline 
\textbf{}  & \textbf{Mean} & \textbf{Std. Dev} & \textbf{$\alpha$} & \textbf{$\beta$}     \\ \hline
\textbf{Regular Demand} & 240 & 15.492 & 240  & 1               \\
\textbf{25\% Increase} & 300  & 15.492              & 375             & 1.25                         \\
\textbf{12.5\% Increase} & 270   & 15.492               & 303.75              & 1.125                       \\
\textbf{12.5\% Decrease} & 210  & 15.492               & 183.75              & 0.875                      \\
\textbf{25\% Decrease} & 180  & 15.492  & 135              & 0.75         \\   \hline \hline
\end{tabular}
\caption{Parameter choices used to generate demand-volume outliers}
\label{tab:params_mags}
\end{table}

In a wide-ranging computational study, we compared the performance of all outlier detection methods described in Section \ref{sec:background}. Appendix B, Table 3 lists the aggregated results from all experiments carried out. For conciseness, the results discussed here focus on the \textit{best} univariate method, {\em parametric (Poisson) tolerance intervals}; the best multivariate method, {\em \(K\)-means clustering with Euclidean distance}; the best functional method, {\em functional depth}; and the best extrapolation method, {\em ARIMA extrapolation combined with functional depth}. 

The settings used for these four methods are as follows:
\begin{itemize}
    \item \emph{Parametric tolerance intervals}: The distribution chosen is Poisson, see Appendix A.1 for details. The coverage proportion is chosen to be \(\beta = 0.95\), and the confidence level is \(\alpha = 0.05\) by default.  \vspace{-0.3cm}
    \item \emph{$K$-means clustering}: The number of clusters, $K$, is chosen to be 2, see Appendix A.1 for reasoning. The default threshold for classifying a booking pattern as an outlier is half the sum of the maximum and minimum distances of the patterns from their cluster centres \citep{DebDay2017}. \vspace{-0.3cm}
    \item \emph{Functional depth}: The number of bootstrap samples for the threshold is chosen to be 1000. The smoothing method is as suggested by \citet{Febrero2008}. Similarly, the percentile chosen for this analysis is the $1^{st}$ percentile, as suggested by \citet{Febrero2008}. \vspace{-0.3cm}
    \item \emph{Functional depth + ARIMA extrapolation}: Thresholds are calculated as in functional depth. The orders of the ARIMA extrapolation are selected using \texttt{auto.arima} in R, based on AICc, with the augmented Dickey-Fuller test used to choose the order of differencing. The parameters are estimated using maximum likelihood with starting values chosen by conditional-sum-of-squares. 
\end{itemize}
We provide further details on the extent of the computational study, including aggregated results, in Appendix B.

\section{Results} \label{sec:results}
To investigate different outlier simulation and detection techniques, we follow a four-step process. We contrast foresight detection performance of different outlier detection methods in Section \ref{sec:within}. This analysis focuses on detection performance across the booking horizon, and evaluates the detection approaches' ability to detect outliers early in the booking horizon. We also quantify the gain in outlier detection performance resulting from the inclusion of the extrapolation step proposed in Section \ref{sec:functional}. Subsequently, Section \ref{sec:types} investigates the effect of different types of outliers on the performance of the outlier detection method. Additionally, Section \ref{sec:db} considers an empirical data set to demonstrate the practical implications of the approach. Finally, Section \ref{sec:improvement} presents a final set of experiments intended to measure the potential increase in revenue generated by analysts correctly taking actions based on alerts from the proposed method of outlier detection. Note that all experiments analysed in this section implement the EMSRb-MR heuristic, which is a better fit with the given demand model. We have investigated the implications of applying the EMSRb heuristic and assessed the revenue generated as well as the effect on identifying outliers in an ancillary study. The results can be found in Appendix C.1.


\subsection{Benchmarking Foresight Detection of Demand-volume Outliers} \label{sec:within}
\begin{figure}[htbp]
    \centering
        \begin{subfigure}[b]{0.47\textwidth}
            \centering
            \includegraphics[width=0.8\textwidth,height=0.8\textwidth]{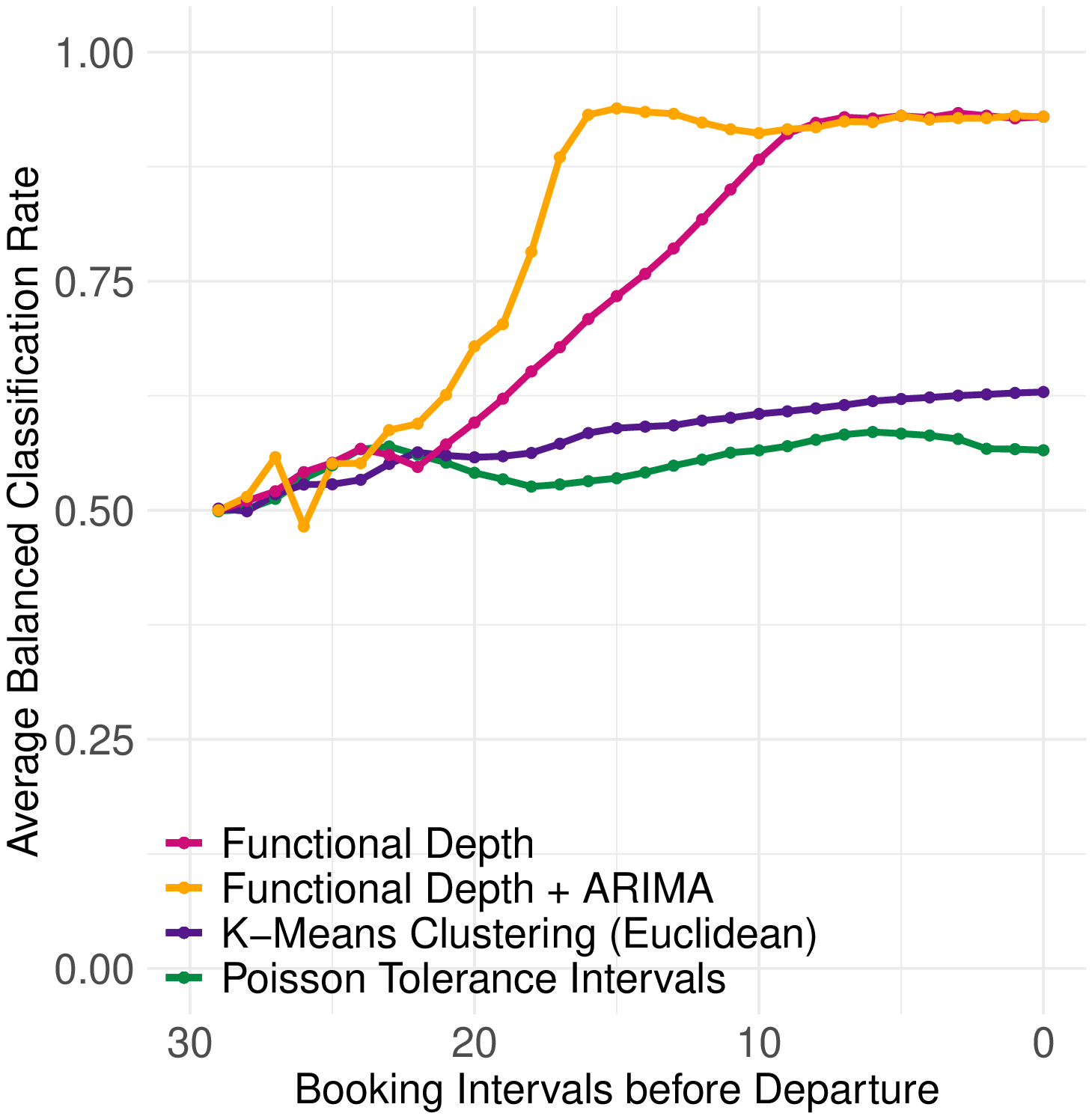}
            \caption{Comparison of best performing outlier detection methods}  
			\label{fig:online}
        \end{subfigure} \hspace{0.4cm}
        \begin{subfigure}[b]{0.47\textwidth}  
            \centering 
            \includegraphics[width=0.8\textwidth,height=0.8\textwidth]{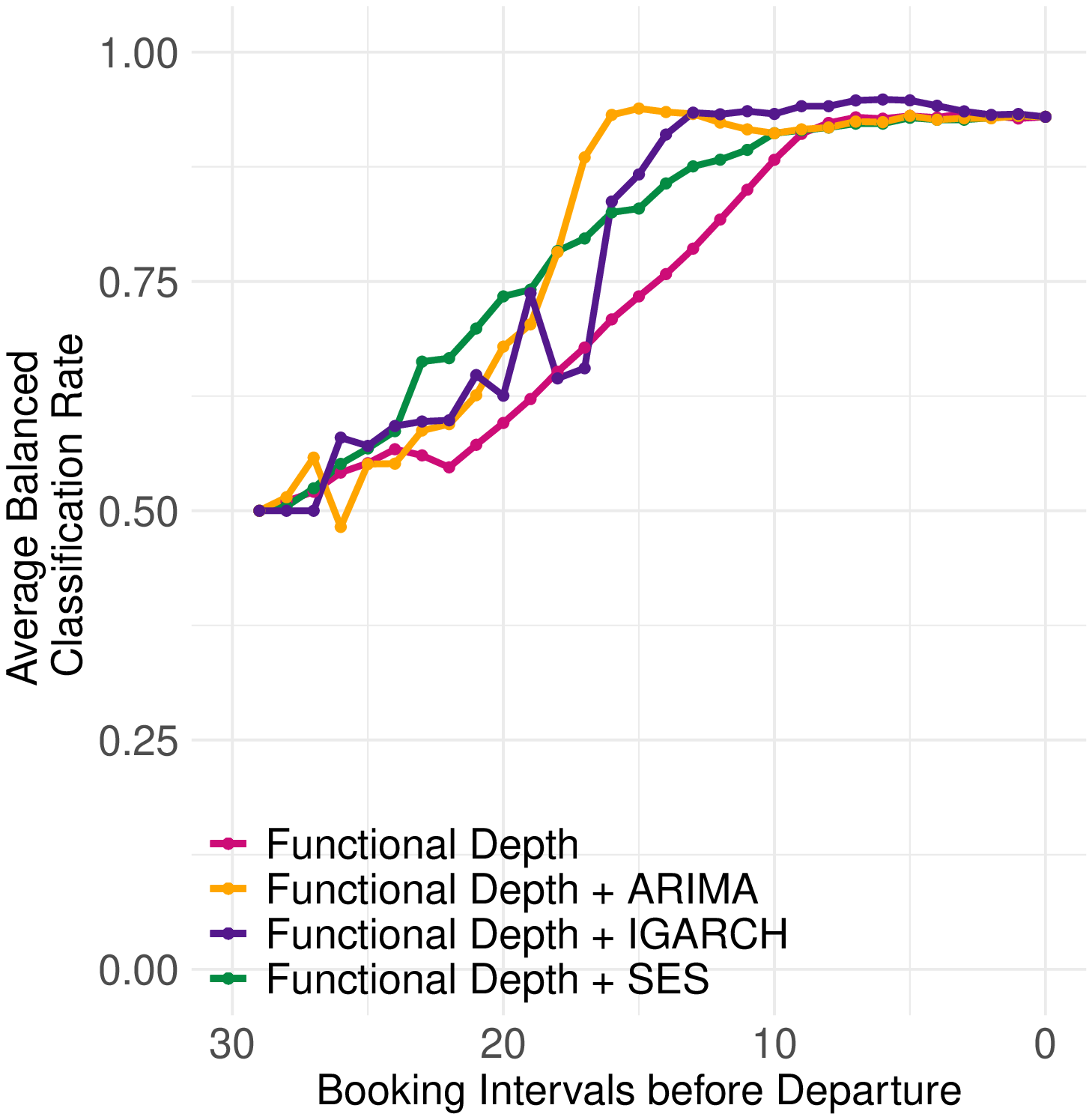}
            \caption{Improvement from incorporating extrapolation}
			\label{fig:extrapolation}
        \end{subfigure}
        \quad
        \caption{Comparison of foresight outlier detection averaged over different magnitudes of demand outliers with 5\% outlier frequency}
        \label{fig:online2}
\end{figure}

To evaluate {\em foresight detection} performance, Figure~\ref{fig:online} displays the average BCR per booking interval. Very early in the booking horizon, all four methods suffer from poor performance but for different reasons -- some suffer from low true positive rates, others from high false positive rates. See Appendices C.4 and C.5 for details. At around 21 booking intervals before departure, the average BCR of functional methods quickly accelerate towards 1, whereas the univariate and multivariate approaches at best only show mild improvements in classification performance.

\begin{figure}[!h]
    \centering
        \begin{subfigure}[b]{0.47\textwidth}
            \centering
            \includegraphics[width=0.8\textwidth,height=0.8\textwidth]{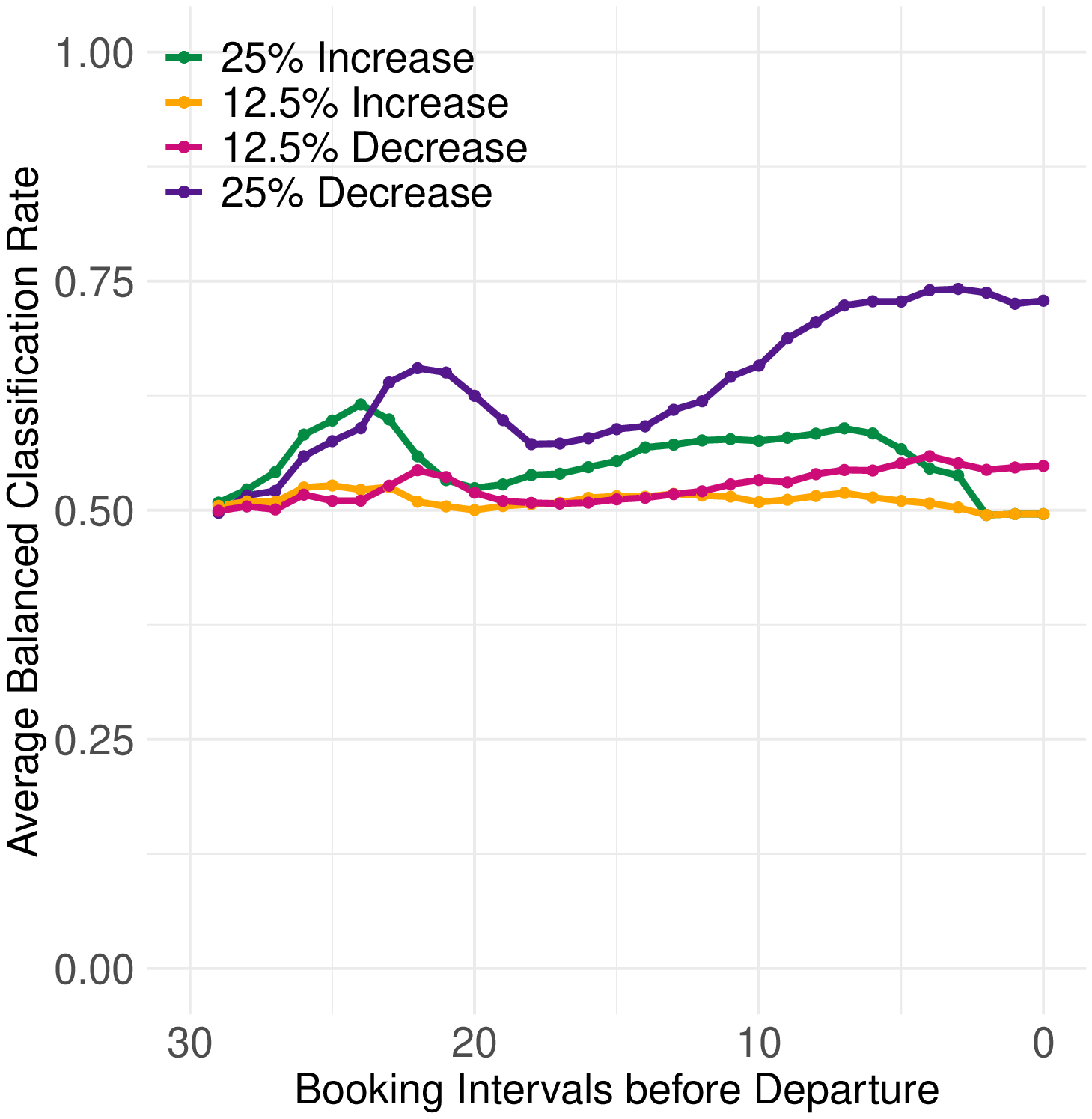}
            \caption{Parametric (Poisson) tolerance intervals outlier detection}  
			\label{fig:pois}
        \end{subfigure} \hspace{0.4cm}
        \begin{subfigure}[b]{0.47\textwidth}  
            \centering 
            \includegraphics[width=0.8\textwidth,height=0.8\textwidth]{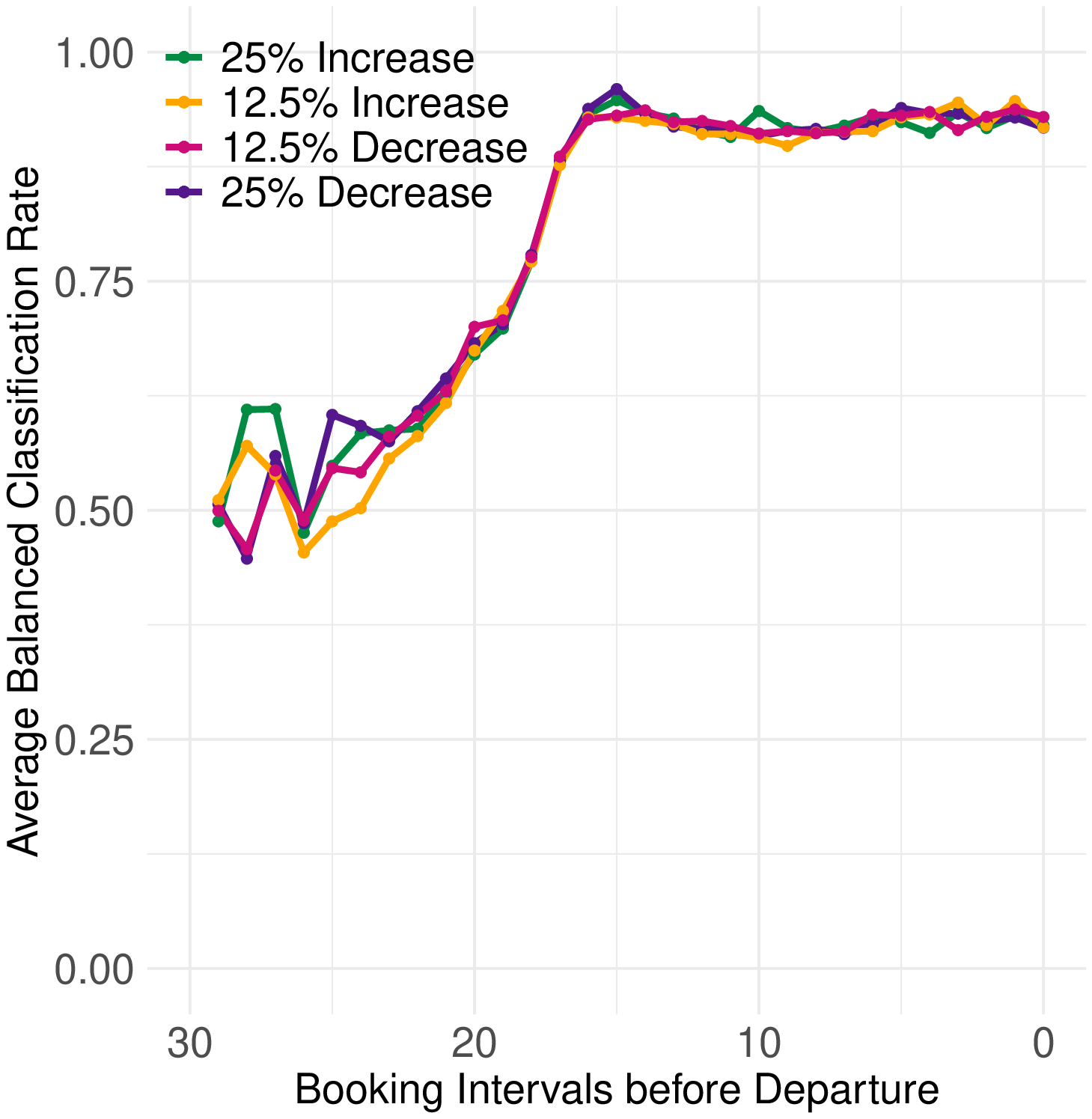}
            \caption{Functional halfspace depth with ARIMA extrapolation outlier detection}
			\label{fig:halfarima}
        \end{subfigure}
        \quad
        \caption{Balanced Classification Rate under different magnitudes of outliers with 5\% outlier frequency}
        \label{fig:types}
\end{figure}

Including ARIMA extrapolation markedly accelerates classification performance, especially between 20 and 10 booking intervals before departure. Note that the aim of extrapolation is not necessarily to increase the overall BCR, but to achieve peak performance earlier in the booking horizon to gain time for analyst intervention. The extrapolation achieves this by increasing the variance of the booking patterns, leading to an increase in the number classified as outliers. See Appendix C.6 for further detail. Additional analysis of Receiver Operating Characteristic (ROC) curves (see Section \ref{sec:roc}), further support the inclusion of ARIMA extrapolation. In Figure \ref{fig:extrapolation}, we also compare functional depth with IGARCH and SES extrapolation, and similar improvements are observed as with ARIMA extrapolation. ARIMA provides larger gains in performance compared to SES and IGARCH. This is likely due to the flexibility of ARIMA in capturing the changing curvature of the booking pattern, and its ability to encapsulate the autocorrelations induced by censoring from the booking limits. In the last third of the booking horizon, the extrapolation makes up a much smaller part of the pattern, i.e. most of the pattern is now made up of observed rather than extrapolated data. Hence, the input data to the outlier detection algorithm with different extrapolations is similar, and so they produce similar results. Further analysis on the relationship between extrapolation accuracy and the resulting improvement in outlier detection is available in Appendix C.8.

As noted in Section \ref{sec:functional}, extrapolation could also be combined with multivariate outlier detection methods such as $K$-means clustering. Given the superior performance of functional depth we focus our main results on this combination, but additional results regarding combining with multivariate techniques are presented in Appendix C.3.

\subsection{Receiver Operating Characteristic (ROC) Curves} \label{sec:roc}
\begin{figure}[!ht]
    \centering
    \begin{subfigure}[h]{0.42\textwidth}
    \centering
        \includegraphics[width=0.6\textwidth,height=0.6\textwidth]{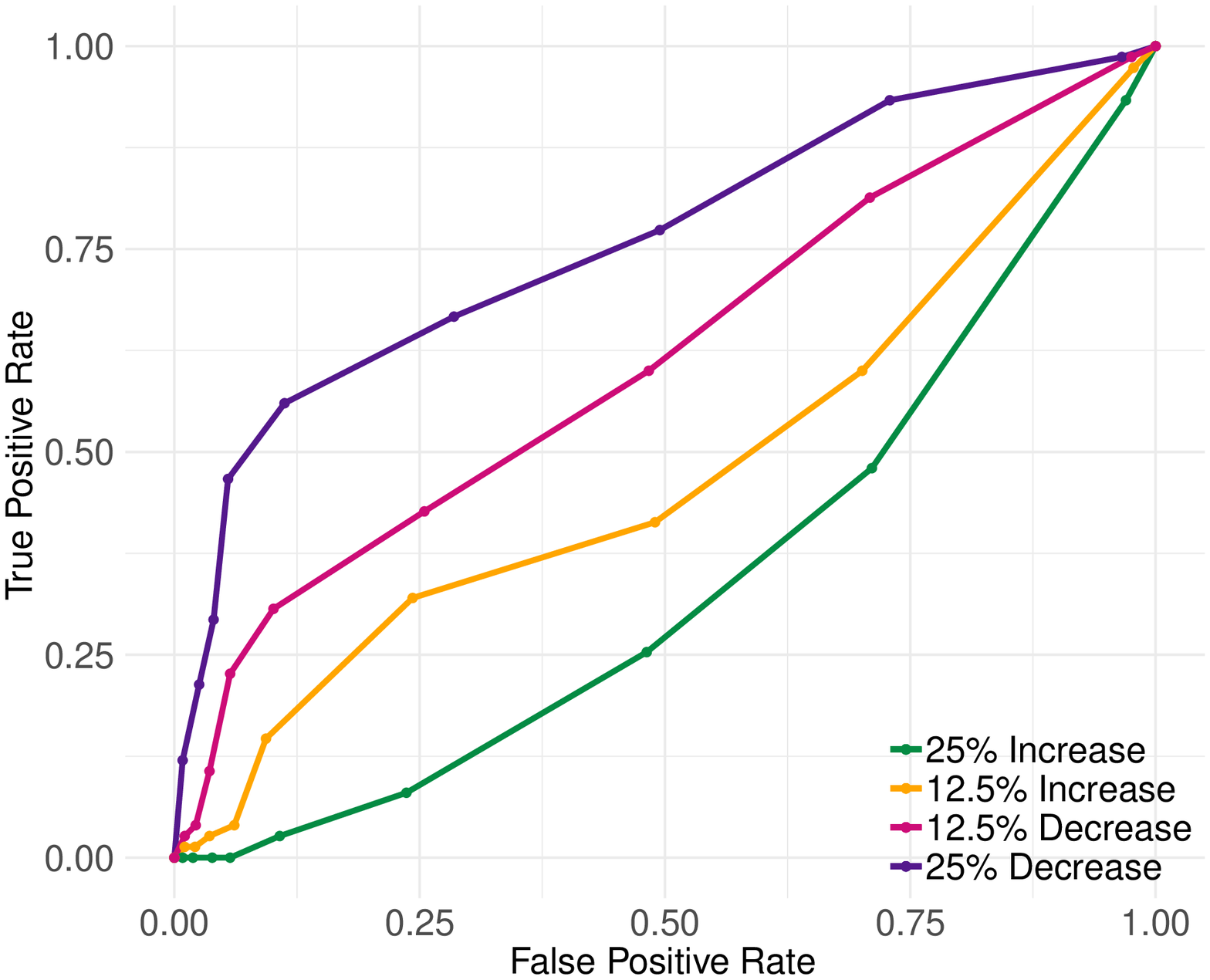}
        \caption{$K$-means clustering outlier detection at 20 booking intervals before departure}  
		\label{fig:roc20KM}
    \end{subfigure} \hspace{0.5cm}
    \begin{subfigure}[h]{0.42\textwidth}  
    \centering 
        \includegraphics[width=0.6\textwidth,height=0.6\textwidth]{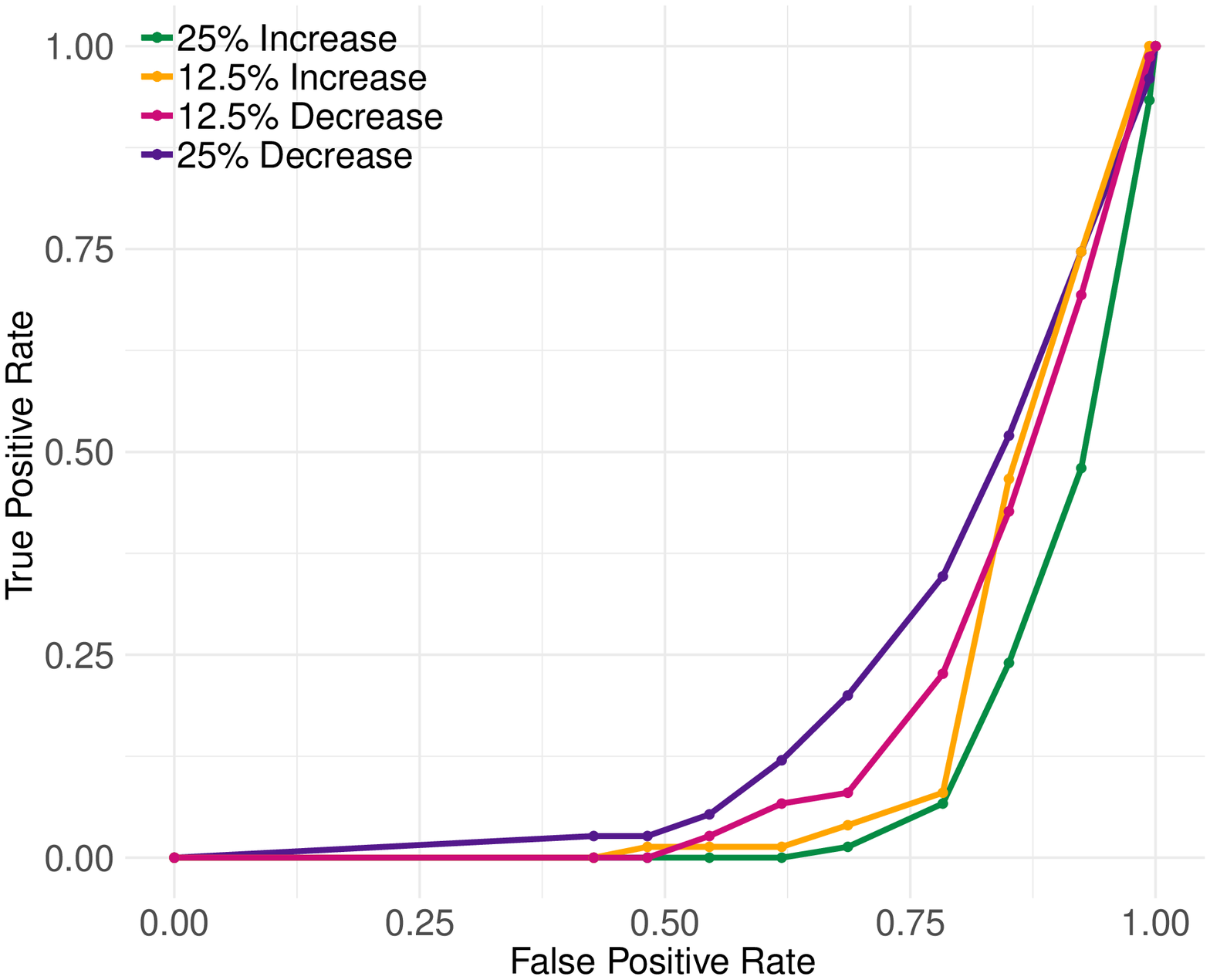}
        \caption{$K$-means clustering  outlier detection at 10 booking intervals before departure}
		\label{fig:roc10KM}
    \end{subfigure}
    \quad
    \begin{subfigure}[h]{0.42\textwidth}
    \centering
        \includegraphics[width=0.6\textwidth,height=0.6\textwidth]{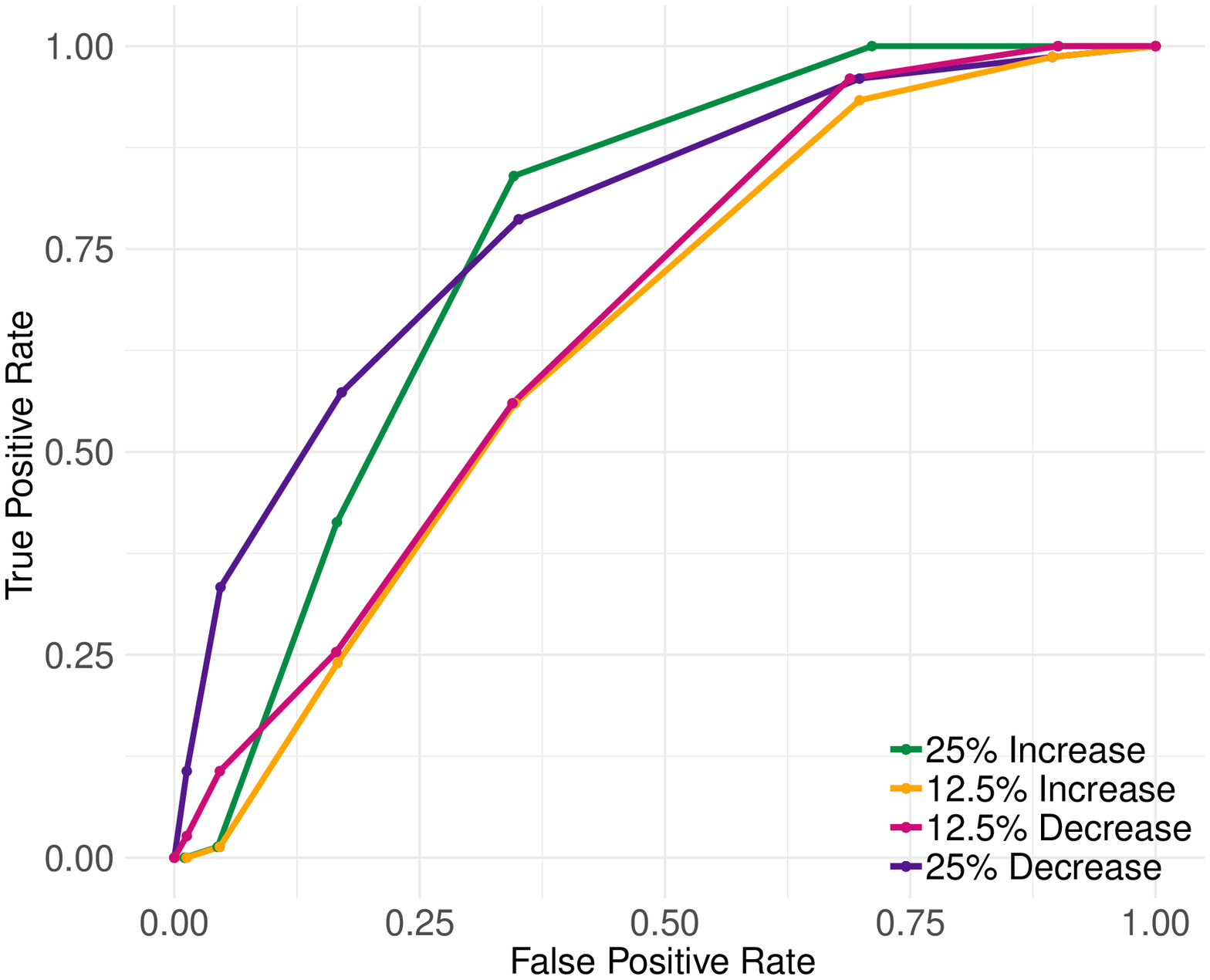}
        \caption{Functional depth outlier detection at 20 booking intervals before departure}  
		\label{fig:roc20}
    \end{subfigure} \hspace{0.5cm}
    \begin{subfigure}[h]{0.42\textwidth}  
    \centering 
        \includegraphics[width=0.6\textwidth,height=0.6\textwidth]{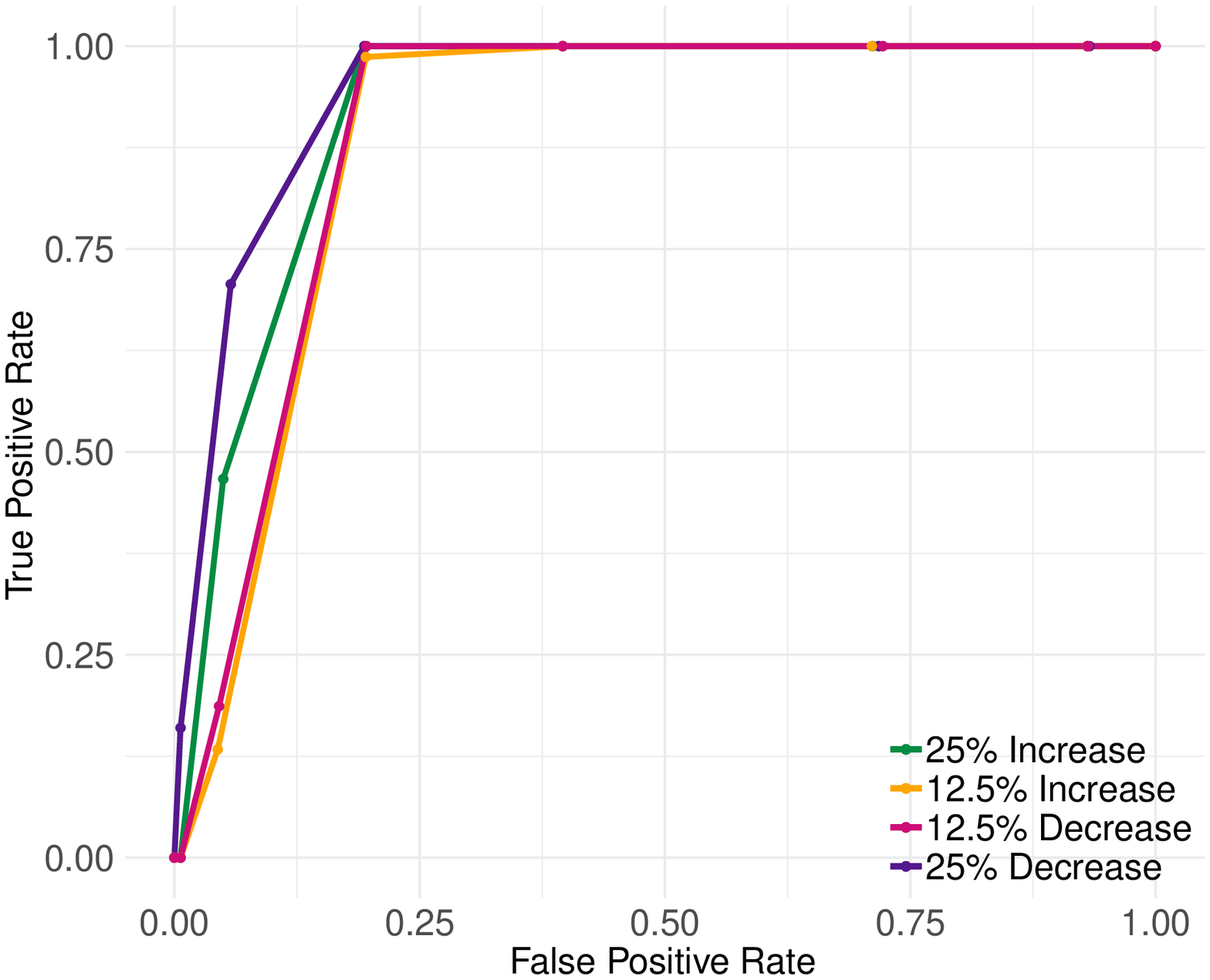}
        \caption{Functional depth outlier detection at 10 booking intervals before departure}
		\label{fig:roc10}
    \end{subfigure}
    \quad
    \begin{subfigure}[h]{0.42\textwidth}
    \centering
        \includegraphics[width=0.6\textwidth,height=0.6\textwidth]{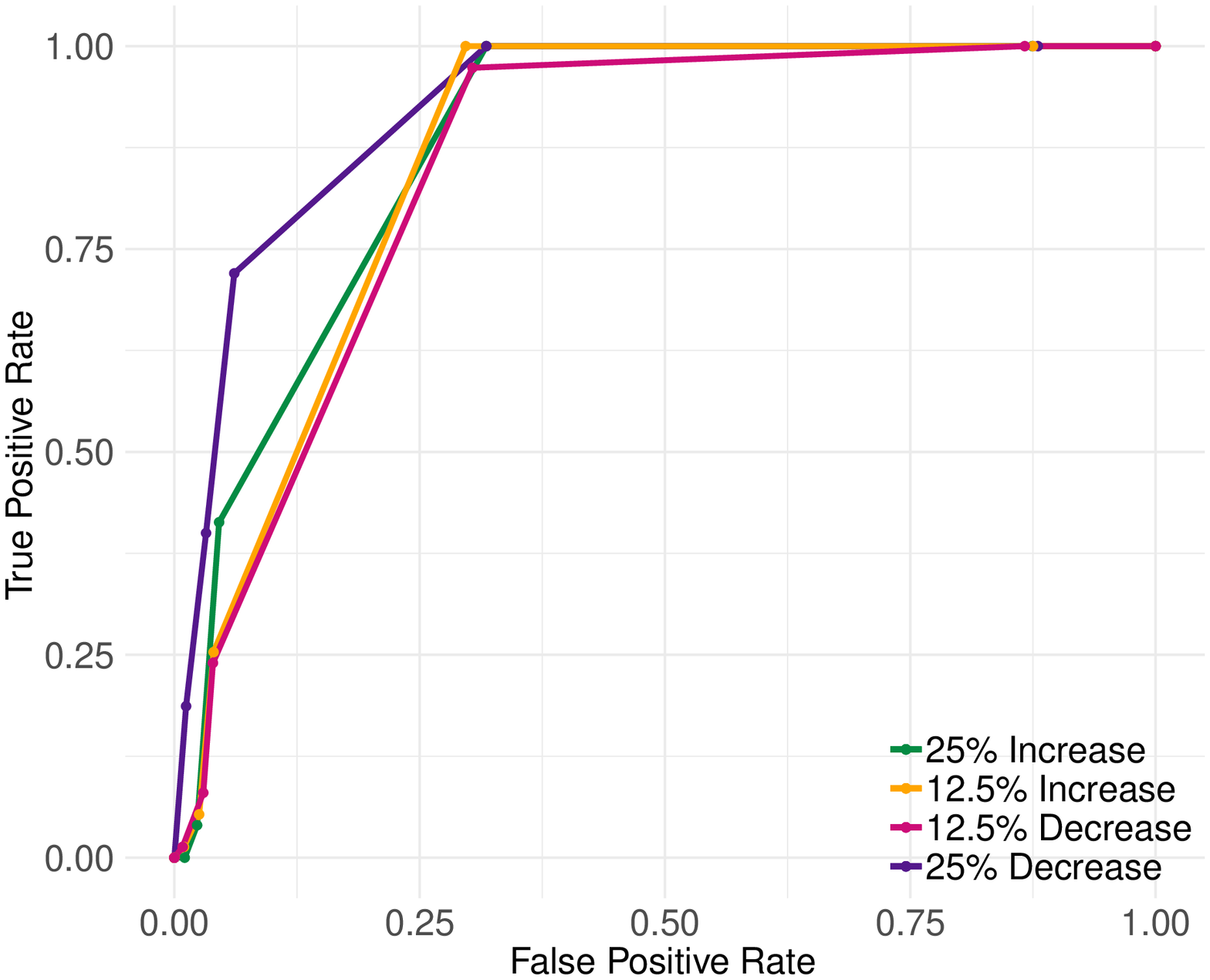}
        \caption{Functional depth with ARIMA extrapolation outlier detection at 20 booking intervals before departure}
		\label{fig:roc20ARIMA}
    \end{subfigure} \hspace{0.5cm}
    \begin{subfigure}[h]{0.42\textwidth}  
    \centering 
        \includegraphics[width=0.6\textwidth,height=0.6\textwidth]{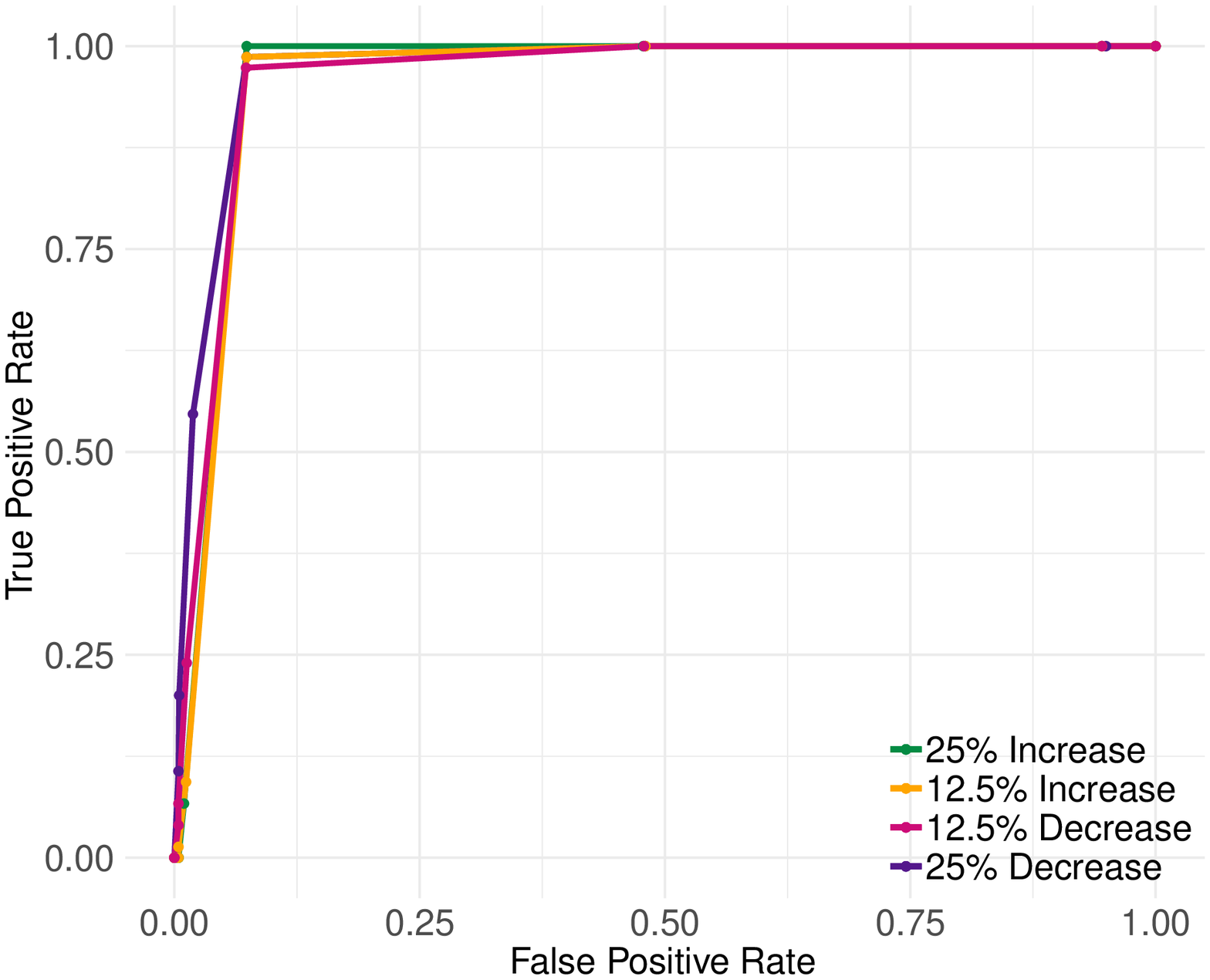}
        \caption{Functional depth with ARIMA extrapolation outlier detection at 10 booking intervals before departure}
		\label{fig:roc10ARIMA}
    \end{subfigure}
    \quad
    \caption{Receiver operating characteristic (ROC) curves}
    \label{fig:roc}
\end{figure}

To show that our conclusions in Section \ref{sec:within} are robust to different parameterisations of the outlier detection settings, we perform an ROC curve analysis by varying the thresholds for $K$-means clustering, functional depth, and functional depth with extrapolation.  Figure \ref{fig:roc} shows the results for two time intervals in the booking horizon: one early at 20 intervals before departure, and one later at 10 intervals before departure.

There are three main conclusions that can be drawn from the results of the ROC analysis. (i) The area under the ROC curve is consistently higher for functional approaches than for $K$-means. Similarly, the area under the curve is even higher when we include extrapolation. (ii) For $K$-means, the area under the ROC curve diminishes as the number of booking intervals increases, suggesting issues with sparsity caused by high dimensionality. Thus, even a better choice of threshold criteria would not result in improved performance for $K$-means. (iii) The improvement between functional depth and functional depth with extrapolation is smaller towards the end of the booking horizon. This is due to the fact that, at this point, the extrapolation makes up a smaller part of the input data and so the two approaches are more similar.

\subsection{Outlier Detection for Diverse Types of Outliers}
\label{sec:types}
Next, we investigate how the average BCR varies depending on the type and magnitude of outliers. All experiments in this section feature an outlier frequency of 5\%. When we tested the sensitivity of approaches to different frequencies of outliers (ranging from 1\% to 10\%, results omitted here), we found little impact on outlier detection performance across methods, such that the conclusions drawn from this section are generally robust. Results on the effect of outlier frequency are available in Appendix C.2.

First, we vary the magnitude of \emph{demand-volume outliers} to \(\pm 12.5\% and \pm 25\%\). Figure \ref{fig:pois} displays the average BCR over time for parametric (Poisson) tolerance intervals. We observe that higher magnitudes of outliers are easier to classify, but also decreases in demand are easier to classify than increases. The latter observation is intrinsic to RM systems: An unexpected decrease in demand causes a decrease in bookings, but an increase in demand does not necessarily result in an increase in bookings if the booking limit for a fare class has been reached, i.e., if the fare class is no longer offered. This censoring leads to the phenomenon of observing a constrained version of demand.

Similar observations arise when testing all other univariate and multivariate outlier detection approaches. In contrast, Figure \ref{fig:halfarima} displays the average BCR over time with functional halfspace depth and ARIMA extrapolation. Here the average BCR is very similar for all four magnitudes of outliers considered. This classification approach therefore appears to be very robust to the magnitude and direction of outliers considered. The robustness to the direction of the outlier demand shift is a result of the choice of depth measure. \citet{Hubert2012} define the multivariate functional halfspace depth for the purposes of identifying curves which are only outlying for a fraction of the time they are observed over. This means that if a booking pattern is affected by censoring, as long as it has still been an outlier before censoring came into effect, it can still be detected later in the horizon. In terms of robustness to magnitude, we hypothesise that much smaller outlier magnitudes would need to be considered before the average BCR decreases. We further consider demand shifts of \(\pm\)1\%, \(\pm\)5\%, \(\pm\)10\%. The results are as expected - for $\pm 10\%$, the performance is only slightly poorer; for $\pm 5\%$, we see a drop in performance with the algorithm at best having a BCR of around 0.75; and a level of $\pm 1\%$ performance is particularly poor with a BCR of close to 0.5. This is behaviour we would expect, given that outliers caused by such a small deviation in demand are unlikely to be considered \textit{outliers} in any real sense. These results are available in Appendix C.7.


\begin{figure}[htbp]
    \centering
    \begin{subfigure}[h]{0.47\textwidth}
    \centering
        \includegraphics[width=0.8\textwidth,height=0.8\textwidth]{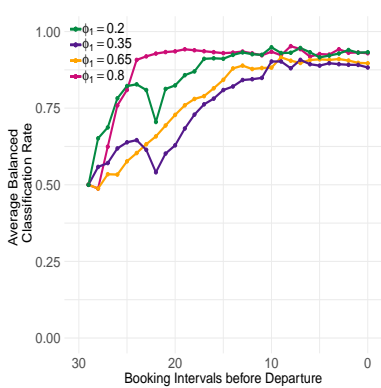}
        \caption{BCR of functional depth outlier detection for changes in $\phi_i$}  
		\label{fig:cat2_func}
    \end{subfigure}
    \begin{subfigure}[h]{0.47\textwidth}  
    \centering 
        \includegraphics[width=0.8\textwidth,height=0.8\textwidth]{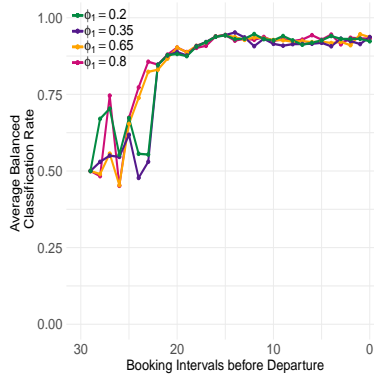}
        \caption{BCR of functional depth with ARIMA extrapolation outlier detection for changes in $\phi_i$}
		\label{fig:cat2_funcARIMA}
    \end{subfigure}
    \quad
    \begin{subfigure}[h]{0.47\textwidth}
    \centering
        \includegraphics[width=0.8\textwidth,height=0.8\textwidth]{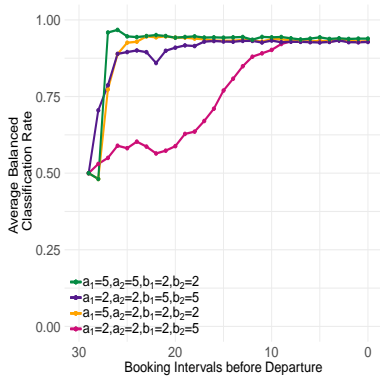}
        \caption{BCR of functional depth outlier \\ detection for changes in \\ $a_1, b_1, a_2, b_2$}
		\label{fig:cat3_func}
    \end{subfigure}
    \begin{subfigure}[h]{0.47\textwidth}  
    \centering 
        \includegraphics[width=0.8\textwidth,height=0.8\textwidth]{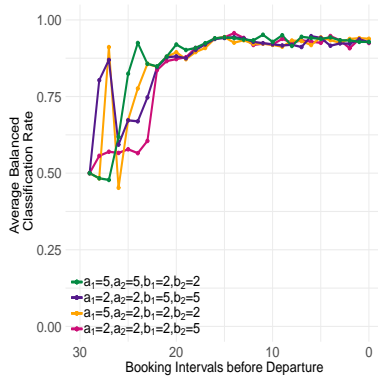}
        \caption{BCR of functional depth with ARIMA extrapolation outlier detection for changes in $a_1, b_1, a_2, b_2$}
		\label{fig:cat3_funcARIMA}
    \end{subfigure}
    \quad
    \caption{Performance of functional depth (with and without ARIMA extrapolation) for different types of outliers}
    \label{fig:outlier_types_bcr}
\end{figure}

Figures \ref{fig:cat2_func} and \ref{fig:cat2_funcARIMA} illustrate effects from \emph{willingness-to-pay outliers}, where the ratio of high-value to low-value arrivals changes. The default value in our simulations is $\phi_1 = \phi_2 = 0.5$ such that there is a 1:1 ratio, but we allow this ratio to change to create outliers. Here, $\phi_1 < 0.5$ creates a higher percentage of total arrivals from low paying, early arriving customers of type 2. Under functional depth outlier detection it is easier to detect this type of outlier when the change in $\phi_1$ is larger. There is a dip in performance around interval 22, as this large number of low-paying arrivals causes censoring when booking limits render cheaper classes unavailable.
Setting $\phi_1 > 0.5$ creates a larger percentage of type 1 customers, who arrive late and are willing to pay more. Again, this is easier to detect under functional depth when the change in $\phi_1$ is larger. Incorporating the ARIMA extrapolation generally improves performance in the last two-thirds of the horizon. However, early in the booking horizon it provides mixed results.

Figures \ref{fig:cat3_func} and \ref{fig:cat3_funcARIMA} demonstrate the performance of functional depth (with and without extrapolation) for detecting \emph{arrival-time outliers}. These outliers are caused by changes in the parameters $a_1, a_2, b_1, b_2$ (resulting in a subset of customer types arriving later or earlier than in the regular case), as outlined in Table \ref{tab:other_types}.

\begin{table}[htbp]
\centering
\resizebox{\textwidth}{!}{%
\begin{tabular}{r|ccccl}
\hline \hline 
\textbf{}             & \textbf{$a_1$} & \textbf{$b_1$} & \textbf{$a_2$} & \textbf{$b_2$} & \textbf{Effect of parameter choices}    \\ \hline
\textbf{Regular Demand} & 5 & 2 & 2  & 5              & low value customers arrive before high value customers \\
\textbf{Setting 1} & 5              & 2              & 5              & 2              & some low value customers arrive a lot later            \\
\textbf{Setting 2} & 2              & 5              & 2              & 5              & some high value customers arrive a lot earlier         \\
\textbf{Setting 3} & 5              & 2              & 2              & 2              & some low value customers arrive a little later         \\
\textbf{Setting 4} & 2              & 2              & 2              & 5              & some high value customers arrive a little earlier  \\   \hline \hline
\end{tabular}}
\caption{Parameter choices used to generate arrival time outliers}
\label{tab:other_types}
\end{table}

Outliers in Settings 1, 2 and 3  are easy to detect even early in the booking horizon using functional depth without extrapolation. This is fairly intuitive - Settings 1 and 3 create almost no bookings early in the horizon, which is very different from regular behaviour. In contrast, Setting 2 creates far more bookings early in the horizon than the regular setting. ARIMA extrapolation is not needed nor beneficial in Settings 1-3, due to the ease of spotting outliers immediately. In contrast, outliers from Setting 4 are more difficult to detect. This is likely due to the fact that for most of the first half of the horizon, outlier booking patterns and regular booking patterns are similar. In the later half of the horizon, booking limits render the cheaper fare classes unavailable, so that arriving customers purchase higher fare classes only slightly earlier in time. In Setting 4 extrapolation is found to significantly help the classification performance in this more challenging setting.


\subsection{Detecting Outliers in Railway Booking Patterns} \label{sec:db}

We demonstrate the proposed outlier detection method by identifying outliers in a data set of 1387 booking patterns obtained from the main German railway company, Deutsche Bahn. This preliminary empirical study can be thought of as a guide to practitioners for how to apply the algorithm. A detailed analysis of the algorithm's performance in practice would require a manually annotated data set or, potentially, a field study. While of significant interest, such an analysis is beyond the scope of this paper.

\begin{figure}[htbp]
    \centering
    \begin{subfigure}[b]{0.47\textwidth}  
            \centering 
            \includegraphics[width=1\textwidth,height=0.8\textwidth]{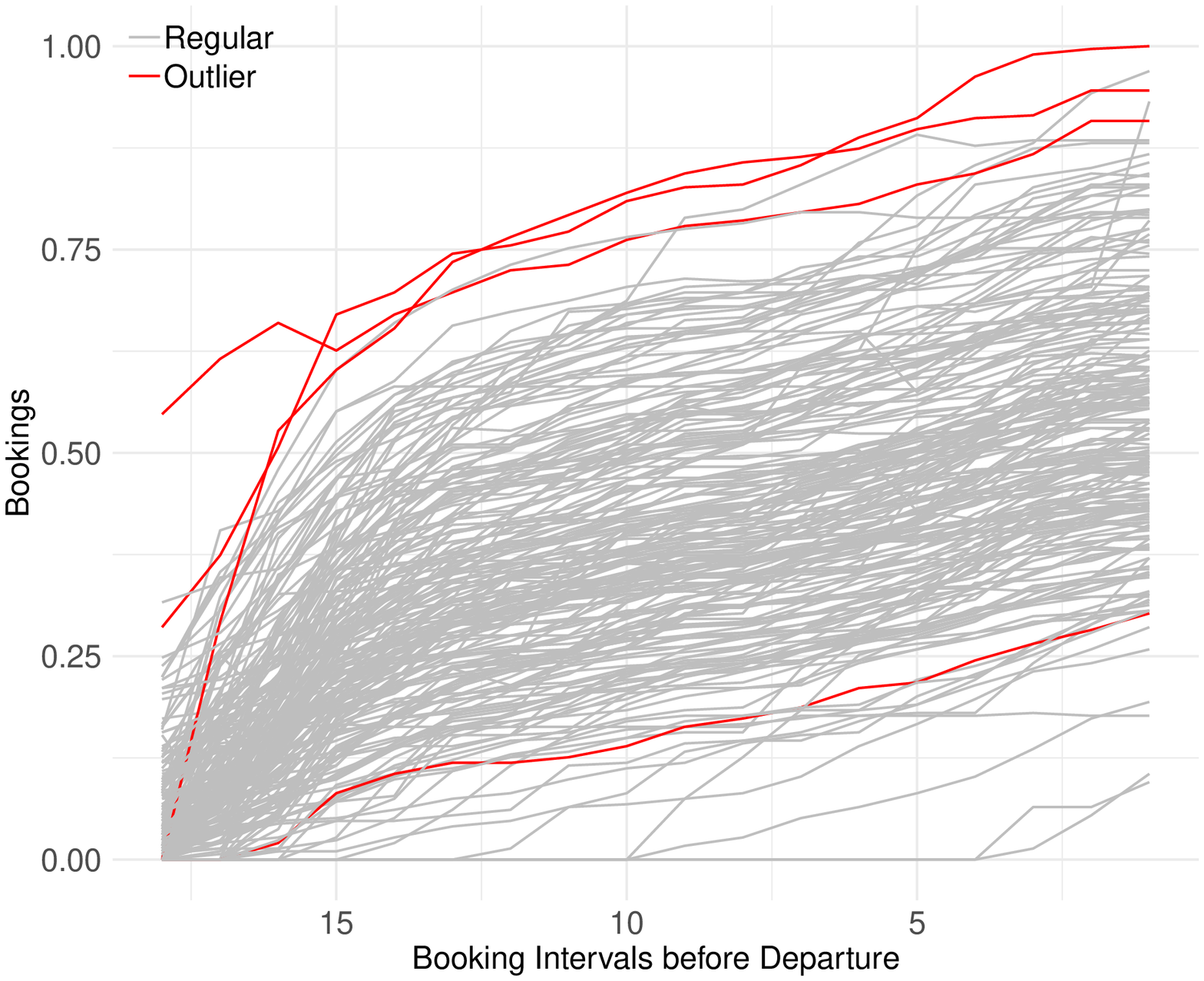}
            \caption{Booking patterns (outliers shown in red)}
			\label{fig:DBdata_outliers}
        \end{subfigure}
    \begin{subfigure}[b]{0.47\textwidth}
            \centering
            \includegraphics[width=1\textwidth,height=0.8\textwidth]{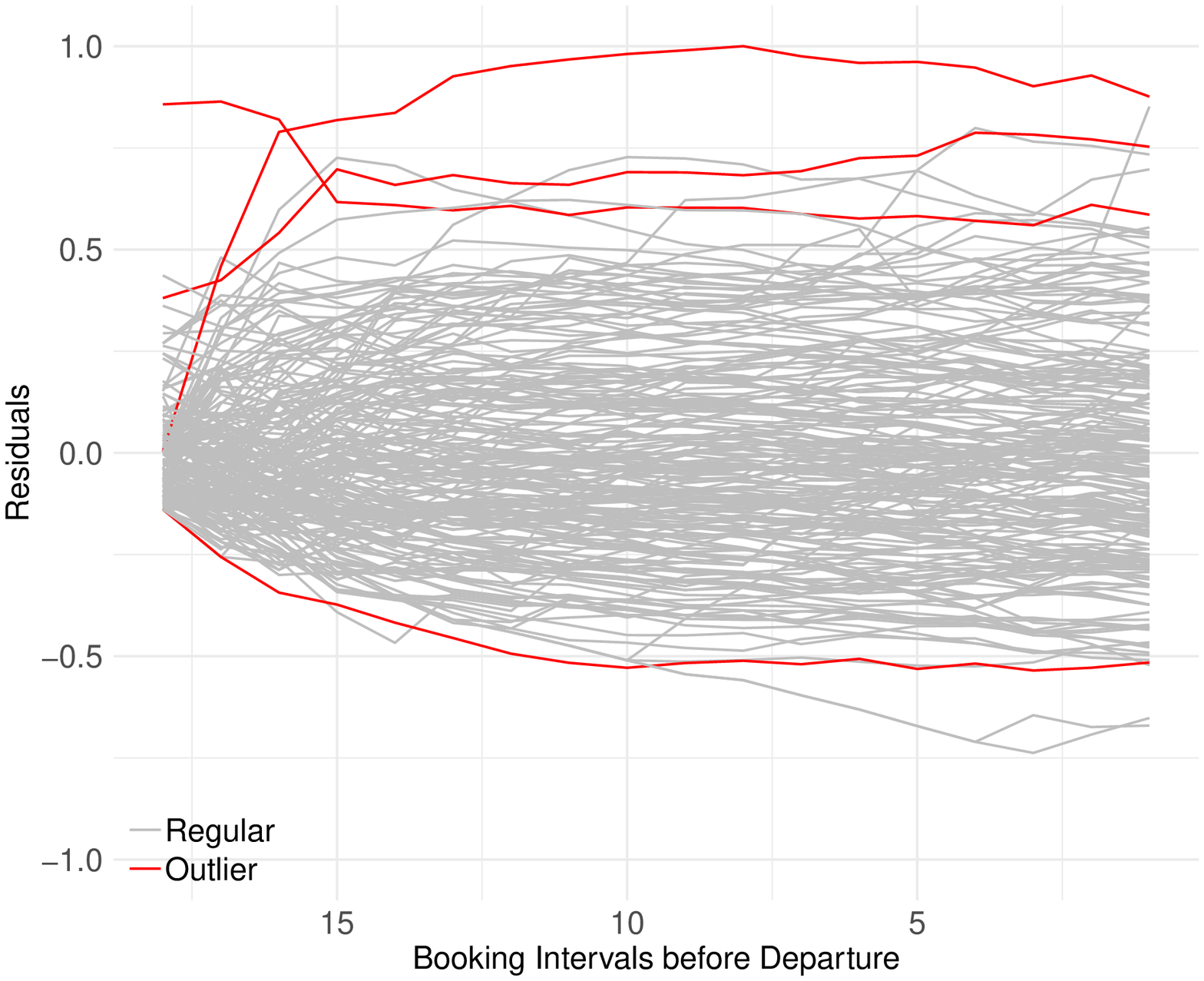}
            \caption{Residuals (outliers shown in red)}
			\label{fig:DBresiduals_outliers}
    \end{subfigure}
        \quad
        \caption{Pre-processing of data}
        \label{fig:DBoutliers}
\end{figure}

We consider booking patterns that were observed for a single departure time every day of the week, for one railway leg that directly connects an origin and a destination. The 1387 booking patterns are observed over 18 booking intervals, where the first booking interval is observed 91 days before departure. Figure \ref{fig:DBdata_outliers} illustrates 148 of these booking patterns, which relate to trains departing on Mondays. For the purposes of Figure \ref{fig:DBoutliers}, we have rescaled the number of bookings to be between 0 and 1. The booking data is generated from an RM system that implements an EMSR variant, which sets and updates booking limits based on forecasted demand and observed bookings.
In order to obtain a homogeneous data set to allow for outlier detection, we must account for two factors: (i) departure days of the week and (ii) shortened booking horizons. We compare booking patterns for different days of the week by applying pairwise functional ANOVA tests \citep{Cuevas2004}. In general, booking patterns for different days of the week are not directly comparable (see Appendix C.10 for details). In addition, shortened booking horizons are a special characteristic of this data set that are caused by the railway service provider's process for implementing schedule changes. As a consequence, some booking horizons are foreshortened and the majority of bookings typically arrive much closer to departure (see Appendix C.10).

To prepare the data for outlier detection, we transform the booking patterns to make them more comparable to each other. To account for both shortened booking horizons and departure days of the week, we apply a functional regression model \citep{Ramsay2009}. This functional regression model accounts for the way in which average booking patterns changes from day to day, and fits a mean function (see Appendix C.10 for details) to the booking patterns for each day of the week. The model is of the form:
\begin{equation} \label{eqn:db_funcreg}
\begin{split}
   bookings_{i}(t) = \beta_0(t) + \beta_1(t)I_{Monday_{i}} + \beta_2(t)I_{Tuesday_{i}} + \beta_3(t)I_{Wednesday_{i}} + \\ \beta_4(t)I_{Thursday_{i}} + \beta_5(t)I_{Friday_{i}} + \beta_6(t)I_{Saturday_{i}} + 
   \beta_7(t)I_{Shorter\mbox{ }Horizon_{i}} + e_i(t)
\end{split}
\end{equation}
where the $\beta_j(t)$ are functions of time. Here, $I_{Monday_{i}} =1$ if booking pattern $i$ relates to a departure on a Monday, $0$ otherwise, and so on. This means that $\beta_1(t)$ accounts for the change in average bookings between Sunday and Monday departures. The purpose of allowing the $\beta_j(t)$ to be functions of time is not to remove the trend from the booking patterns but rather to allow the relationship between different days of the week to change over the course of the booking horizon. Since every departure belongs to a single day of the week, $\beta_0(t)$ represents the average bookings for Sunday departures. In this model, $I_{Shorter\mbox{ }Horizon_{i}} = 1$ if the booking horizon has been shortened due to scheduling changes (affecting departures from mid-December to mid-March), 0 otherwise.

We run the functional depth outlier detection routine on the residuals, as shown in Figure \ref{fig:DBresiduals_outliers}, with detected outliers shown in red. We also show these corresponding outliers in red in Figure \ref{fig:DBdata_outliers}. Of the 1387 booking patterns in the data set, we classify 66 ($\approx 5\%$) as outliers. Note that the frequency of outliers is not an assumption provided to the outlier detection routine, and coincides with the frequency of outliers used in the simulation setup (5\%), thus justifying this choice in our earlier simulations.

For validation, we provided the labelled data set back to Deutsche Bahn. The company's domain experts have confirmed that the relative proportion of outliers is appropriate to support analyst work on improving demand forecast and booking controls. Furthermore, their hindsight analysis has confirmed that most automatically identified outliers would have benefitted from such corrections.

In addition, we compared the dates of the booking patterns classified as outliers with a list of known holidays and events. Of the 66 booking patterns classified as outliers, 30 could be attributed to known events e.g. public holidays. This leaves 36 outlying booking patterns which would otherwise have gone undetected. However, we do not aim to solely identify already known events, as there would be little point to only confirming known information. Therefore, the additionally identified outliers are not necessarily false positives - they are in fact the very booking patterns we are attempting to identify.


\subsection{Revenue Improvement Under Outlier Detection of Demand-volume Outliers}

To evaluate the effect of demand deviating from the forecasts used by EMSRb and EMSRb-MR, we now introduce a best-case scenario where the RM system anticipates outliers and generates accurate demand forecasts (as opposed to implementing booking controls based on the initial erroneous forecasts). The percentage change in revenue,  when switching from erroneous to correct forecasts, under four demand changes is shown in Table \ref{tab:revenuepotential}. Results show the impact of detecting and correcting outliers in demand depends on the demand factor, the choice of booking control heuristic, and the magnitude of the demand deviation. 

Under EMSRb, the effect on revenue is asymmetric across positive and negative outliers. When the outlier is caused by a decrease in demand, correcting the forecast and updating controls leads to significant increases in revenue, particularly at higher demand factors. Conversely, when the outlier is caused by an increase in demand, correcting the forecast and updating controls has a negative impact on revenue. Although counter-intuitive at first glance, this agrees with previous findings. EMSRb is known to be too conservative \citep{Belobaba2002} and reserve too many  units of capacity for high fare classes, thereby rejecting an excessive number of requests from customers with a lower willingness to pay. In consequence, there is left-over capacity at the end of the booking horizon. Hence, under-forecasting can be beneficial under EMSRb.

Under EMSRb-MR booking controls, the results are more symmetric across positive and negative outliers, in that correctly adjusting forecasts increases revenue regardless of whether the initial forecast was too high or too low. Under both types of heuristic, the magnitude of the change in revenue (either positive or negative) is generally larger when the change in demand from the forecast is larger.
\begin{table}[htbp]
\centering
\begin{tabular}{ c|c|cccc } 
 \hline \hline
 \textbf{Optimisation} & \textbf{Forecasted} & \multicolumn{4}{c}{\textbf{\% Change in Demand from Forecast}} \\
 \cline{3-6} 
 \textbf{Heuristic} & \textbf{Demand Factor} & \textbf{\thead{-25\%}} & \textbf{\thead{-12.5\%}} & \textbf{\thead{+12.5\%}} & \textbf{\thead{+25\%}} \\ \hline
& 0.90 						& \textcolor{cadmiumgreen}{+0.1\%}		& \textcolor{cadmiumgreen}{+0.1\%}	& \textcolor{carnelian}{-0.9\%}	& \textcolor{carnelian}{-3.6\%}	\\
\textbf{EMSRb} & 1.20 						& \textcolor{cadmiumgreen}{+10.2\%}	 	& \textcolor{cadmiumgreen}{+6.4\%}	& \textcolor{carnelian}{-2.3\%} 	& \textcolor{carnelian}{-2.3\%} 	\\
& 1.50	 					& \textcolor{cadmiumgreen}{+12.2\%}		& \textcolor{cadmiumgreen}{+4.4\%}	& \textcolor{carnelian}{-4.5\%}	& \textcolor{carnelian}{-6.8\%}	\\
& Avg.						& \textcolor{cadmiumgreen}{\textbf{+7.5\%}} & \textcolor{cadmiumgreen}{\textbf{+3.6\%}}	& \textcolor{carnelian}{\textbf{-2.5\%}}	& \textcolor{carnelian}{\textbf{-4.2\%}}	\\ 
 \hline 
& 0.90 						& \textcolor{cadmiumgreen}{+2.3\%}      	& \textcolor{cadmiumgreen}{+1.3\%}  & \textcolor{cadmiumgreen}{+0.4\%}	   &  \textcolor{cadmiumgreen}{+2.9\%}	\\      
\textbf{EMSRb-MR} & 1.20 						& \textcolor{cadmiumgreen}{+2.0\%}       	& \textcolor{cadmiumgreen}{+4.1\%}  & \textcolor{cadmiumgreen}{+4.4\%}     &  \textcolor{cadmiumgreen}{+9.9\%}	\\ 
& 1.50	 					& \textcolor{cadmiumgreen}{+16.2\%}   	& \textcolor{cadmiumgreen}{+7.7\%}  & \textcolor{cadmiumgreen}{+5.0\%}     &  \textcolor{cadmiumgreen}{+9.5\%} 	\\ 
& Avg.						& \textcolor{cadmiumgreen}{\textbf{+6.9\%}} & \textcolor{cadmiumgreen}{\textbf{+4.4\%}}	& \textcolor{cadmiumgreen}{\textbf{+3.3\%}}	& \textcolor{cadmiumgreen}{\textbf{+7.4\%}}	\\  
 \hline \hline
\end{tabular}
\caption{\% Change in revenue resulting from correcting inaccurate demand forecasts}
\label{tab:revenuepotential}
\end{table}

Furthermore, we compared the performance of different outlier detection methods under the two different heuristics. The results (omitted for space considerations) under EMSRb and EMSRb-MR were found to be very similar regardless of the outlier detection method used. Given the similarity in performance between the two heuristics, and that EMSRb-MR accounts for the more realistic demand model of customers choosing the cheapest class offered, the remainder of the results in Section \ref{sec:results} relate to those from EMSRb-MR.

\label{sec:improvement}
\begin{figure}[htbp]
    \centering
    \includegraphics[width=0.4\textwidth, height=0.4\textwidth]{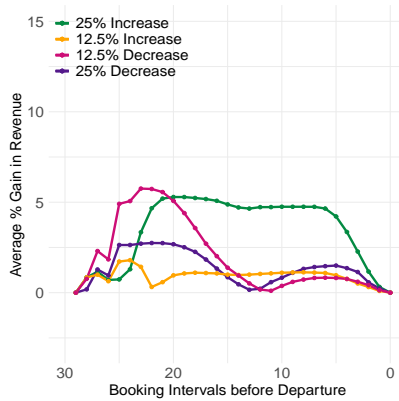}
	\caption{Gain in revenue under different magnitudes of outliers using functional depth with ARIMA extrapolation}    
    \label{fig:updating}
\end{figure}
Figure \ref{fig:updating} shows the average percentage gain in revenue, at each point in the booking horizon, from analysts correcting forecasts for those booking patterns identified as outliers. The percentage gain is in comparison to the analyst making no changes and using the incorrect forecast for the entirety of the booking horizon. 

The outlier detection method of choice in Figure \ref{fig:updating} is functional depth with ARIMA extrapolation. We consider an idealised scenario, in that when a booking pattern is flagged as an outlier, if it is a true positive (genuine outlier) then analysts adjust the forecast according to the correct distribution. Similarly, if the flagged outlier is a false positive, analysts do not make any changes to the forecast. Although idealised, the results here highlight the potential gains in revenue from analyst intervention, as well as the utility of using functional outlier detection in detecting true positives and avoiding false negatives (missed outliers).

Results show the use of our method creates a peak early in the booking horizon, when the potential revenue gain is highest. This peak is caused by a combination of being far enough into the booking horizon such that some bookings have occurred and the outlier detection method is able to identify outliers, but being early enough in the horizon such that any actions taken still have time to make an impact.

\section{Conclusion and Outlook} \label{sec:conclusion}
In conclusion, the work presented in this paper gives rise to several insights. 

We benchmarked a set of outlier detection techniques and find that the functional outlier detection approach offers the best performance and the most scope for further extensions. Our results show that combining functional outlier detection with our proposed extrapolation step significantly improves performance overall, and accelerates the correct identification of outliers earlier in the booking horizon. We do note however that all methods perform poorly very early in the booking horizon where very little data has been gathered, and clearly at this stage analyst expertise or prior information is needed rather than relying on booking data alone.

By analysing an empirical railway booking data set, we demonstrated that such data is similar in shape as the data generated by the simulation model. Furthermore, the frequency of outliers detected via applying functional outlier detection to the empirical data was similar to what was observed on simulation data. As, in contrast to the simulation setting, the empirical data does not provide information on the labelling of actual outliers, it was not possible to compute detection rates for that data. However, we validated our findings by presenting them to domain experts.

Outliers in demand diminish revenue when they go undetected. The exact effect depends on the combination of outlier and optimisation method, as shown in Section \ref{sec:improvement}. Nevertheless, we argue that using a heuristic with an intrinsic bias that is then compensated by undetected outliers (as observed for EMSRb and undetected positive demand outliers) cannot be desirable for an automated system.

We have demonstrated that identifying outlier booking curves and adjusting the demand forecast accurately early in the booking horizon supports revenue optimisation. Currently, revenue management analysts decide on which booking patterns are outliers based on their previous experience of observing demand and their knowledge about special events. Automated outlier detection routines provide another procedure of alerting analysts to unusual patterns. If the detection algorithm identifies a booking pattern as an outlier, the RM system alerts the responsible analyst. When the system and the analyst agree that a booking pattern is critical and that it requires intervention, an analyst must decide which action(s) to take. Specifically, they need to decide whether to increase or decrease the forecast or inventory controls, and by how much. Further work could investigate methods to adjust the initial forecast to account for outliers. 

Within the context of RM, thoroughly examining the effects across further outlier situations, e.g. outliers affecting only part of the booking pattern, and optimisation solutions, e.g. dynamic programming, appears to be a valid topic for further research. Furthermore, future research might consider more differentiated forecasting situations, featuring trends and seasonalities. Beyond RM, other paradigms of offer optimisation, such as mark-down pricing or the pricing of Veblen products, might offer different challenges with regards to outlier detection. Given that the resulting sales observations should also take the format of time series, we consider it interesting to find out whether the same methods would broadly apply in such different settings.

\section*{Acknowledgements}
We gratefully acknowledge the support of the Engineering and Physical Sciences Research Council funded EP/L015692/1 STOR-i Centre for Doctoral Training. The authors also acknowledge Deutsche Bahn for the provision of data, and are grateful to Philipp Bartke and Valentin Wagner for helpful discussions and suggestions.

\bibliographystyle{apalike}

\newpage
\appendix
\appendixpage
\appendix
\appendixpage

\section{Technical Description of Methodologies}
\label{sec:appout}
\subsection{Outlier Detection Approaches}
Let \(N\) be the number of booking patterns. We observe the cumulative number of bookings for each booking pattern at \(T\) time points over a booking horizon of length \(t_T\): \(t_1, \hdots, t_{\tau}, \hdots, t_T\). Note that \(t_1, \hdots, t_{\tau}, \hdots, t_T\) do not necessarily need to be equally spaced. \sloppy Then \(\bm{y}_n(t_{\tau})\) is a time series of bookings for pattern \(n\), up to time \(t_{\tau}\): \(\bm{y}_n(t_{\tau}) = \left(y_{n}(t_1), y_{n}(t_2), \hdots, y_{n}(t_{\tau}) \right)\).
\subsubsection{Nonparametric Percentiles}
Let \(\bm{y}(t_{\tau}) = (y_1(t_{\tau}), \hdots, y_N(t_{\tau}))\) be the cumulative number of bookings for patterns \(1, \hdots, N\) at time \(t_{\tau}\). Find the lower and upper (2.5\% and 97.5\%) percentiles of the ordered sample, \(L\) and \(U\). For any booking pattern \(n\), if the number of bookings at time \(t_{\tau}\), \(y_n(t_{\tau})\) is less than \(L\) or greater than \(U\), it is defined as an outlier at time \(t_{\tau}\). Note that an alternative (parametric) approach would be to fit a distribution to the data and use the lower and upper percentiles of the fitted distribution. 
\subsubsection{Tolerance Intervals}
For \(Y(t_{\tau})_1, \hdots, Y(t_{\tau})_n\), a random sample from a population with distribution \(F(Y(t_{\tau}))\), if:
\begin{equation}
\Pro \left(F(U(t_{\tau}))- F(L(t_{\tau})) > \beta \right) = 1 - \alpha,
\end{equation}
then the interval \((L(t_{\tau}), U(t_{\tau}))\) is called a \(\left(\beta, 1 - \alpha\right)\) two-sided tolerance interval \citep{Hahn1981}. 
At each booking interval, a tolerance interval for the number of bookings until that point in time, can be defined. If the number of bookings lies outside of this tolerance interval, the booking pattern can be deemed an outlier. 
\begin{itemize}
\item \textbf{Nonparametric Tolerance Intervals}: Let \(Y(t_{\tau})_{(1)}, \hdots, Y(t_{\tau})_{(n)}\) be the ordered observations of the sample \(Y(t_{\tau})_1, \hdots, Y(t_{\tau})_n\). \citet{Wilks1941} details that a \(\left(\beta, 1 - \alpha\right)\) tolerance interval can be calculated as follows: 
\begin{enumerate}
\item Let \(B \sim Binomial(n, \beta)\), then let \(k\) be the smallest integer such that:
\begin{equation}
\Pro\left(B \leq k - 1 \right) \geq 1 - \alpha 
\end{equation}
\item Letting \(k = s-r\), where \(1 \leq r < s \leq n\), then \(\left(Y(t_{\tau})_{(r)}, Y(t_{\tau})_{(s)}\right)\) is a tolerance interval, for any such \(r\) and \(s\). It is common to choose:
\begin{equation}
r = \left\lfloor \frac{n-k+1}{2} \right\rfloor,
\end{equation}
then \(s = k + r\) i.e. \(s = n - r + 1\).
\end{enumerate}
\item \textbf{Parametric Tolerance Intervals}:  Given the discrete, count nature of the data, an obvious first choice for the number of bookings at time \(t_{\tau}\), is a Poisson distribution. Supposing \(y(t_{\tau})\) is the observed value of a random variable \(Y(t_{\tau})\) which has a Poisson distribution, \(Po(n\lambda)\), a \(\left(\beta, 1 - \alpha\right)\) tolerance interval based on \(y(t_{\tau})\) is constructed in two steps, as described by \citet{Hahn1981}:
\begin{enumerate}
\item Calculate a two-sided \(\left(1 - \alpha\right)\)-level confidence interval, \(\left(l(t_{\tau}), u(t_{\tau})\right)\) for \(\lambda\), such as:
\begin{equation}
\left(l(t_{\tau}),u(t_{\tau})\right) = \left(\frac{\chi^2_{(\alpha/2;2y(t_{\tau}))}}{2n},\frac{\chi^2_{(1-\alpha/2;2y(t_{\tau})+2)}}{2n}\right)
\end{equation}
\item Find the minimum number \(U(t_{\tau})\), and the maximum number \(L(t_{\tau})\) such that:
\begin{eqnarray}
\Pro\left(Y(t_{\tau}) < U(t_{\tau}) | \lambda = u(t_{\tau}) \right) &\geq& \frac{1+\beta}{2} \\ 
\mbox{ and } \Pro\left(Y(t_{\tau}) > L(t_{\tau}) | \lambda = l(t_{\tau})\right) &\geq& \frac{1+\beta}{2}.
\end{eqnarray}
\end{enumerate}
Given the desire for a general method, the presence of differing mean-variance relationships between fare classes and over time, suggests that assuming a Poisson distribution may not be appropriate, given the fixed (equal) mean-variance relationship of this distribution. Alternative distributions which could be tested include the Negative Binomial, which has two parameters for mean and variance (although only allows the variance to be larger than the mean), or the Generalised Poisson distribution, which has an additional parameter allowing the variance to change. 
\end{itemize}
\subsubsection{Robust Z-Score}
 Let \(y_n(t_{\tau})\) be the cumulative number of bookings for flight \(n\) at time \(t_{\tau}\). The robust Z-score can be calculated as \citep{Iglewicz1993}:
\begin{equation}
\tilde{Z}_n = \frac{0.6745\left(y_n(t_{\tau}) - \tilde{y}(t_{\tau})\right)}{MAD(t_{\tau})},
\end{equation}
where \(\tilde{y}(t_{\tau})\) is the median number of bookings at time \(t_{\tau}\) across all booking patterns, and the Median Absolute Deviation at time \(t_{\tau}\), \((MAD(t_{\tau}))\), is given by:
\begin{equation}
MAD(t_{\tau}) = median\left\{\left| y_n(t_{\tau}) -  \tilde{y}(t_{\tau})\right|\right\}
\end{equation}
A booking pattern, \(n\), can be classified as an outlier at time \(t_{\tau}\), if the number of bookings at time \(t_{\tau}\), \(y_n(t_{\tau})\), has a modified Z-score with magnitude above 3.5, as described by \citet{Iglewicz1993}.

\subsubsection{Distance}
Given that a time series of length \(\tau\) can be thought of as a point in a \(\tau\)-dimensional space, the distance between two time series can be calculated and used as a measure of the difference between them. In particular, for a time series \(\bm{y}_n(t_{\tau}) = \left(y_{n}(t_1), y_{n}(t_2), \hdots, y_{n}(t_{\tau}) \right)\), we define: 
\begin{equation} 
D_n(t_{\tau}) = \frac{1}{N-1}\sum_{m=1}^{N} \mathcal{D}(\bm{y}_n(t_{\tau}), \bm{y}_m(t_{\tau}))
\end{equation}
where \(\mathcal{D}(\bm{y}_n(t_{\tau}), \bm{y}_m(t_{\tau}))\) is the distance between two booking patterns, \(n\) and \(m\), up to time \(t_{\tau}\), and \(N\) is the total number of booking patterns being considered. Here the distance-based outlier score is given as the mean distance of a point to its \(k\)-nearest neighbours, and we set \(k = N-1\), all other points. Hence, for some given threshold, all booking patterns whose mean distance is larger than the threshold can be marked as an outlier.  Booking pattern \(n\) can be defined as an outlier, at time \(t_{\tau}\), if:
\begin{equation}
D_n(t_{\tau}) \geq \mu_d + 3\sigma_d
\end{equation}
where \(\mu_d\) is the mean of the mean distances, and \(\sigma_d\) the standard deviation. We consider both Euclidean and Manhattan distance metrics: 
\begin{itemize}
    \item \textbf{Euclidean}:
    \begin{equation}
        \mathcal{D}_E(\bm{y}_n(t_{\tau}), \bm{y}_m(t_{\tau})) = \left(\sum_{u=1}^{\tau} \left( y_n(t_u) - y_m(t_u)  \right)^2 \right)^{\frac{1}{2}}
    \end{equation}
    \item \textbf{Manhattan}:
    \begin{equation}
        \mathcal{D}_M(\bm{y}_n(t_{\tau}), \bm{y}_m(t_{\tau})) = \sum_{u=1}^{\tau} \left| y_n(t_u) - y_m(t_u)  \right|
    \end{equation}
\end{itemize}

\begin{figure}[!h]
    \centering
    \begin{subfigure}[b]{0.47\textwidth}
        \centering
        \includegraphics[width=0.7\textwidth,height=0.7\textwidth]{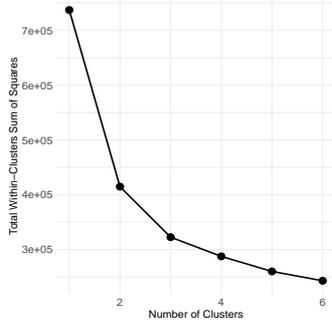}
        \caption{Within cluster sum of squares for choosing \(K\).}
		\label{fig:kmeans}
    \end{subfigure} \hspace{0.4cm}
    \begin{subfigure}[b]{0.47\textwidth}  
        \centering 
        \includegraphics[width=0.7\textwidth,height=0.7\textwidth]{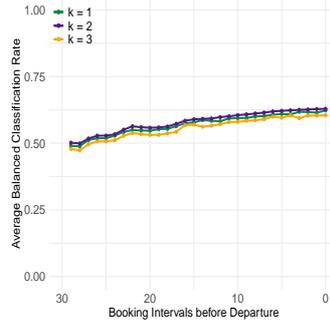}
        \caption{BCR under different values of $K$}
		\label{fig:kmeans_bcr}
    \end{subfigure}
    \quad
    \caption{Choosing $K$}
    \label{fig:kmeans_k}
\end{figure}

\subsubsection{\(K\)-Means Clustering}
It should be noted that clustering algorithms, such as \(K\)-means clustering, are optimised to determine clusters instead of outliers meaning that the success of the outlier detection relies on an algorithm's ability to accurately determine the structure of the clusters. The distance threshold at which a point is classified as an outlier also needs to be specified. \citet{DebDay2017} describe a global threshold distance, at which point those observations which are further away from their cluster centre are classed as outliers, as being half the sum of the maximum and minimum distances. The procedure for identifying booking patterns observed up to time $t_{\tau}$ as outliers is as follows:
\begin{enumerate}
    \item Choose $K$, the number of clusters.
    \item Randomly assign $K$ booking patterns to be the initial cluster centres.
    \item Calculate the $\tau$-dimensional distance (Euclidean or Manhattan) from each booking pattern in the data set to each cluster centre, and assign each booking pattern to the cluster centre from which it is the smallest distance.
    \item Recalculate the centre of each cluster based on the booking patterns assigned to it.
    \item Repeat steps (3) and (4) until the assignment of booking patterns to clusters no longer changes.
\end{enumerate}
\begin{figure}[!h]
    \centering
    \begin{subfigure}[b]{0.47\textwidth}
        \centering
        \includegraphics[width=0.7\textwidth,height=0.7\textwidth]{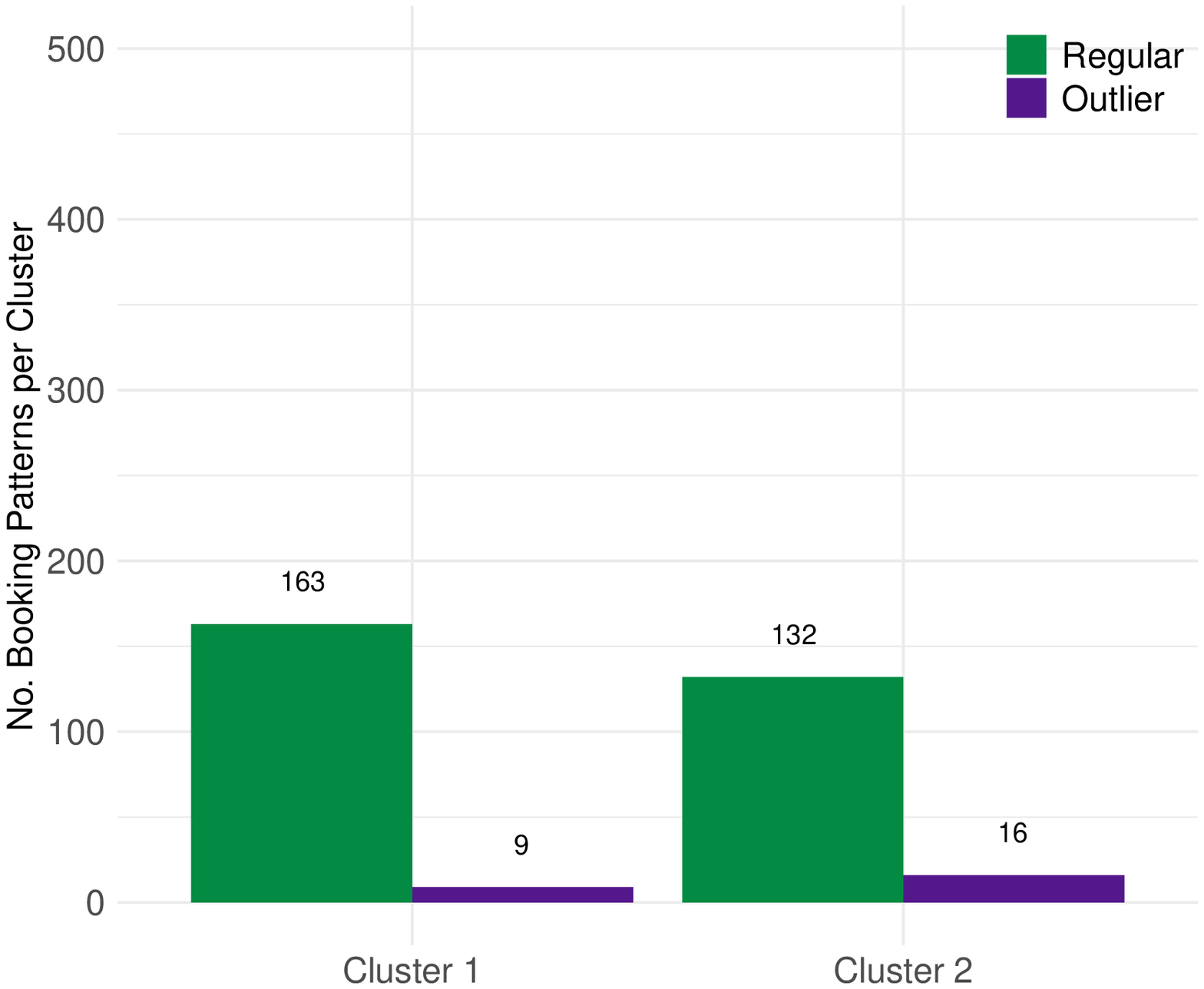}
        \caption{$K$ = 2}
		\label{fig:k2}
    \end{subfigure}
    \begin{subfigure}[b]{0.47\textwidth}  
        \centering 
        \includegraphics[width=0.7\textwidth,height=0.7\textwidth]{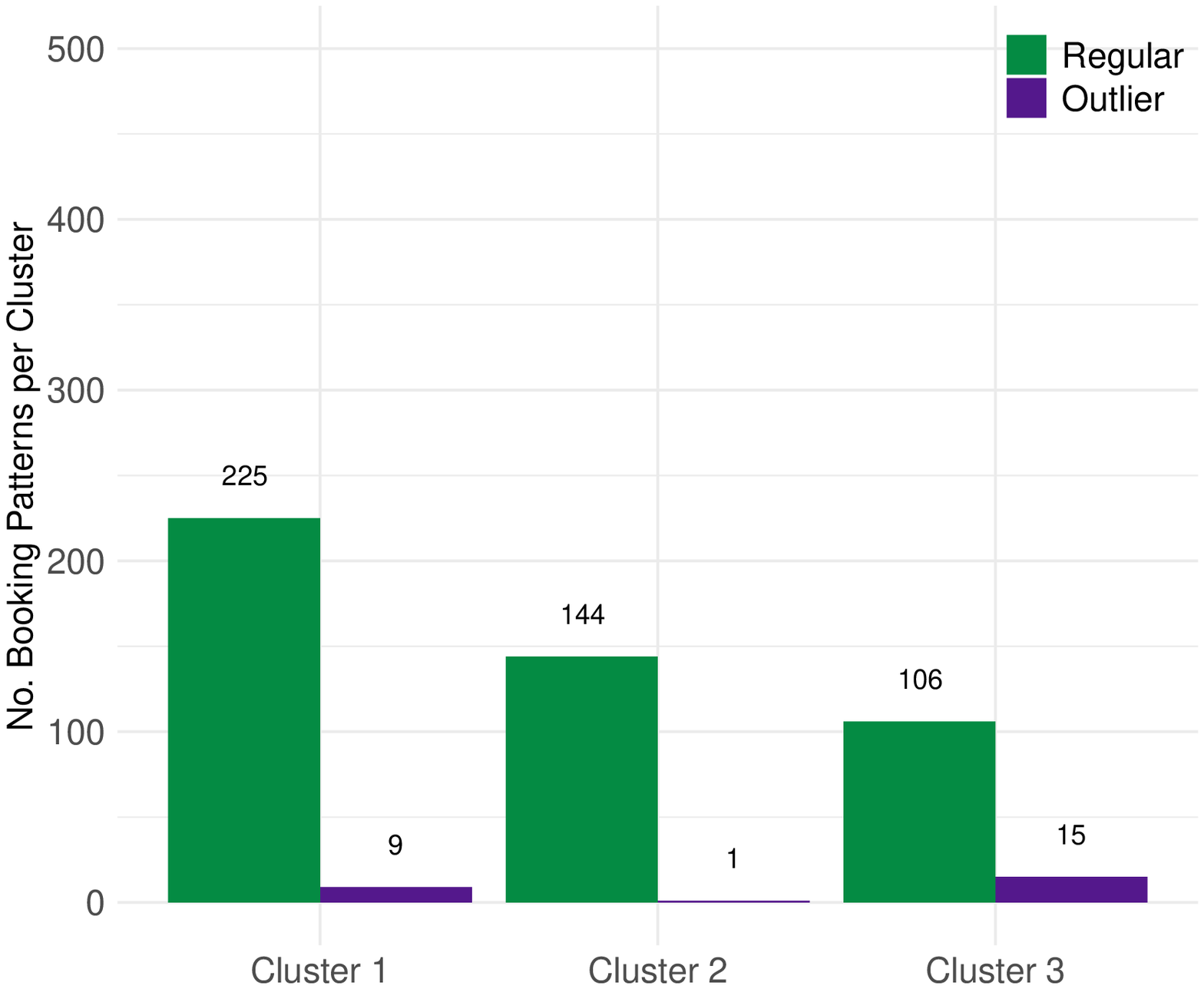}
        \caption{$K$ = 3}
		\label{fig:k3}
    \end{subfigure}
    \quad
    \caption{Distribution of Genuine Outliers Across Clusters}
    \label{fig:outlier_clust_dist}
\end{figure}

\(K\)-means clustering relies on specifying the number of clusters in advance. The optimal number of clusters should seek to minimise the within cluster sum of squares without overfitting. Choosing \(k\) is a difficult problem as it requires fitting \(k\)-means with multiple values of \(k\) and choosing the best one. Figure \ref{fig:kmeans} demonstrates the within cluster sum of squares for multiple values of \(k\), where the optimal number of clusters is chosen as the {\em elbow} of the plot, \(k = 2\). To investigate the impact that the choice of $K$ has on the outlier detection performance, we compare the balanced classification for $K$ = 1, 2, and 3. These results are shown in Figure \ref{fig:kmeans_bcr}. There is little difference between the different values of $K$, suggesting that the choice of $K$ is not what drives the poor performance of $K$-means clustering.

It may be surprising that the optimal number of clusters is chosen as 2 rather than 1, given that the regular demand is generated from a single distribution. It raises the question of whether the algorithm is clustering the booking patterns into regular and outlying patterns. This would mean that it would fail to detect the outlying booking patterns as they have their own cluster and so distance to their cluster centre is small. However, upon investigation of the distribution of outlying booking patterns across clusters (shown in Figure \ref{fig:outlier_clust_dist}), it was found not to be the case. It is possible that the booking limits introduce some element of bi-modality.

\subsubsection{Multivariate Functional Halfspace Depth}
The general procedure for detecting outliers at time \(\tau\) using functional depth, as described by \citet{Febrero2008} and \citet{Hubert2015}, is as follows:
\begin{enumerate}
    \item Define \(\mathcal{D}_n(\bm{y}_n(t_{\tau}))\) to be the functional depth of the \(\bm{y}_n(t_{\tau}) = \left(y_{n}(t_1), y_{n}(t_2), \hdots, y_{n}(t_{\tau}) \right)\), booking pattern \(n\) at time \(t_{\tau}\). 
    \item Define a threshold, \(C\), for the functional depth. 
    \item Those booking patterns with functional depths, \(\mathcal{D}_n(\bm{y}_n(t_{\tau}))\), below the threshold are classified as outliers, delete them from the sample. 
    \item Recalculate functional depths on the new sample, and remove further outliers. Repeat until no more outliers are found.
\end{enumerate} 
As described by \citet{Febrero2008}, the threshold, \(C\), is ideally chosen such that:
\begin{equation}
\Pro(\mathcal{D}_n(\bm{y}_n(t_{\tau})) \leq C) = 0.01,\mbox{ } n = 1,\hdots,N,
\end{equation} 
when there are no genuine outliers present in the sample. However, this would require knowing the distribution of functional depths when there are no outliers. \citet{Febrero2008} discuss two bootstrapping-based procedures for estimating \(C\). The general idea of the bootstrapping method used in this paper, as described by \citet{Febrero2008}, is to (i) resample the booking patterns, with probability proportional to their functional depths (such that any outlying patterns are less likely to be resampled), (ii) smooth the bootstrap samples, then (iii) set \(C\) as the median value of the 1\% percentiles of the empirical distributions of the depths of the bootstrapped samples. More specifically:
\begin{enumerate}
    \item Calculate the functional depths for each booking pattern, \(\mathcal{D}_n(\bm{y}_1(t_{\tau})), \hdots, \mathcal{D}_n(\bm{y}_n(t_{\tau}))\).
    \item Resample the original booking patterns to obtain $B$ bootstrap samples, where each booking pattern is sampled with probability proportional to its functional depth. Denote the $n^{th}$ booking curve in the $b^{th}$ bootstrap sample as $\bm{x}^{b}_{n}$.
    \item Smooth the bootstrap samples to obtain $\bm{s}^{b}_{n} = \bm{x}^{b}_{n} + \bm{z}^{b}_{n}$, where $\bm{z}^{b}_{n} = \left(z_{n}(t_1), z_{n}(t_2), \hdots, z_{n}(t_{\tau}) \right)$ is normally distributed with mean 0 and covariance matrix $\gamma \Sigma$. $\gamma$ is a smoothing parameter, and  $\Sigma$ is the covariance matrix of the original sample. 
    \item Calculate the functional depths for the resampled booking patterns in each of the smoothed bootstrap samples. Let $C^b$ be the empirical $1^{st}$ percentile of the distribution of these depths for the $b^{th}$ sample.
    \item Choose the threshold $C$ as the median of the values of $C^b$, for $b = 1,\hdots,B$.
\end{enumerate}
For full details, see \citet{Febrero2008}.

In this paper, we restrict our attention to halfspace depth. In the case of one-dimensional random variables, the halfspace depth of a point \(y_n\) with respect to a sample \(y1, \hdots, y_N\) drawn from distribution \(F\) is:
    \begin{equation}
        HD(y_n) = min\left\{F_N(y_n) , 1-F_N(y_n) \right\}
    \end{equation}
where \(F_N\) is the empirical cumulative distribution of the sample \(y_1, \hdots, y_N\) \citep{Febrero2008}. This definition has been extended to the functional data setting, see \citet{Hubert2012} and \citet{Claeskens2014}. Let \(\bm{y}_n(t_{\tau}) = \left(y_{n}(t_1), y_{n}(t_2), \hdots, y_{n}(t_{\tau}) \right)\) be booking pattern \(n\) up to time \(t_{\tau}\), where \(n = 1, \hdots, N\), and each \(y_{n}(t_i)\) is a \(K\)-variate vector. In the functional setting, the multivariate functional halfspace depth of a pattern \(\bm{y}_n(t_{\tau}) = \left(y_{n}(t_1), y_{n}(t_2), \hdots, y_{n}(t_{\tau}) \right)\) is given by: 
    \begin{equation}
        MFHD_{N, \tau} (\bm{y}_n(t_{\tau}); \alpha) = \sum_{j=1}^{\tau} w_{\alpha, N}(t_j) HD_{N,j}(\bm{y}_{n}(t_{j}))
    \end{equation}
where, using \(t_{\tau+1} = t_{\tau} + 0.5(t_{\tau} - t_{\tau-1})\), the weights, \(w_{\alpha, N}(t_j)\), are, according to \citet{Hubert2012}: 
    \begin{equation}
        w_{\alpha, N}(t_j) = \frac{(t_{j+1} - t_j) \mbox{vol}\left[\left\{\bm{x} \in \R^k : HD_{N,j}(\bm{x}) \geq \alpha \right\}\right]}{\sum_{j = 1}^{\tau} (t_{j+1} - t_j) \mbox{vol}\left[\left\{\bm{x} \in \R^k : HD_{N,j}(\bm{x}) \geq \alpha \right\}\right]}
    \end{equation}
and the sample halfspace depth of a \(K\)-variate vector \(x\) at time \(t_j\) is given by \citep{Hubert2012}:
    \begin{equation}
        HD_{N,j}(x) = \frac{1}{N} min_{\bm{u}, \left||\bm{u}\right|| = 1} \# \left\{ y_{n}(t_j), n = 1, \hdots, N : \bm{u}^Ty_{n}(t_j) \geq \bm{u}^T\bm{x}\right\} 
    \end{equation}
In this paper, we are considering a univariate, \(K = 1\), functional halfspace depth since we choose to monitor booking patterns only. However, the definition of a multivariate functional halfspace depth opens up the possibility of jointly monitoring booking patterns and revenue patterns, for example. As described by \citet{Hubert2012}, computing the multivariate functional halfspace depth can be done with fast algorithms, and in this paper we use the R-package \texttt{mrfDepth} to do so. 

\subsection{Univariate Forecasting Techniques for Extrapolation}
Although an important element of a revenue management system is forecasting, there are multiple reasons why we create new forecasts to extrapolate rather than using the existing ones generated by the RM system. Three particular reasons are (i) depending on the optimisation routine used to set booking limits, forecasts of how demand builds up over time may not have been calculated. Some methods only require forecasts of final demand, and so the type of forecasts we wish to use for extrapolation may not exist. (ii) In the event that forecasts of how demand builds up over time do exist, historical forecasts may not be stored. In terms of identifying critical booking patterns in historical data, this also means the forecasts used for extrapolation are not available. (iii) Forecasts for how demand accumulates over time are typically based on data from similar historical booking patterns. The use of data from other booking patterns to extrapolate has the potential to mask outliers by normalising behaviour. Hence, at each time point we wish to create a forecast based solely on the data for an individual booking pattern, with the goal not being to accurately predict demand, but rather to amplify the differences between booking patterns. 
\subsubsection{Simple Exponential Smoothing (SES)}
SES works on the principle of averaging whilst down-weighting older observations. Further details can be found in \citet{Chatfield1975}. Given a time series \(y_{n}(t_1), y_{n}(t_2), \hdots, y_{n}(t_{\tau})\), a forecast for time \(t_{\tau +1}\), \(\hat{y}_{n}(t_{\tau+1})\) is given by:
    \begin{equation}
    \hat{y}_{n}(t_{\tau+1})  = \alpha y_{n}(t_{\tau}) + (1-\alpha) \hat{y}_{n}(t_{\tau}), 
    \end{equation}
for some smoothing constant, \(\alpha\). Note that this results in a constant forecast for the bookings from time \(t_{\tau +1}, \hdots, t_T \). Due to the inability of SES to cope with trend, we apply SES to the time series of demand per booking interval, rather than the time series of cumulative demand.

\subsubsection{Autoregressive Integrated Moving Average (ARIMA)}
ARIMA models incorporate a trend component, and assume that future observations are an additive, weighted combination of previous observations and previous errors. Let \(\bm{x}_{n}(t_{\tau})\) be the \(d^{th}\) differenced time series relating to \(\bm{y}_{n}(t_{\tau})\). See \citet{BoxJenkins} for an overview of differencing procedures, and \citet{Chatfield1975} for a description of ARIMA processes. The one-step ahead forecast \(\hat{x}_{n}(t_{\tau +1})\) is given by:
    \begin{equation}
    \hat{x}_{n}(t_{\tau +1})  = \mu + \phi_1 x_{n}(t_{\tau}) + \hdots + \phi_p x_n(t_{\tau -p+1}) - \theta_1 \epsilon(t_{\tau}) - \hdots - \theta_q \epsilon(t_{\tau -q+1})
    \end{equation}
for some constant mean \(\mu\), parameters \(\phi_1, \hdots, \phi_p, \theta_1, \hdots, \theta_q\) and white noise process \(\left(\epsilon_{t_j}\right)\). We use AIC and Dickey-Fuller tests, in combination with visual inspection, to select the orders \(p\), \(q\), and \(d\). See \citet{BoxJenkins}, and the R package \texttt{forecast}.

\subsubsection{Integrated Generalised Autoregressive Conditional Heteroskedasticity (IGARCH)}
IGARCH models incorporate a trend component and assume that the variance structure follows an autoregressive moving average model. Again, let \(\bm{x}_{n}(t_{\tau})\) be the \(d^{th}\) differenced time series relating to \(\bm{y}_{n}(t_{\tau})\). See \citet{Tsay2002} for further details on IGARCH processes. IGARCH(1,d,1) models assume the following structure:
    \begin{eqnarray}
    x_{n}(t_{\tau +1}) &=& \mu + \epsilon_{n}(t_{\tau +1}) \\ 
    \epsilon_{n}(t_{\tau +1}) &=& z_{n}(t_{\tau +1}) \sigma_{n}(t_{\tau +1}) \\ 
    \sigma^2_{n}(t_{\tau +1}) &=& w + \alpha \epsilon^2_{n}(t_{\tau +1}) + \beta  \sigma^2_{n}(t_{\tau})
    \end{eqnarray}
We assume that the order of the IGARCH model is \((1,d,1)\) to reduce computational time.


\newpage
\section{Details of Simulation-based Framework}
\label{sec:appsim}
\subsection{Forecasts}
\begin{table}[ht]
\centering
\begin{tabular}{ ccccccccc } 
 \hline \hline
& \textbf{Fare Class} 	& \textbf{Fare (\euro)} &	\multicolumn{2}{c}{\textbf{\thead{\(f_{D}\) = 0.9}}} 	& \multicolumn{2}{c}{\textbf{\thead{\(f_{D}\) = 1.2}}} 	& \multicolumn{2}{c}{\textbf{\thead{\(f_{D}\) = 1.5}}} 	\\ \hline
\(j\) &							& 		\(r_j\)			& \textbf{\(\hat{\mu_j}\)} 	& \textbf{\(\hat{\sigma_j}^2\)} 	& \textbf{\(\hat{\mu_j}\)} 	& \textbf{\(\hat{\sigma_j}^2\)} 	& \textbf{\(\hat{\mu_j}\)} 	& \textbf{\(\hat{\sigma_j}^2\)} 	\\ \hline
 1 & A	& 400	&	31.9 &	23.0 & 46.2	& 25.3	&	52.7	& 32.2 \\					 
 2 & O	& 300	&	17.5 &	14.2 & 24.2	& 18.8	&	28.3	& 30.5 \\
 3 & J	& 280	&	20.0 &	14.2 & 28.6	& 25.5	&	33.6	& 31.8 \\
 4 & P	& 240	&	16.8 &	16.1 & 22.9	& 26.6	&	26.1	& 23.8 \\ 
 5 & R	& 200	&	13.4 &	11.5 & 18.5	& 16.5	&	21.6	& 18.8 \\
 6 & S	& 185	&	12.3 &	14.3 & 16.9	& 11.2	&	21.0	& 21.1 \\
 7 & M	& 175   &	52.6 &	19.2 & 69.8	& 28.2	&	81.8	& 33.8 \\
 \hline \hline
\end{tabular}
\caption{Forecasts of mean and variance of demand for each fare class}
\label{tab:meanvaremsr}
\end{table}
In terms of choosing the number of replications of the simulation, \(N\), to use in the calculations of the forecasts, we consider the standard errors of the estimates. The standard error of the mean is given by:
\begin{equation}
    se(\hat{\mu_j}) = \frac{\hat{\sigma_j}}{\sqrt{N}},
\end{equation}
such that it is typically in the range of 0.3 - 0.6 when \(N=100\). The standard error of the variance is given by:
\begin{equation}
    se(\hat{\sigma^2_j}) = \hat{\sigma^2_j} \sqrt{\frac{2}{N-1}},
\end{equation}
and is typically in the range of 2 - 5 when \(N=100\). Therefore the number of simulations provides reasonable estimates of the demand mean and variance forecasts for each fare class.

\subsection{Optimisation Heuristics to Compute Booking Limits}
\subsubsection{Expected Marginal Seat Revenue-b (EMSRb)}
It is assumed that demand for each fare class, \(d_i\), is independent and normally distributed:
\begin{equation}
d_i \sim \mathcal{N}\left(\mu_i, \sigma^2_i\right),
\end{equation}
where \(\mu_i\) and \(\sigma^2_i\) are forecasted as described above. The protection level for fare class \(j\) is given by \citep{Belobaba1992}:
\begin{equation} \label{eqn:emsrb1}
PL_j = F_j^{-1}\left(1 - \frac{r_{j+1}}{\tilde{r}_j}\right) \mbox{    for }j = 1,\ldots, |\mathcal{J}|-1
\end{equation}
where \(F_j\) is the (Gaussian) distribution of demand for fare class \(j\), and \(r_j\) is the fare in fare class \(j\). \(\tilde{r}_j\) is the weighted-average revenue from classes \(1, \hdots, j\): 
\begin{equation}
\tilde{r}_j = \frac{\sum_{k=1}^j r_k \mu_k}{\sum_{k=1}^j \mu_k}.
\end{equation}
Note that the protection level for all fare classes, \(PL_{|\mathcal{J}|}\), is simply equal to the capacity, \(C\). As stated by \citet{Talluri2004}, Equation \eqref{eqn:emsrb1} becomes:
\begin{equation}
PL_j = \mu + \Phi^{-1}\left(1 - \frac{r_{j+1}}{\tilde{r}_j}\right)\sigma \mbox{    for }j = 1,\ldots, |\mathcal{J}|-1,
\end{equation}
where \(\mu = \sum_{k=1}^j \mu_k\) is the mean, and \(\sigma^2 = \sum_{k=1}^j \sigma^2_k\) is the variance, of the aggregated demand. Hence, the booking limit for class \(j\) is given by the capacity minus the protection level for classes \(j-1\) and higher:
\begin{equation}
BL_j = C - PL_{j-1}.
\end{equation}

\subsubsection{Expected Marginal Seat Revenue-b with Marginal Revenue Transformation (EMSRb-MR)}
The following marginal revenue transformation, described by \citet{Fiig2010}, assumes that customers only buy the lowest available fare, even if they would be willing to pay more. In this setting, letting \(k\) be the lowest available fare product, the demand for all other fare products becomes zero:
\begin{equation}
    \mu_j = 0 \mbox{  } \forall j \neq k.
\end{equation}
Therefore the adjusted demand for fare class \(j\) becomes:
\begin{equation}
    \mu^{'}_j = \mu_j - \mu_{j-1}.
\end{equation}
The adjusted fares are given by:
\begin{equation}
    r^{'}_j = \frac{r_j\mu_j - r_{j-1}\mu_{j-1}}{\mu_j - \mu_{j-1}}.
\end{equation}
An alternative method of calculating adjusted fares without explicitly forecasting demand for each fare class is to assume that: 
\begin{equation}
    \mu_j = \mu_n psup_j,
\end{equation}
the demand for a particular fare class is the baseline demand for the lowest fare class, \(\mu_n\), multiplied by a sell-up probability, \(psup_j\). In practice, these sell-up probabilities can be forecasted instead of the fare class demand assuming an independent model. In our case, due to comparing EMSRb with EMSRb-MR, we have the fare class forecasts already. The two methods are equivalent.

The booking controls under EMSRb and EMSRb-MR are shown in Table \ref{tab:bookingcontrols}, where the demand factor, \(f_{D}\), is defined as the ratio of demand, \(D\), to capacity, \(C\).
\begin{table}[ht]
\centering
\begin{tabular}{ ccccccc } 
 \hline \hline
\textbf{Fare Class}  &	\multicolumn{2}{c}{\textbf{\thead{\(f_{D}\) = 0.9}}} 	& \multicolumn{2}{c}{\textbf{\thead{\(f_{D}\) = 1.2}}} 	& \multicolumn{2}{c}{\textbf{\thead{\(f_{D}\) = 1.5}}} 	\\ \hline
			& \textbf{\thead{EMSRb}} 	& \textbf{\thead{EMSRb-MR}} 	& \textbf{\thead{EMSRb}} 	& \textbf{\thead{EMSRb-MR}} 	& \textbf{\thead{EMSRb}} 	& \textbf{\thead{EMSRb-MR}} 	\\ \hline
 A	&	200	&	200     & 200	& 200  &	200  & 200		\\
 O	&	171	&	165	    & 157	& 151  &	151  & 144		\\
 J	&	155	&	155	    & 134	& 134  &	125  & 125		\\
 P	&	134	&	125     & 105	& 95   &	90   & 79		\\ 
 R	&	117	&	109	    & 81	& 72   &	62   & 52	    \\
 S	&	104	&	109	    & 62	& 72   &	39   & 52		\\
 M	&	91	&	96	    & 45	& 51   &	18   & 24		\\
 \hline \hline
\end{tabular}
\caption{Booking limits under EMSRb and EMSRb-MR}
\label{tab:bookingcontrols}
\end{table}

\begin{table}[htbp]
\resizebox{\textwidth}{!}{\begin{tabular}{lllllllllllllll}
\hline \hline
\multicolumn{1}{c}{\textbf{\rotatebox{90}{\multirow{2}{3cm}{Optimisation \\}}}} & \hspace{0.5cm}\textbf{\rotatebox{90}{\multirow{2}{3cm}{Magnitude of Outliers}}} & \hspace{0.1cm}\textbf{\rotatebox{90}{\multirow{2}{3cm}{Frequency of Outliers}}} & \cellcolor{gray!20}\hspace{-0.3cm}\rotatebox{90}{\thead{\multirow{2}{3cm}{Nonparametric Percentiles}}} & \hspace{-0.3cm}\rotatebox{90}{\thead{\multirow{2}{3cm}{Nonparametric \\Tolerance Intervals}}} & \cellcolor{gray!20}\hspace{-0.3cm}\rotatebox{90}{\thead{\multirow{2}{3cm}{Poisson Tolerance Intervals}}} & \hspace{-0.3cm}\rotatebox{90}{\thead{\multirow{2}{3cm}{Robust \\ Z-Score}}} & 
\cellcolor{gray!20}\hspace{-0.3cm}\rotatebox{90}{\thead{\multirow{2}{3cm}{Euclidean \\ Distance}}} & 
\hspace{-0.3cm}\rotatebox{90}{\thead{\multirow{2}{3cm}{Manhattan \\Distance}}} & 
\cellcolor{gray!20}\hspace{-0.3cm}\rotatebox{90}{\thead{\multirow{2}{3cm}{\(k\)-Means Clustering  (Euclidean)}}} & \hspace{-0.3cm}\rotatebox{90}{\thead{\multirow{2}{3cm}{\(k\)-Means Clustering \\ (Manhattan)}}} & \cellcolor{gray!20}\hspace{-0.3cm}\rotatebox{90}{\thead{\multirow{2}{3cm}{Functional\\Depth}}} & 
\hspace{-0.3cm}\rotatebox{90}{\thead{\multirow{2}{3cm}{Functional Depth \\ + SES}}} & \cellcolor{gray!20}\hspace{-0.3cm}\rotatebox{90}{\thead{\multirow{2}{3cm}{Functional Depth \\ + ARIMA}}} & \hspace{-0.3cm}\rotatebox{90}{\thead{\multirow{2}{3cm}{Functional Depth \\ + IGARCH}}} \\
\hline
\multirow{12}{*}{\rotatebox{90}{EMSRb}}                                                             & \multirow{3}{*}{-25\%}    & 1\%  & \cellcolor{gray!20}    &     & \cellcolor{gray!20}\hspace{0.1cm}0.73 &    &    \cellcolor{gray!20}&   & \cellcolor{gray!20}\hspace{0.1cm}0.68     &     & \cellcolor{gray!20}\hspace{0.1cm}0.94    &     & \cellcolor{gray!20}\hspace{0.1cm}0.93    &    \\
&     & 5\%     & \cellcolor{gray!20}\hspace{0.1cm}0.69 & \hspace{0.1cm}0.63 & \cellcolor{gray!20}\hspace{0.1cm}0.74  & \hspace{0.1cm}0.50     & \cellcolor{gray!20}\hspace{0.1cm}0.69 & \hspace{0.1cm}0.68      & \cellcolor{gray!20}\hspace{0.1cm}0.67  & \hspace{0.1cm}0.68     & \cellcolor{gray!20}\hspace{0.1cm}0.93  & \hspace{0.1cm}0.94     & \cellcolor{gray!20}\hspace{0.1cm}0.94  & \hspace{0.1cm}0.93   \\
&  & 10\%  &   \cellcolor{gray!20} &  & \cellcolor{gray!20}\hspace{0.1cm}0.70 &    &   \cellcolor{gray!20} &  & \cellcolor{gray!20}\hspace{0.1cm}0.66   &  & \cellcolor{gray!20}\hspace{0.1cm}0.93   &   & \cellcolor{gray!20}\hspace{0.1cm}0.93   &    \\ \cline{2-15}
& \multirow{3}{*}{-12.5\%}   & 1\%   & \cellcolor{gray!20}  &  & \cellcolor{gray!20}\hspace{0.1cm}0.55 &  &  \cellcolor{gray!20}  &  & \cellcolor{gray!20}\hspace{0.1cm}0.58   &  & \cellcolor{gray!20}\hspace{0.1cm}0.93  & & \cellcolor{gray!20}\hspace{0.1cm}0.95  &  \\
&  & 5\%   & \cellcolor{gray!20}\hspace{0.1cm}0.56  & \hspace{0.1cm}0.55             & \cellcolor{gray!20}\hspace{0.1cm}0.56  & \hspace{0.1cm}0.50 & \cellcolor{gray!20}\hspace{0.1cm}0.56   & \hspace{0.1cm}0.55   & \cellcolor{gray!20}\hspace{0.1cm}0.57   & \hspace{0.1cm}0.57   & \cellcolor{gray!20}\hspace{0.1cm}0.92   & \hspace{0.1cm}0.92   & \cellcolor{gray!20}\hspace{0.1cm}0.93   & \hspace{0.1cm}0.93   \\
 &   & 10\% &   \cellcolor{gray!20}  &  & \cellcolor{gray!20}\hspace{0.1cm}0.54  & & \cellcolor{gray!20}  &  & \cellcolor{gray!20}\hspace{0.1cm}0.56  &  & \cellcolor{gray!20}\hspace{0.1cm}0.93 &  & \cellcolor{gray!20}\hspace{0.1cm}0.93 &  \\ \cline{2-15}
& \multirow{3}{*}{+12.5\%}  & 1\% & \cellcolor{gray!20} & & \cellcolor{gray!20}\hspace{0.1cm}0.53 &  & \cellcolor{gray!20} & & \cellcolor{gray!20}\hspace{0.1cm}0.59 & & \cellcolor{gray!20}\hspace{0.1cm}0.92  &  & \cellcolor{gray!20}\hspace{0.1cm}0.93  & \\ 
& & 5\%  & \cellcolor{gray!20}\hspace{0.1cm}0.56 & \hspace{0.1cm}0.55  & \cellcolor{gray!20}\hspace{0.1cm}0.53  & \hspace{0.1cm}0.51  & \cellcolor{gray!20}\hspace{0.1cm}0.53  & \hspace{0.1cm}0.53  & \cellcolor{gray!20}\hspace{0.1cm}0.58  & \hspace{0.1cm}0.56  & \cellcolor{gray!20}\hspace{0.1cm}0.93  & \hspace{0.1cm}0.92  & \cellcolor{gray!20}\hspace{0.1cm}0.93  & \hspace{0.1cm}0.93   \\
&  & 10\% &  \cellcolor{gray!20} & & \cellcolor{gray!20}\hspace{0.1cm}0.51 & & \cellcolor{gray!20} &  & \cellcolor{gray!20}\hspace{0.1cm}0.59 &  & \cellcolor{gray!20}\hspace{0.1cm}0.92  &  & \cellcolor{gray!20}\hspace{0.1cm}0.92  & \\ \cline{2-15}
& \multirow{3}{*}{+25\%}  & 1\% & \cellcolor{gray!20} &  & \cellcolor{gray!20}\hspace{0.1cm}0.59 & & \cellcolor{gray!20} & & \cellcolor{gray!20}\hspace{0.1cm}0.65 & & \cellcolor{gray!20}\hspace{0.1cm}0.94 & & \cellcolor{gray!20}\hspace{0.1cm}0.92 & \\
& & 5\% & \cellcolor{gray!20}\hspace{0.1cm}0.65 & \hspace{0.1cm}0.60 & \cellcolor{gray!20}\hspace{0.1cm}0.62 & \hspace{0.1cm}0.54  & \cellcolor{gray!20}\hspace{0.1cm}0.61 & \hspace{0.1cm}0.61  & \cellcolor{gray!20}\hspace{0.1cm}0.69 & \hspace{0.1cm}0.68  & \cellcolor{gray!20}\hspace{0.1cm}0.92 & \hspace{0.1cm}0.93  & \cellcolor{gray!20}\hspace{0.1cm}0.94 & \hspace{0.1cm}0.94  \\
& & 10\% & \cellcolor{gray!20} & & \cellcolor{gray!20}\hspace{0.1cm}0.60 & &  \cellcolor{gray!20} &  & \cellcolor{gray!20}\hspace{0.1cm}0.68 & & \cellcolor{gray!20}\hspace{0.1cm}0.92 & & \cellcolor{gray!20}\hspace{0.1cm}0.93 & \\ \cline{1-15} \\ \hline
\multirow{12}{*}{\rotatebox{90}{EMSRb-MR}} & \multirow{3}{*}{-25\%}                      & 1\% & \cellcolor{gray!20} & & \cellcolor{gray!20}\hspace{0.1cm}0.73 & & \cellcolor{gray!20} & & \cellcolor{gray!20}\hspace{0.1cm}0.65 & & \cellcolor{gray!20}\hspace{0.1cm}0.92 & & \cellcolor{gray!20}\hspace{0.1cm}0.92 &  \\
& & 5\% & \cellcolor{gray!20}\hspace{0.1cm}0.68 & \hspace{0.1cm}0.62 & \cellcolor{gray!20}\hspace{0.1cm}0.77 & \hspace{0.1cm}0.50 & \cellcolor{gray!20}\hspace{0.1cm}0.68 & \hspace{0.1cm}0.68 & \cellcolor{gray!20}\hspace{0.1cm}0.66 & \hspace{0.1cm}0.67 & \cellcolor{gray!20}\hspace{0.1cm}0.92 & \hspace{0.1cm}0.93 & \cellcolor{gray!20}\hspace{0.1cm}0.93 & \hspace{0.1cm}0.92  \\
& & 10\% & \cellcolor{gray!20} & & \cellcolor{gray!20}\hspace{0.1cm}0.71  &  &  \cellcolor{gray!20} &  & \cellcolor{gray!20}\hspace{0.1cm}0.67 &  & \cellcolor{gray!20}\hspace{0.1cm}0.91  &  & \cellcolor{gray!20}\hspace{0.1cm}0.92 &  \\ \cline{2-15}
& \multirow{3}{*}{-12.5\%} & 1\% & \cellcolor{gray!20} & & \cellcolor{gray!20}\hspace{0.1cm}0.55 &  & \cellcolor{gray!20} & & \cellcolor{gray!20}\hspace{0.1cm}0.57 &  & \cellcolor{gray!20}\hspace{0.1cm}0.92 & & \cellcolor{gray!20}\hspace{0.1cm}0.92  & \\
& & 5\% & \cellcolor{gray!20}\hspace{0.1cm}0.56 & \hspace{0.1cm}0.54 & \cellcolor{gray!20}\hspace{0.1cm}0.57 & \hspace{0.1cm}0.50 & \cellcolor{gray!20}\hspace{0.1cm}0.55 & \hspace{0.1cm}0.55 & \cellcolor{gray!20}\hspace{0.1cm}0.57 & \hspace{0.1cm}0.56 & \cellcolor{gray!20}\hspace{0.1cm}0.93 & \hspace{0.1cm}0.93 & \cellcolor{gray!20}\hspace{0.1cm}0.93 & \hspace{0.1cm}0.92 \\
& & 10\% &  \cellcolor{gray!20} & & \cellcolor{gray!20}\hspace{0.1cm}0.59 & & \cellcolor{gray!20} & & \cellcolor{gray!20}\hspace{0.1cm}0.60 & & \cellcolor{gray!20}\hspace{0.1cm}0.93 & & \cellcolor{gray!20}\hspace{0.1cm}0.92 & \\ \cline{2-15}
& \multirow{3}{*}{+12.5\%} & 1\% & \cellcolor{gray!20} & & \cellcolor{gray!20}\hspace{0.1cm}0.51 &  & \cellcolor{gray!20} & & \cellcolor{gray!20}\hspace{0.1cm}0.56 &  & \cellcolor{gray!20}\hspace{0.1cm}0.93 & & \cellcolor{gray!20}\hspace{0.1cm}0.93 &  \\
& & 5\% & \cellcolor{gray!20}\hspace{0.1cm}0.56 & \hspace{0.1cm}0.55  & \cellcolor{gray!20}\hspace{0.1cm}0.54  & \hspace{0.1cm}0.51  & \cellcolor{gray!20}\hspace{0.1cm}0.53 & \hspace{0.1cm}0.53  & \cellcolor{gray!20}\hspace{0.1cm}0.57 & \hspace{0.1cm}0.55 & \cellcolor{gray!20}\hspace{0.1cm}0.92 & \hspace{0.1cm}0.93 & \cellcolor{gray!20}\hspace{0.1cm}0.93  & \hspace{0.1cm}0.92   \\
& & 10\% & \cellcolor{gray!20} & & \cellcolor{gray!20}\hspace{0.1cm}0.51 & & \cellcolor{gray!20} & & \cellcolor{gray!20}\hspace{0.1cm}0.59  &  & \cellcolor{gray!20}\hspace{0.1cm}0.93 & & \cellcolor{gray!20}\hspace{0.1cm}0.92  & \\ \cline{2-15}
& \multirow{3}{*}{+25\%} & 1\% & \cellcolor{gray!20} & & \cellcolor{gray!20}\hspace{0.1cm}0.63 & & \cellcolor{gray!20} & & \cellcolor{gray!20}\hspace{0.1cm}0.68 & & \cellcolor{gray!20}\hspace{0.1cm}0.94 & & \cellcolor{gray!20}\hspace{0.1cm}0.93  & \\
& & 5\% & \cellcolor{gray!20}\hspace{0.1cm}0.66 & \hspace{0.1cm}0.61 & \cellcolor{gray!20}\hspace{0.1cm}0.65 & \hspace{0.1cm}0.54 & \cellcolor{gray!20}\hspace{0.1cm}0.62 & \hspace{0.1cm}0.62 & \cellcolor{gray!20}\hspace{0.1cm}0.69 & \hspace{0.1cm}0.69 & \cellcolor{gray!20}\hspace{0.1cm}0.92 & \hspace{0.1cm}0.93 & \cellcolor{gray!20}\hspace{0.1cm}0.94 & \hspace{0.1cm}0.92 \\
& & 10\% & \cellcolor{gray!20} & & \cellcolor{gray!20}\hspace{0.1cm}0.61 & & \cellcolor{gray!20} & & \cellcolor{gray!20}\hspace{0.1cm}0.70 & & \cellcolor{gray!20}\hspace{0.1cm}0.93 & & \cellcolor{gray!20}\hspace{0.1cm}0.92 & 
\\ \hline \hline
\end{tabular}}
\caption{Balanced classification rate (offline) results for extended simulation study}
\label{tab:experiments_bcr}
\end{table}


\newpage
\section{Additional Results}
\label{app:results} \vspace{-0.2cm}
\subsection{Comparison of Booking Limit Heuristics} \label{app:optim} 
\begin{figure}[!h]
    \centering
    \includegraphics[width=0.4\textwidth, height=0.4\textwidth]{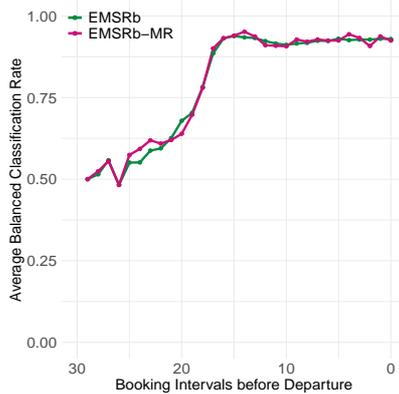}
	\caption{EMSRb vs. EMSRb-MR under functional depth with ARIMA extrapolation}  
	\label{fig:compoptim}
\end{figure}

\vspace{-0.1cm}
Table \ref{tab:revenue} shows the resulting revenue under EMSRb and EMSRb-MR booking limits with different demand factors, as compared to accepting bookings on a first-come-first-served basis (FCFS). Both heuristics offer an improvement over FCFS. Given the presence of buy-down in the demand model, EMSRb-MR outperforms EMSRb, particularly in situations that feature a high demand-to-capacity ratio.
\begin{table}[htbp]
\centering
\begin{tabular}{ cccc } 
 \hline \hline
 \textbf{\thead{Demand Factor}} & \textbf{\thead{FCFS Revenue (\euro)}} & \textbf{\thead{EMSRb as \\ Factor of FCFS}} & \textbf{\thead{EMSRb-MR as \\ Factor of FCFS}}  \\ \hline
0.90 			& 28948.50			& 1.03  		& 	1.06      \\
1.20 			& 34835.50			& 1.04  	 	& 	1.08      \\
1.50	 		& 35000.00			& 1.05  		& 	1.09      \\ 
 \hline \hline
\end{tabular}
\caption{Revenue generated under EMSRb vs EMSRb-MR booking controls}
\label{tab:revenue}
\end{table} \vspace{-0.5cm}
Given the significant impact of heuristic choice on revenue, we investigate whether the superior performance of EMSRb-MR also results in a change in outlier detection performance. Figure \ref{fig:compoptim} shows the balanced classification rate of functional depth with ARIMA extrapolation outlier detection, under EMSRb and EMSRb-MR heuristics. There is no significant impact on outlier detection performance.

\newpage
\subsection{Sensitivity to Frequency of Outliers}
\begin{figure}[h!]
    \centering
    \includegraphics[width=0.45\textwidth, height=0.45\textwidth]{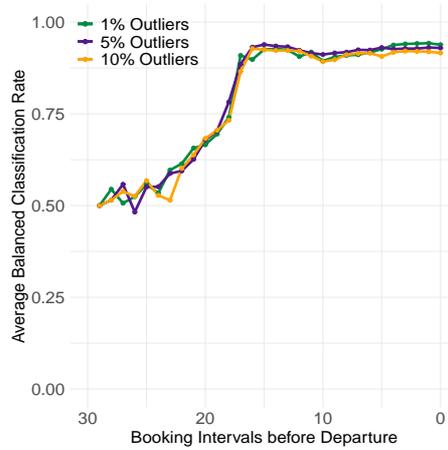}
	\caption{Balanced  Classification  Rate  under  different  frequencies  of  outliers for functional depth with ARIMA extrapolation}
    \label{fig:percentages}
\end{figure}

We test the sensitivity of the functional depth (with and without extrapolation) to the different frequencies of outliers i.e. the proportion of booking patterns considered which are genuine outliers. There is no significant change in performance as the frequency of outliers changes, shown in Figure \ref{fig:percentages}. This consistent performance of the functional depth-based methods is down to the fact that it does not classify a specific proportion of the data as outlying. Given this, our simulation study considers the case where 5\% of the $N = 500$ booking patterns are genuine outliers.

\newpage
\subsection{$K$-means clustering with ARIMA extrapolation}
\begin{figure}[h!]
    \centering
    \includegraphics[width=0.52\textwidth, height=0.52\textwidth]{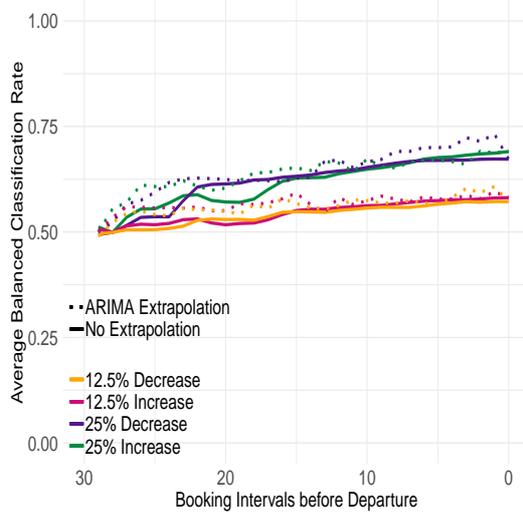}
	\caption{Balanced Classification Rate for $K$-means clustering with ARIMA extrapolation for 5\% outlier frequency over different magnitudes of demand outliers.}
    \label{fig:km_extrap}
\end{figure}
As noted in Section 4, extrapolation could also be used with the multivariate outlier detection approaches. Although in this paper we have chosen to focus on combining the extrapolation with the most promising outlier detection method (functional depth), we also present results here (see Figure \ref{fig:km_extrap}) on combining extrapolation with $K$-means clustering. As when combining extrapolation with functional depth (see Appendix C.6), the extrapolation increases the number of booking patterns classified as outliers. Extrapolation does provide an improvement in outlier detection performance, though the increase in performance is smaller in comparison when combined with functional depth. This is unsurprising given the poor performance of $K$-means clustering even when the curves are fully observed. The overall performance is still not as good as combining extrapolation with functional depth (or even functional depth without extrapolation).

\newpage
\subsection{Motivating the Use of Functional Analysis}
\subsubsection{Importance of Time-Ordered Observations}
\begin{figure}[!h]
    \centering
        \includegraphics[width=0.52\textwidth,height=0.52\textwidth]{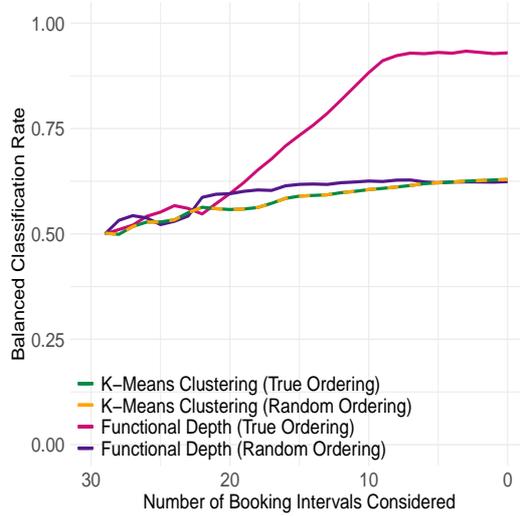}
    \caption{Comparison of correct time-ordering vs. random time-ordering}
    \label{fig:random_order}
\end{figure}
To stress the importance of the time-ordering of the observations, we benchmark $K$-means and functional depth on two cases where (i) the observations per pattern are correctly ordered, and (ii) the observations are shuffled randomly. As shown in Figure \ref{fig:random_order}, the performance of $K$-means clustering is independent of this change, as the observations stay the same and only their order, which $K$-means does not consider, changes. However, Figure \ref{fig:random_order} shows that the performance of functional depth improves when the data is ordered, as this method can exploit this characteristic of the booking patterns. In other words, if the booking patterns were not time-ordered, multivariate approaches could indeed outperform functional depth. However, this temporal dependency does exist and motivates our use of functional analysis.

\newpage
\subsubsection{Dealing with High Dimensionality}
The ROC analysis in Section 6.2 shows that the performance of $K$-means clustering becomes poorer as the dimensionality increases. The use of principal component analysis (PCA) was considered as a method to reduce the dimensionality for both distance measures and $K$-means clustering approaches. However, preliminary results were poor. When selecting the principal components to maximise the proportion of variance explained, it was deemed best to chose the time points early in the booking horizon. This is due to the censoring caused by the booking controls, i.e. when no booking limits have yet been reached, demand is more varied. This means that if we choose the principal components which best explain the variance, we no longer take into account the newest information. As such, detection accuracy was inferior to using the full vector of bookings, despite the dimensionality issue.

\newpage
\subsection{True Positive Rates}
\begin{figure}[!ht]
    \centering
    \begin{subfigure}[h]{0.47\textwidth}
    \centering
        \includegraphics[width=0.8\textwidth,height=0.8\textwidth]{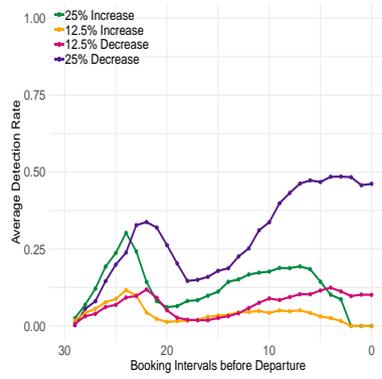}
        \caption{TPR for (Poisson) tolerance \\ intervals}  
		\label{fig:tpr_ptol}
    \end{subfigure}
    \begin{subfigure}[h]{0.47\textwidth}  
    \centering 
        \includegraphics[width=0.8\textwidth,height=0.8\textwidth]{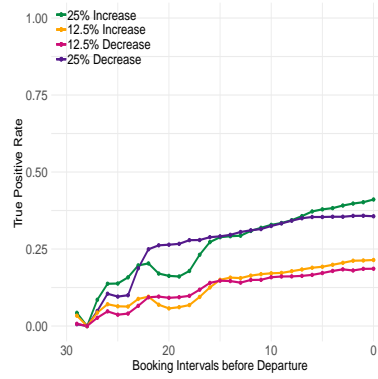}
        \caption{TPR for $K$-means clustering \\ (Euclidean distance)}
		\label{fig:tpr_km}
    \end{subfigure}
    \quad
    \begin{subfigure}[h]{0.47\textwidth}
    \centering
        \includegraphics[width=0.8\textwidth,height=0.8\textwidth]{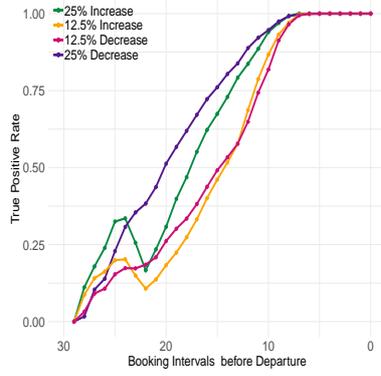}
        \caption{TPR for functional \\ depth}  
		\label{fig:tpr_func}
    \end{subfigure}
    \begin{subfigure}[h]{0.47\textwidth}  
    \centering 
        \includegraphics[width=0.8\textwidth,height=0.8\textwidth]{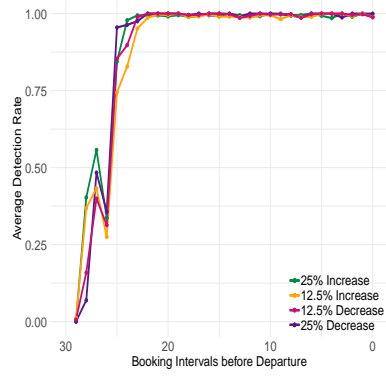}
        \caption{TPR for functional depth \\ with ARIMA extrapolation}
		\label{fig:tpr_ARIMA}
    \end{subfigure}
    \quad
    \caption{True positive rates for various outlier detection methods}
    \label{fig:tpr_results}
\end{figure}

\newpage
\subsection{False Positive Rates}
\begin{figure}[!ht]
    \centering
    \begin{subfigure}[h]{0.47\textwidth}
    \centering
        \includegraphics[width=0.8\textwidth,height=0.8\textwidth]{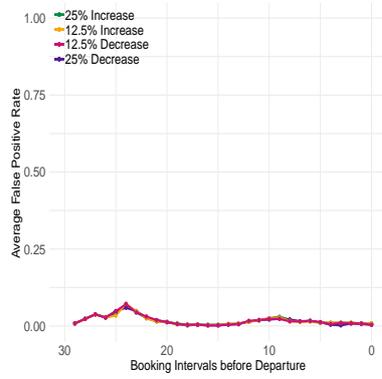}
        \caption{FPR for (Poisson) tolerance \\ intervals}  
		\label{fig:fpr_ptol}
    \end{subfigure}
    \begin{subfigure}[h]{0.47\textwidth}  
    \centering 
        \includegraphics[width=0.8\textwidth,height=0.8\textwidth]{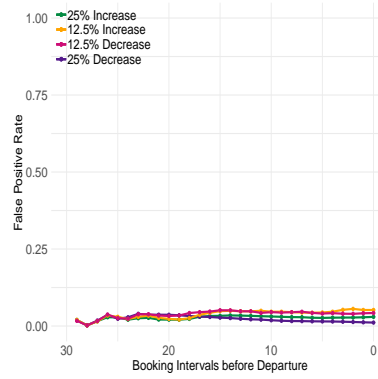}
        \caption{FPR for $K$-means clustering \\ (Euclidean distance)}
		\label{fig:fpr_km}
    \end{subfigure}
    \quad
    \begin{subfigure}[h]{0.47\textwidth}
    \centering
        \includegraphics[width=0.8\textwidth,height=0.8\textwidth]{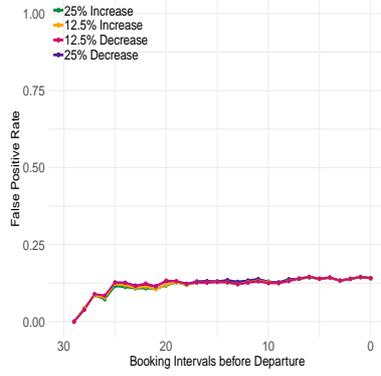}
        \caption{FPR for functional \\ depth}  
		\label{fig:fpr_func}
    \end{subfigure}
    \begin{subfigure}[h]{0.47\textwidth}  
    \centering 
        \includegraphics[width=0.8\textwidth,height=0.8\textwidth]{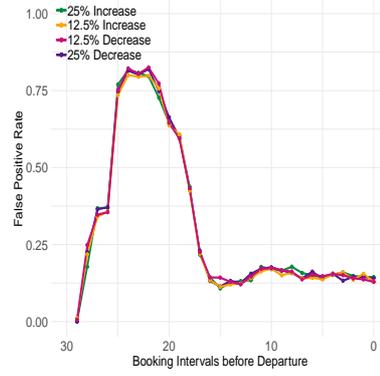}
        \caption{FPR for functional depth \\ with ARIMA extrapolation}
		\label{fig:fpr_ARIMA}
    \end{subfigure}
    \quad
    \caption{False positive rates for various outlier detection methods}
    \label{fig:fpr_results}
\end{figure}

\newpage
We suggest that early in the booking horizon, all methods perform poorly but for different reasons -- some suffer from low true positive rates, others from high false positive rates. The balanced classification rate (BCR) does not allow us to easily compare these two situations. In order to test this hypothesis, and investigate the spike in false positives early in the booking horizon when incorporating extrapolation, we additionally consider the Positive Likelihood Ratio (LR+) \citep{Habibzadeh2019}. That is, the ratio between the true positive rate, and the false positive rate:
\begin{equation}
    LR+ = \frac{TP/(TP+FN)}{FP/(FP+TN)}
\end{equation}
A higher LR+ (specifically those greater than 1), represents the fact that a booking pattern classified as an outlier is more likely to be a genuine outlier.
\begin{figure}[!ht]
    \centering
    \begin{subfigure}[h]{0.47\textwidth}
    \centering
        \includegraphics[width=0.8\textwidth,height=0.8\textwidth]{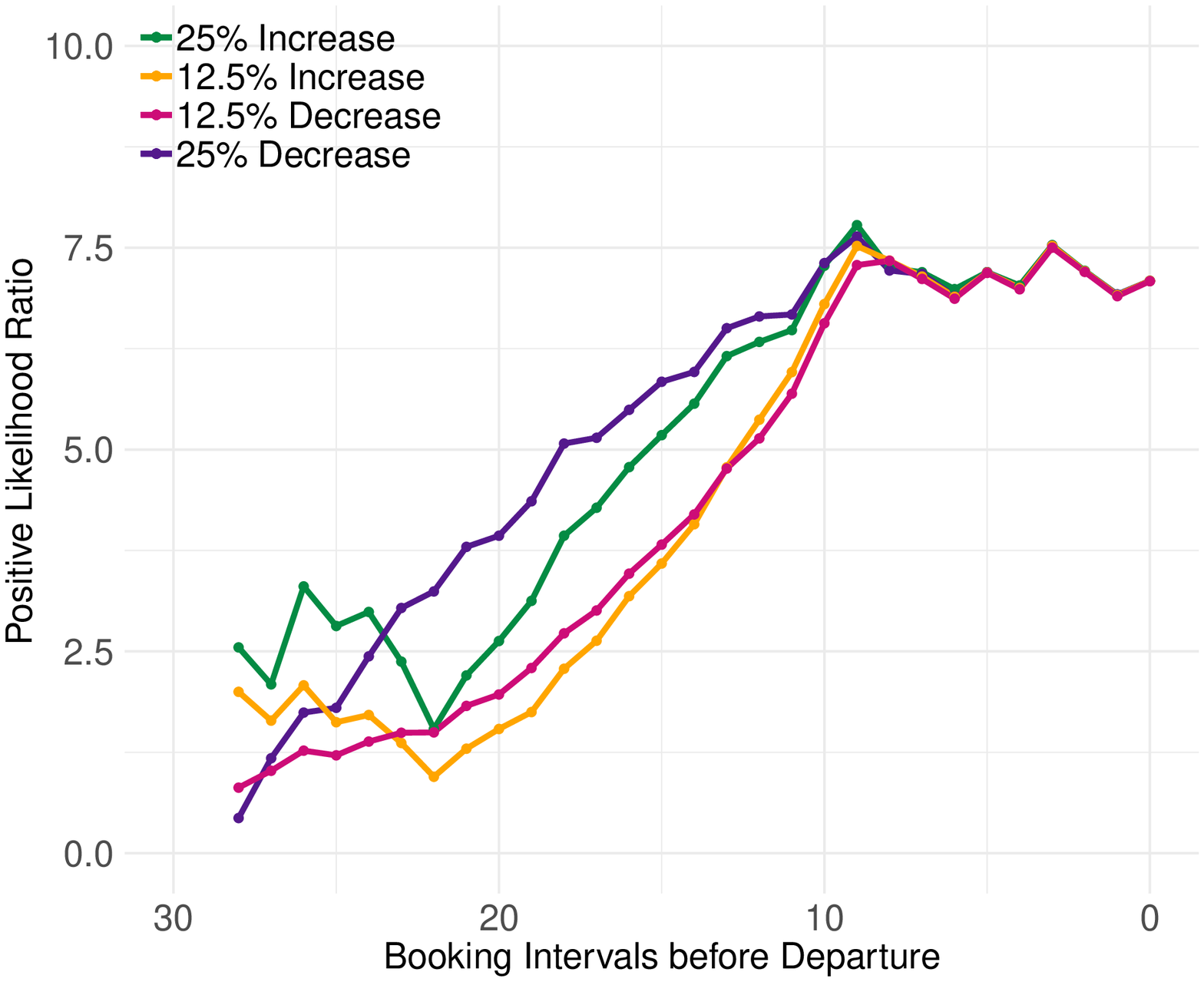}
        \caption{LR+ for functional \\ depth}  
		\label{fig:plr_func}
    \end{subfigure}
    \begin{subfigure}[h]{0.47\textwidth}  
    \centering 
        \includegraphics[width=0.8\textwidth,height=0.8\textwidth]{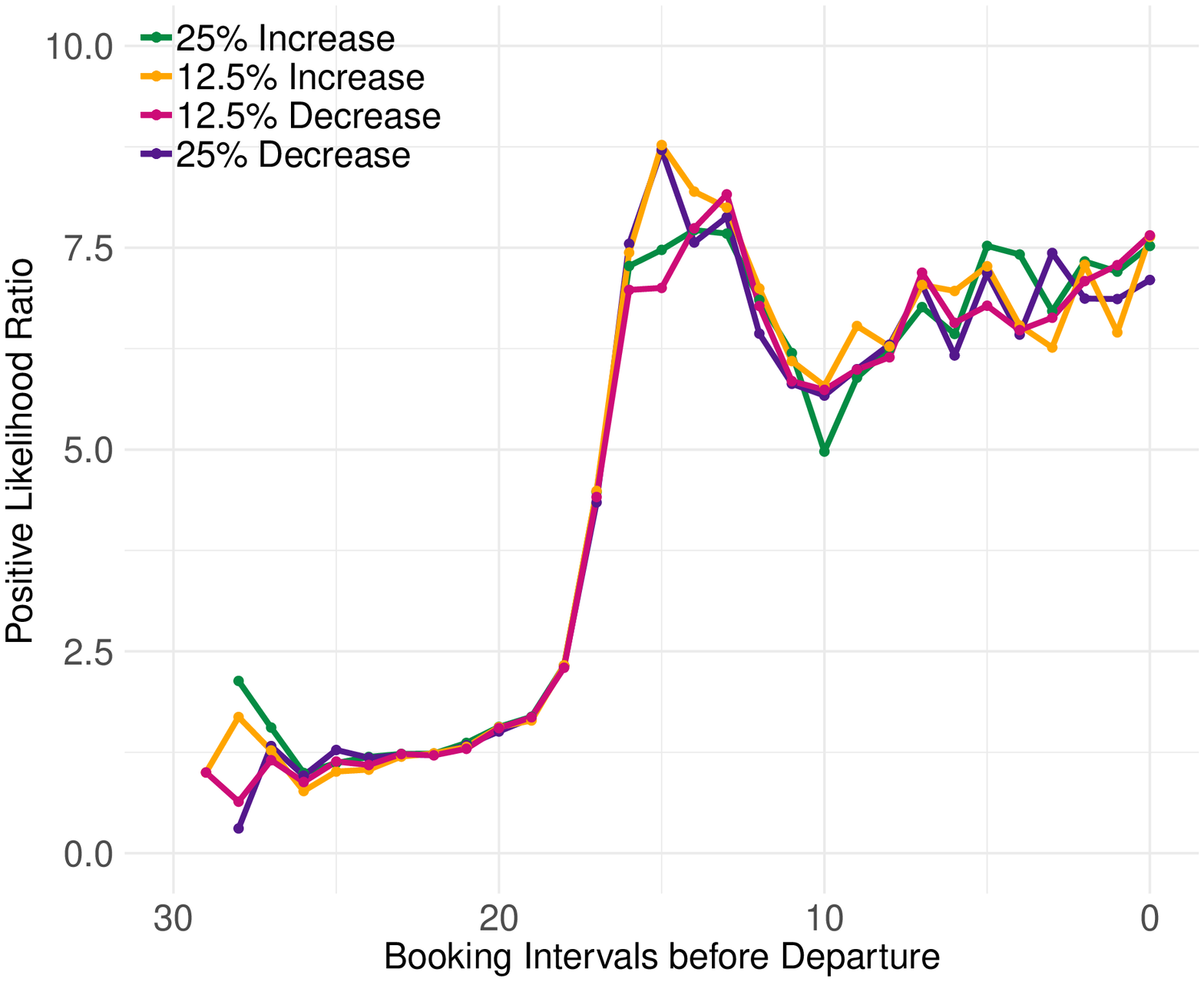}
        \caption{LR+ for functional depth with \\ ARIMA extrapolation}
		\label{fig:plr_ARIMA}
    \end{subfigure}
    \quad
    \caption{Positive likelihood ratio for functional depth with and without ARIMA extrapolation}
    \label{fig:plr_results}
\end{figure}
The results, shown in Figure \ref{fig:plr_results}, show that functional depth both with and without extrapolation performs poorly early in the horizon -- with a LR+ just slightly above 1. Due to the high false positive rate, functional depth without extrapolation may even be marginally better early in the horizon. However, outlier detection with the inclusion of extrapolation reaches its peak LR+ around 16 intervals before departure, compared to 9 intervals for functional depth alone. This peak in functional depth with ARIMA extrapolation corresponds to both the sharp increase in true positives (Figure \ref{fig:tpr_ARIMA}), and sharp drop-off in false positives (Figure \ref{fig:fpr_ARIMA}). A similar impact was observed when $K$-means clustering was combined with extrapolation, that is, the number of booking patterns classified as outliers increased. These results, in addition with the ROC curves shown in Section 6.2, show that, on balance, it is still beneficial to include extrapolation into the outlier detection, especially in the middle portion of the booking horizon. However, classification results from all methods should be treated with caution very early in the booking horizon for the reasons outlined.

\begin{figure}[!ht]
    \centering
    \begin{subfigure}[h]{0.47\textwidth}
    \centering
        \includegraphics[width=0.8\textwidth,height=0.8\textwidth]{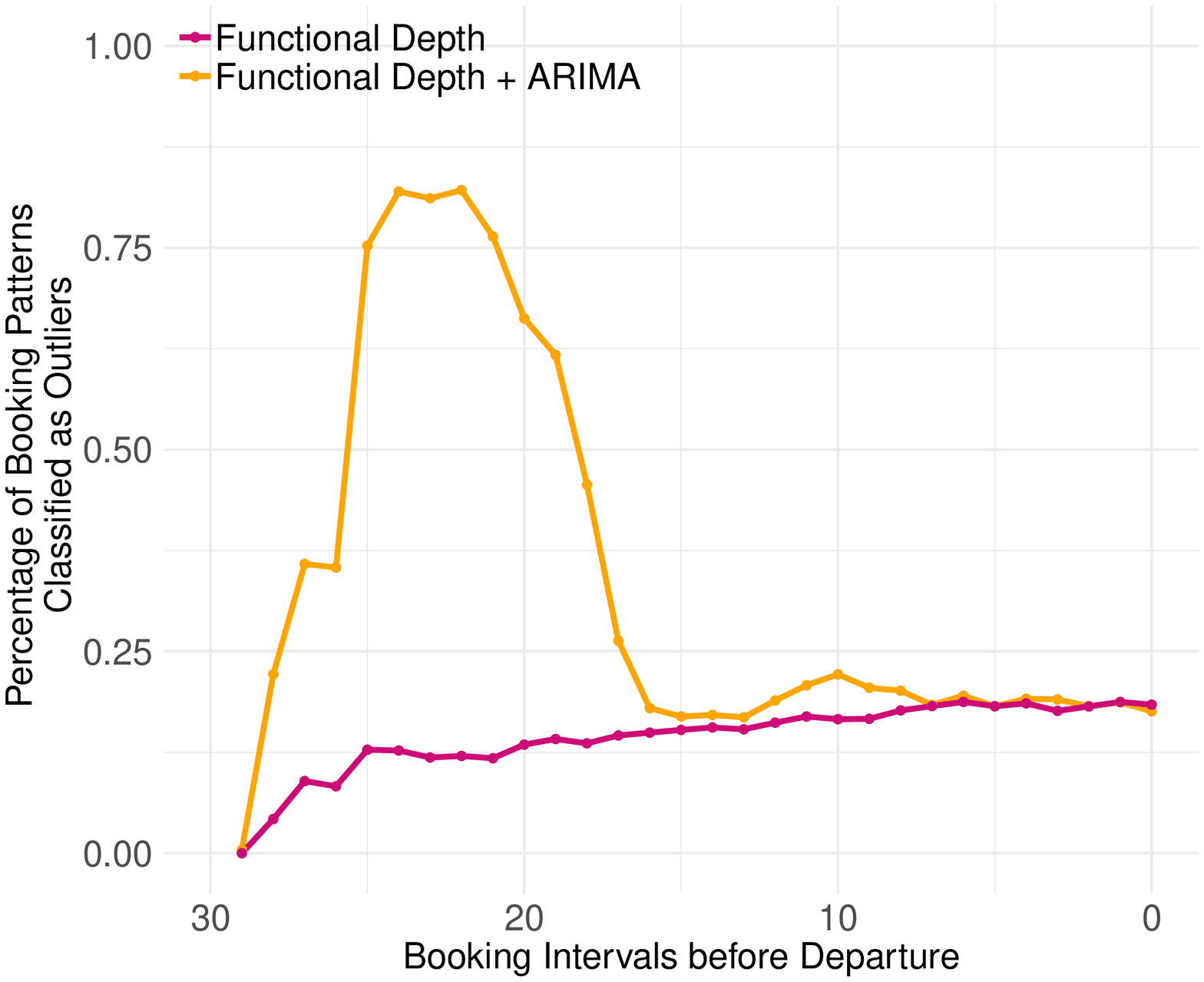}
        \caption{Percentage of booking patterns classified as outliers}
		\label{fig:num_outs}
    \end{subfigure} \hspace{0.4cm}
    \begin{subfigure}[h]{0.47\textwidth}  
    \centering 
        \includegraphics[width=0.8\textwidth,height=0.8\textwidth]{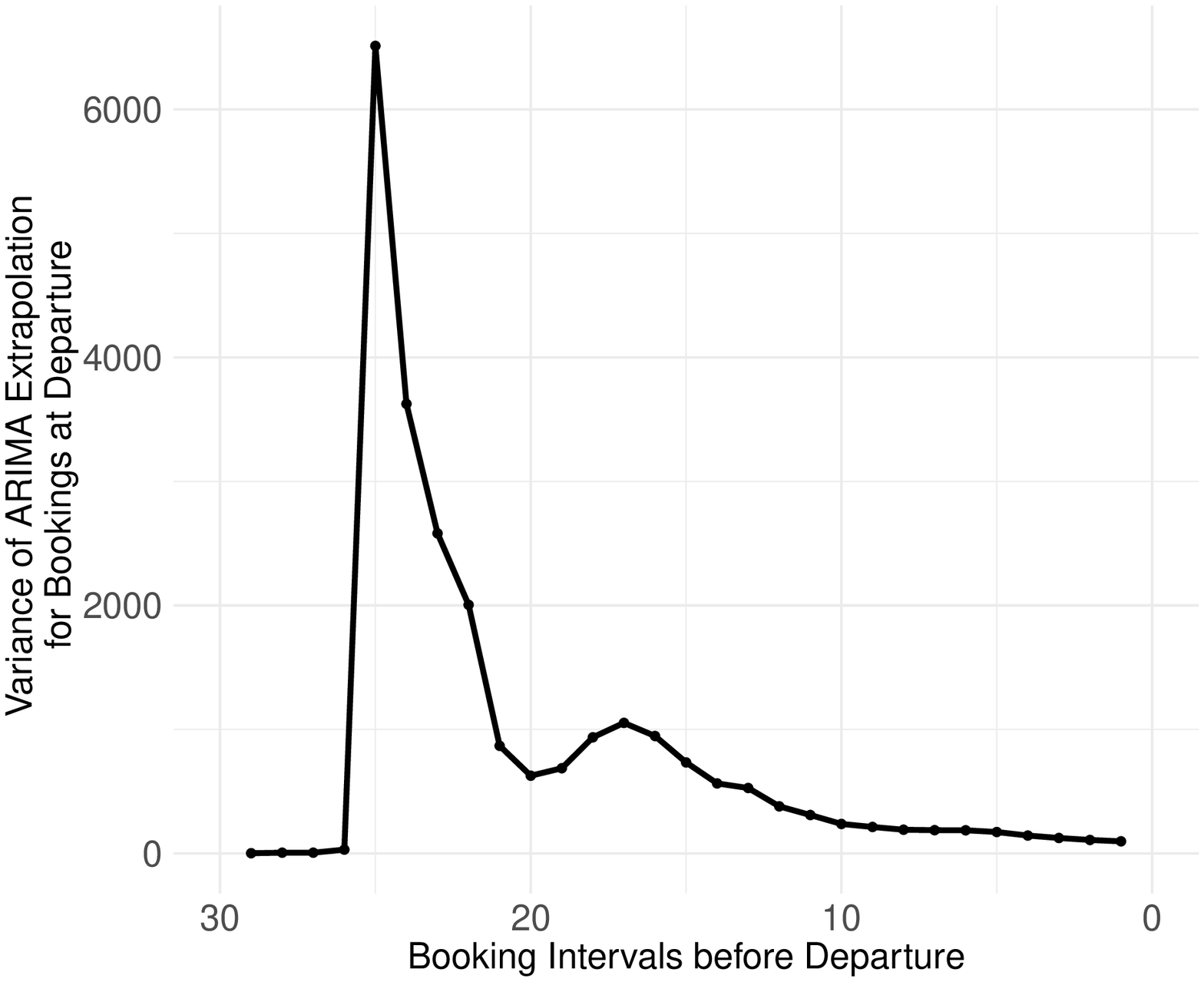}
        \caption{Variance of ARIMA extrapolation for bookings at departure}
		\label{fig:ARIMA_var}
    \end{subfigure}
    \quad
    \caption{Relationship between variance and number of classified outliers}
    \label{fig:ARIMA_var_num}
\end{figure}

Given the superior performance across a range of thresholds (evidence by the ROC curves in Section 6.2), of functional depth with extrapolation, we consider whether using the same parameters to calculate the threshold across the booking horizon (following those implemented by \citet{Febrero2008}) is the best approach. We compare the percentage of booking patterns classified as outliers by functional depth with and without extrapolation (Figure \ref{fig:num_outs}). In addition, Figure \ref{fig:ARIMA_var} the variance, across the booking horizon, of the ARIMA extrapolation at time \(t_{T}\). We see that there is a relationship between the variance of the ARIMA extrapolation the number of patterns classified as outliers, and therefore the false positives. It may perhaps be possible to vary the threshold parameters according to the functional variance, as it changes across the booking horizon with extrapolation. We see this as an opportunity for further work.

In practice, companies have a limited number of analysts to respond to outlier detection-based alerts. Hence, the threshold would likely be left variable. That is, if an analyst is receiving too many alerts (caused by the high false positive rate), they can reduce the threshold. In this case, given the results from the consideration of the ROC curves (Section 6.2) where the threshold changes, extrapolation would still be preferred.


\newpage
\subsection{Effect of Magnitudes of Outliers}
\begin{figure}[!ht]
    \centering
    \begin{subfigure}[h]{0.47\textwidth}
    \centering
        \includegraphics[width=0.8\textwidth,height=0.8\textwidth]{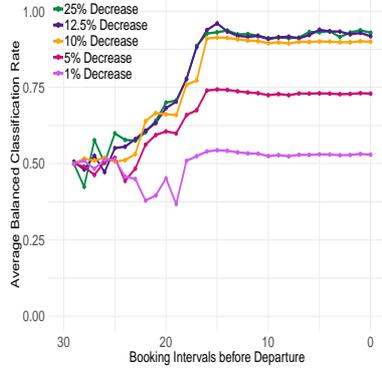}
        \caption{BCR of functional depth with ARIMA extrapolation over a range of (negative) outlier magnitudes}  
		\label{fig:mag_dec}
    \end{subfigure} \hspace{0.4cm}
    \begin{subfigure}[h]{0.47\textwidth}  
    \centering 
        \includegraphics[width=0.8\textwidth,height=0.8\textwidth]{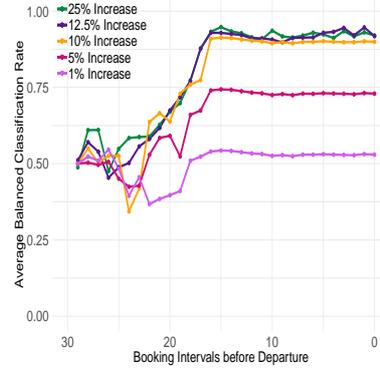}
        \caption{BCR of functional depth with ARIMA extrapolation over a range of (positive) outlier magnitudes}
		\label{fig:mag_inc}
    \end{subfigure}
    \quad
    \begin{subfigure}[h]{0.47\textwidth}
    \centering
        \includegraphics[width=0.8\textwidth,height=0.8\textwidth]{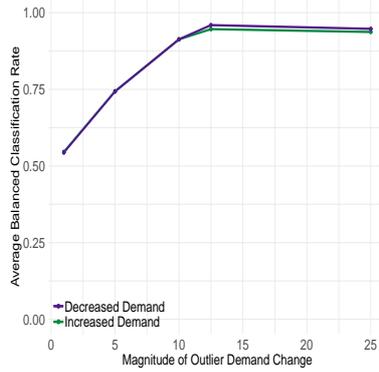}
        \caption{Average hindsight performance of functional depth with ARIMA extrapolation over a range of outlier magnitudes}  
		\label{fig:elbow}
    \end{subfigure} \hspace{0.4cm}
    \begin{subfigure}[h]{0.47\textwidth}  
    \centering 
        \includegraphics[width=0.8\textwidth,height=0.8\textwidth]{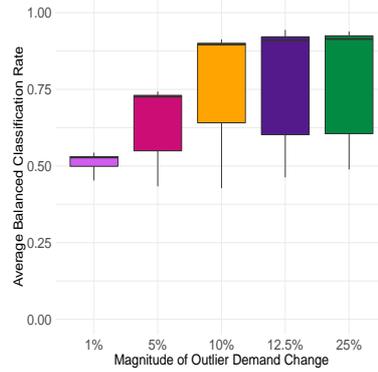}
        \caption{Average foresight performance of functional depth with ARIMA extrapolation over a range of outlier magnitudes}
		\label{fig:box}
    \end{subfigure}
    \quad
    \caption{Effects of magnitude of demand outliers on functional depth with ARIMA extrapolation outlier detection}
    \label{fig:demand_mag}
\end{figure}



\newpage
\subsection{Relationship Between Extrapolation Accuracy and Outlier Detection Improvement}
\begin{figure}[h!]
   \centering 
   \includegraphics[width=0.5\textwidth,height=0.45\textwidth]{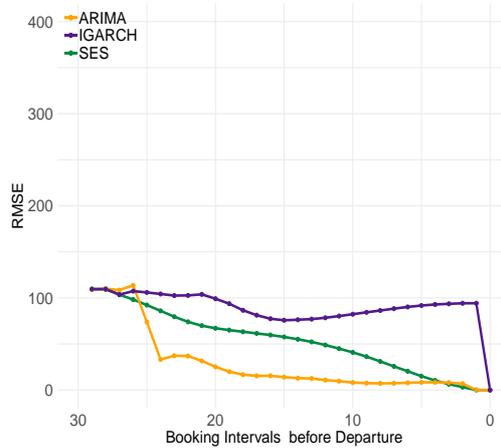}
   \caption{RMSE of different extrapolation methods}
\label{fig:rmse}
\end{figure}

To investigate the relationship between the accuracy of the extrapolation and the improvement in outlier detection from an extrapolation method, we computed the average root mean square error (RMSE) of each of the extrapolation methods across the booking horizon -- see Figure \ref{fig:rmse}. The RMSE of each individual method means little on its own. As we have increased data available to input to our forecast and are forecasting fewer steps ahead, it is of little surprise that the RMSE decreases over time. However, from the comparison of the RMSE of the different extrapolation methods, we gain some insight into the performance of the outlier detection when using that method. Generally, ARIMA forecasts have the lower RMSE of the methods, and also provide the largest gain in performance overall when used as the extrapolation method. The exception to this is the IGARCH model where the RMSE has a slight increase in the later part of the booking horizon. This is most likely due to the fact that we have fixed the order of the IGARCH model to be (1,d,1) for computational reasons and are therefore imposing a variance structure in the forecast that does not exist in the data. It is interesting to note that despite the poorer performance of the IGARCH forecast, it still provides a reasonable improvement in outlier detection performance as an extrapolation method.


\newpage
\subsection{Comparison of Methods for Hindsight Detection of Demand-volume Outliers}
\begin{figure}[h!]
    \centering
    \includegraphics[width=0.7\textwidth]{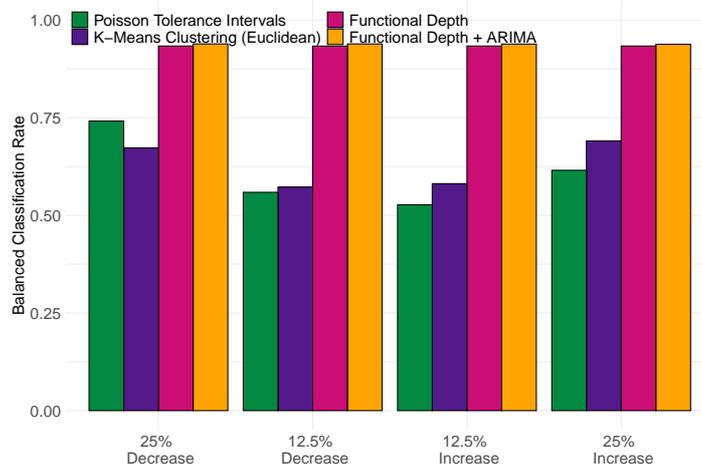}
	\caption{Comparison of hindsight outlier detection under different magnitudes of demand outliers with 5\% outlier frequency}    
    \label{fig:hindsight}
\end{figure}
For {\em hindsight detection} performance, we rely on the BCR averaged across all booking intervals. As shown in Figure \ref{fig:hindsight}, hindsight detection performance typically increases as the complexity of the outlier detection method increases across all categories of outliers tested. These results are consistent with those for foresight detection. Figure \ref{fig:hindsight} shows that including the extrapolation step induces only a small improvement in hindsight detection performance. However, outliers are detected early in the horizon, meaning any actions taken as a result of their identification will have a significant positive impact in terms of revenue overall, both within and beyond the booking horizon. 

Within the revenue management process, identifying outliers and adjusting controls as early as possible provides the most benefit. Nevertheless, even detecting outliers in hindsight promises some advantages over not identifying them at all. 


\newpage
\subsection{Additional Analysis of Railway Booking Patterns}
\begin{figure}[!ht]
    \centering
    \begin{subfigure}[h]{0.47\textwidth}
    \centering
        \includegraphics[width=0.8\textwidth,height=0.8\textwidth]{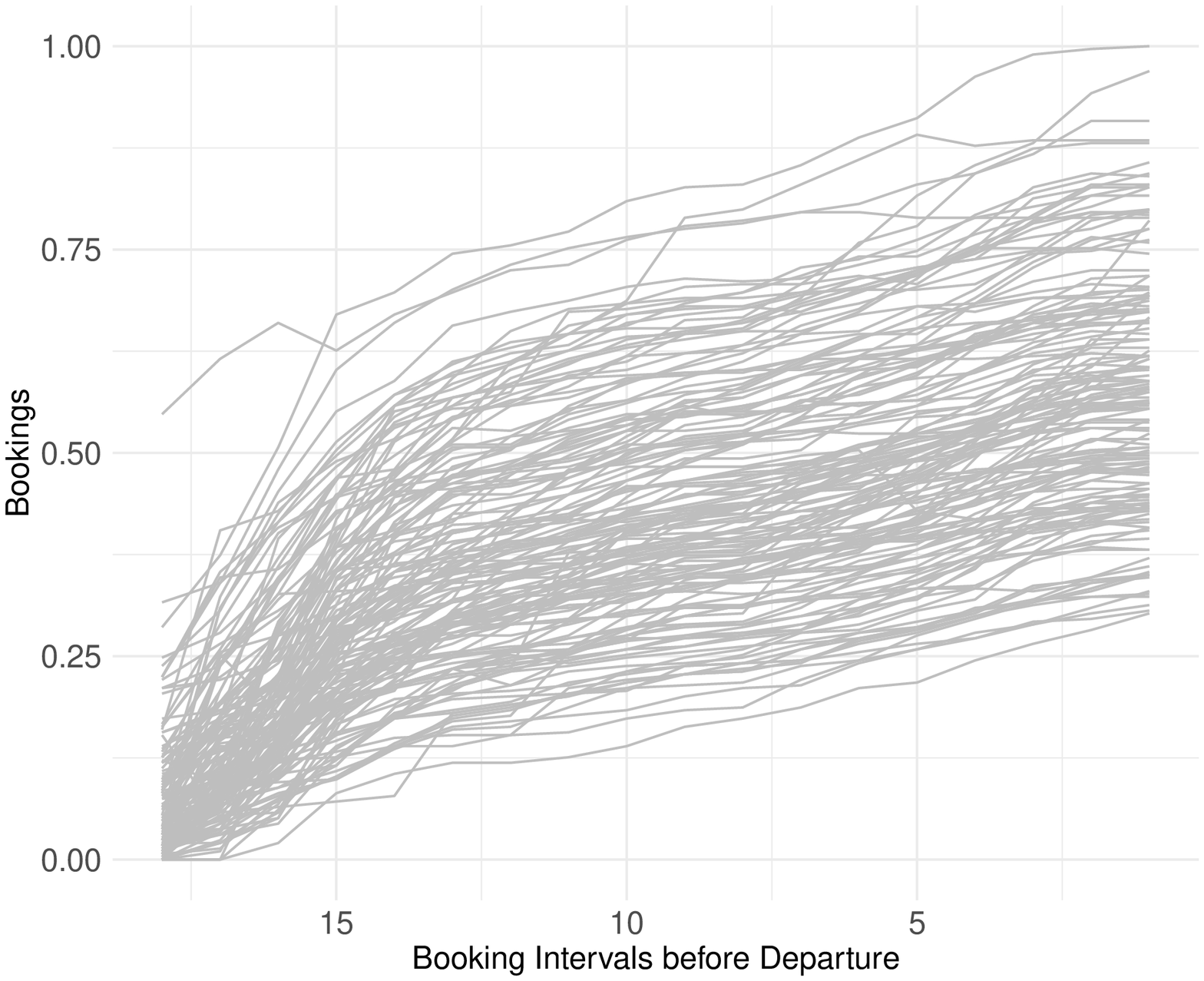}
        \caption{Railway booking patterns}
		\label{fig:rail_bp}
    \end{subfigure}
    \begin{subfigure}[h]{0.47\textwidth}  
    \centering 
        \includegraphics[width=0.8\textwidth,height=0.8\textwidth]{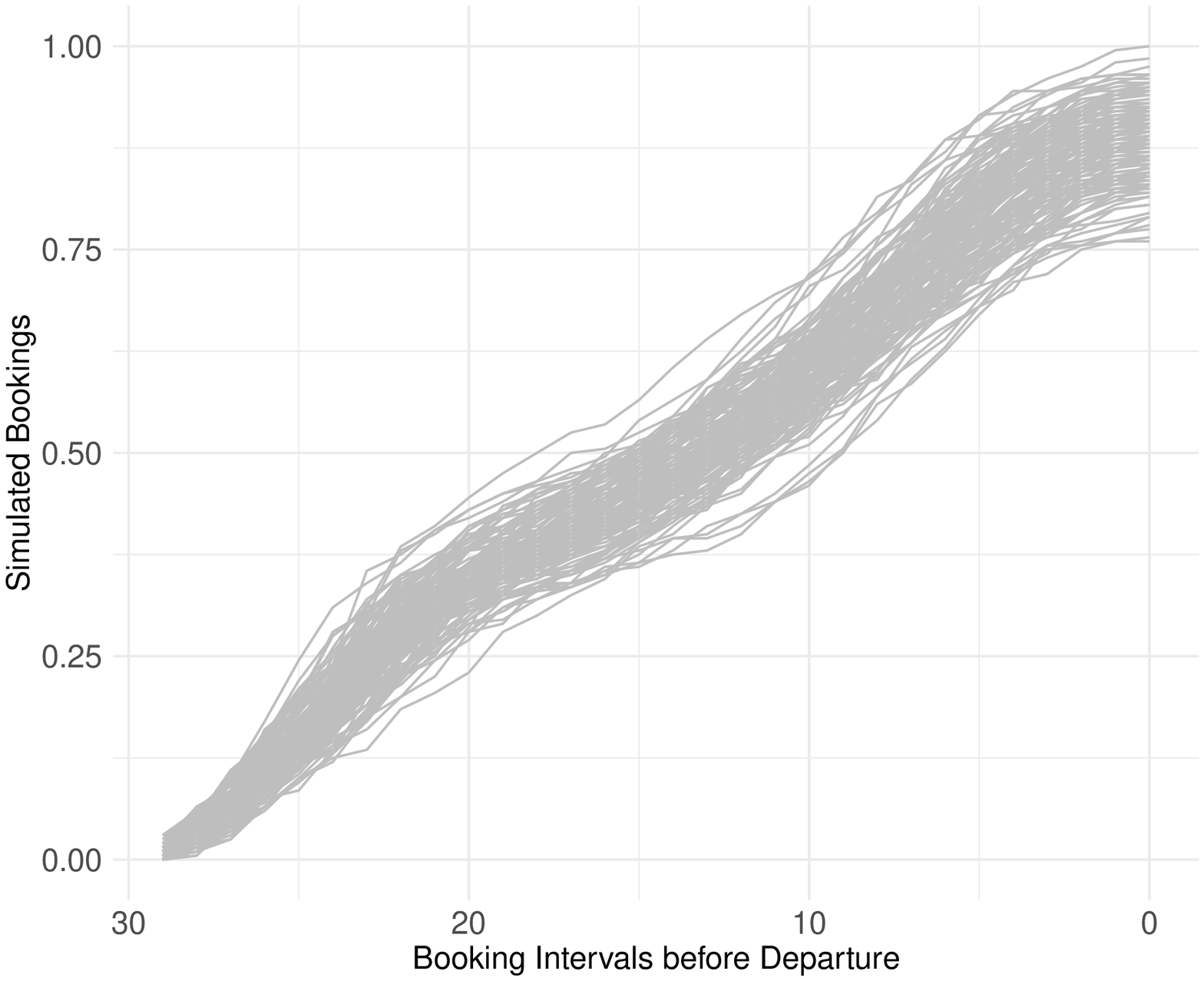}
        \caption{Simulated booking patterns}
		\label{fig:sim_bp}
    \end{subfigure}
    \quad
    \caption{Railway vs simulated booking patterns}
    \label{fig:comp_sim}
\end{figure}
Here, we compare the simulated booking patterns with those from the railway company. Note that both the railway and simulated booking patterns in Figure \ref{fig:comp_sim} have been rescaled to be between 0 and 1. Therefore, although it may appear that the variance of the railway booking patterns is much higher than that of the simulated patterns, it is not necessarily significant (given the rescaling only transforms the mean, not the variance of the booking patterns). The main takeaway from Figure \ref{fig:comp_sim} is the similar shape of the booking patterns -- starting with a steep increase, followed by a slight flattening out, then another increase.

\begin{figure}[!ht]
    \centering
    \begin{subfigure}[h]{0.47\textwidth}
    \centering
        \includegraphics[width=0.8\textwidth,height=0.8\textwidth]{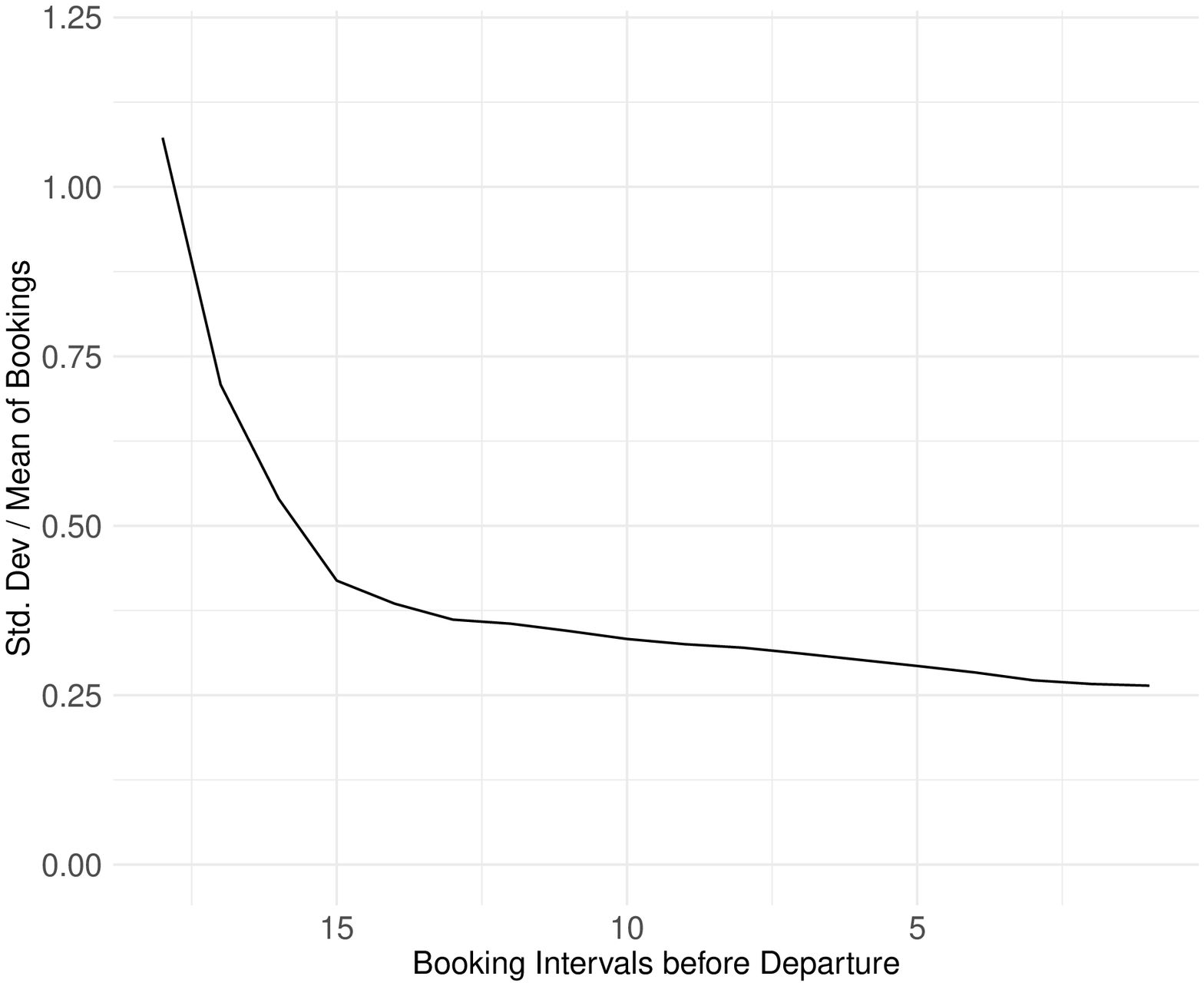}
        \caption{Railway booking patterns}  
		\label{fig:rail_sm}
    \end{subfigure}
    \begin{subfigure}[h]{0.47\textwidth}  
    \centering 
        \includegraphics[width=0.8\textwidth,height=0.8\textwidth]{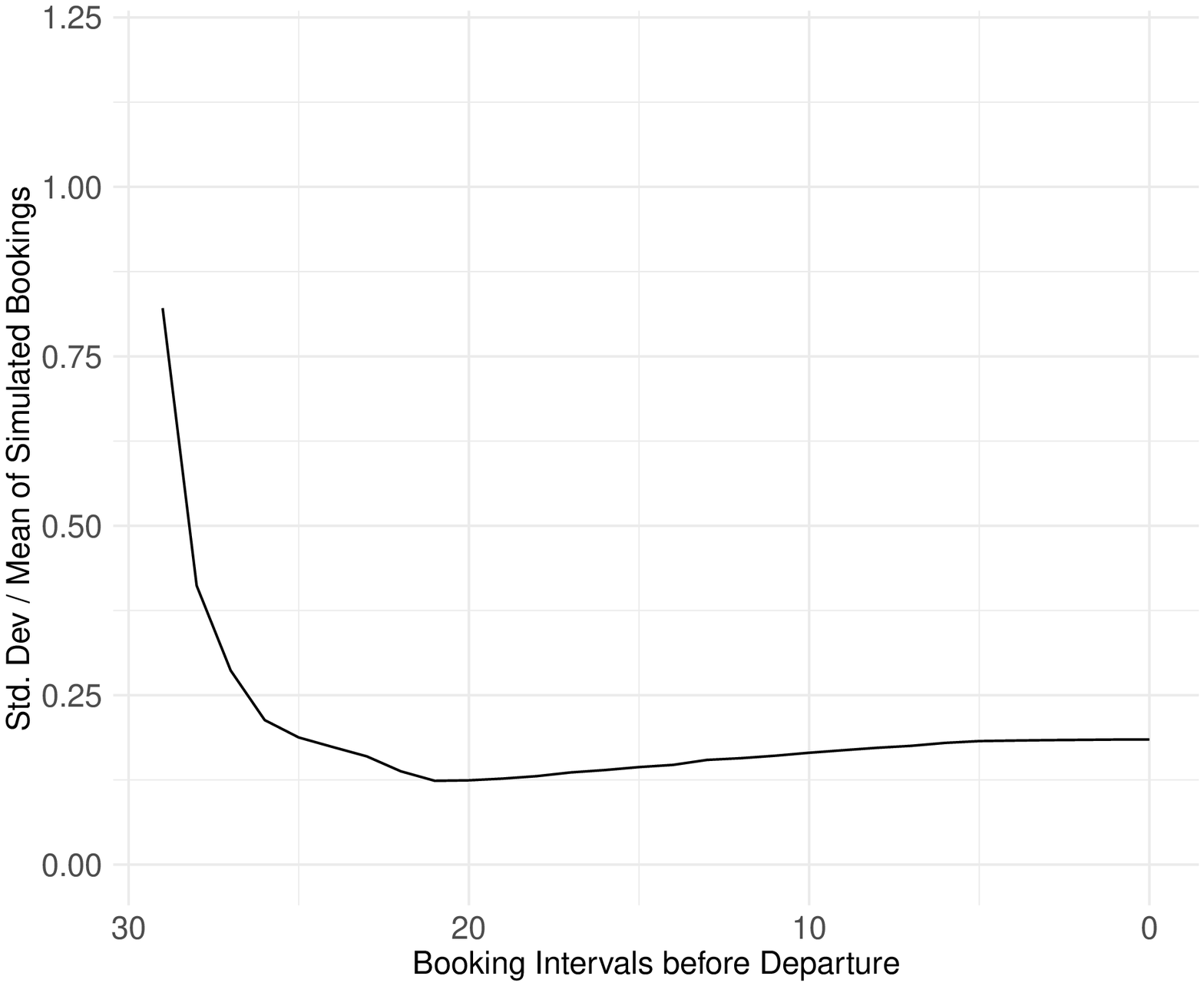}
        \caption{Simulated booking patterns}
		\label{fig:sim_sm}
    \end{subfigure}
    \quad
    \caption{Standard deviation / Mean of railway vs simulated booking patterns}
    \label{fig:s_dev_mean}
\end{figure}

In order to compare the simulated booking patterns with the railway booking patterns, we analyse the relationship between the mean and standard deviation of bookings across the horizon. Figure \ref{fig:rail_sm} shows the standard deviation divided by the mean number of bookings in the railway booking data, and Figure \ref{fig:sim_sm} analogously for the simulated booking patterns. The two figures show a similar shape -- higher at the start of the horizon, then quickly flattening out. The values of the standard deviation / mean are also of a similar magnitude.

\begin{figure}[!ht]
    \centering
    \begin{subfigure}[h]{0.47\textwidth}
    \centering
        \includegraphics[width=0.8\textwidth,height=0.8\textwidth]{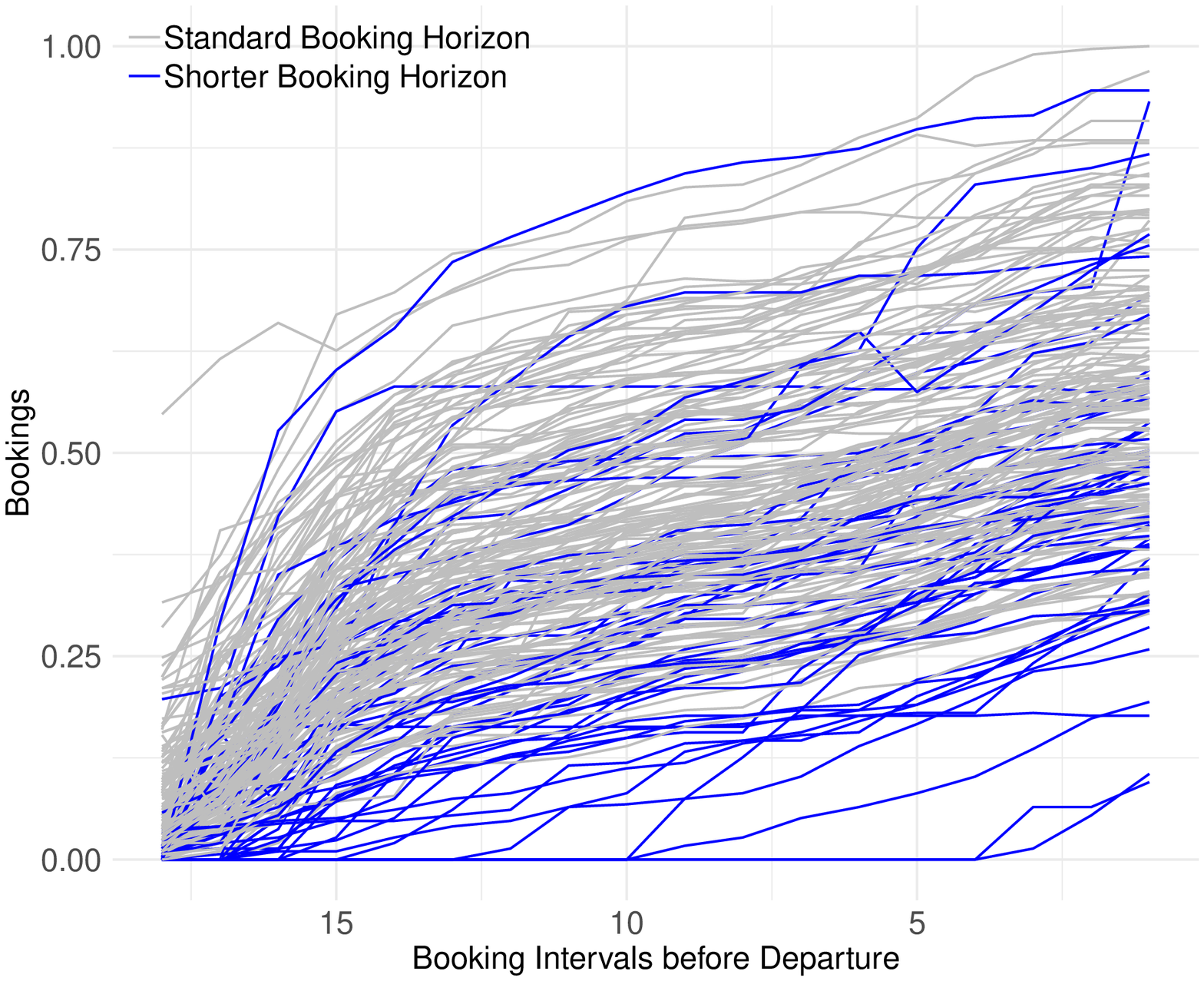}
        \caption{Shorter booking horizons in railway booking patterns}
		\label{fig:shorter_bh}
    \end{subfigure} \hspace{0.4cm}
    \begin{subfigure}[h]{0.47\textwidth}  
    \centering 
        \includegraphics[width=0.8\textwidth,height=0.8\textwidth]{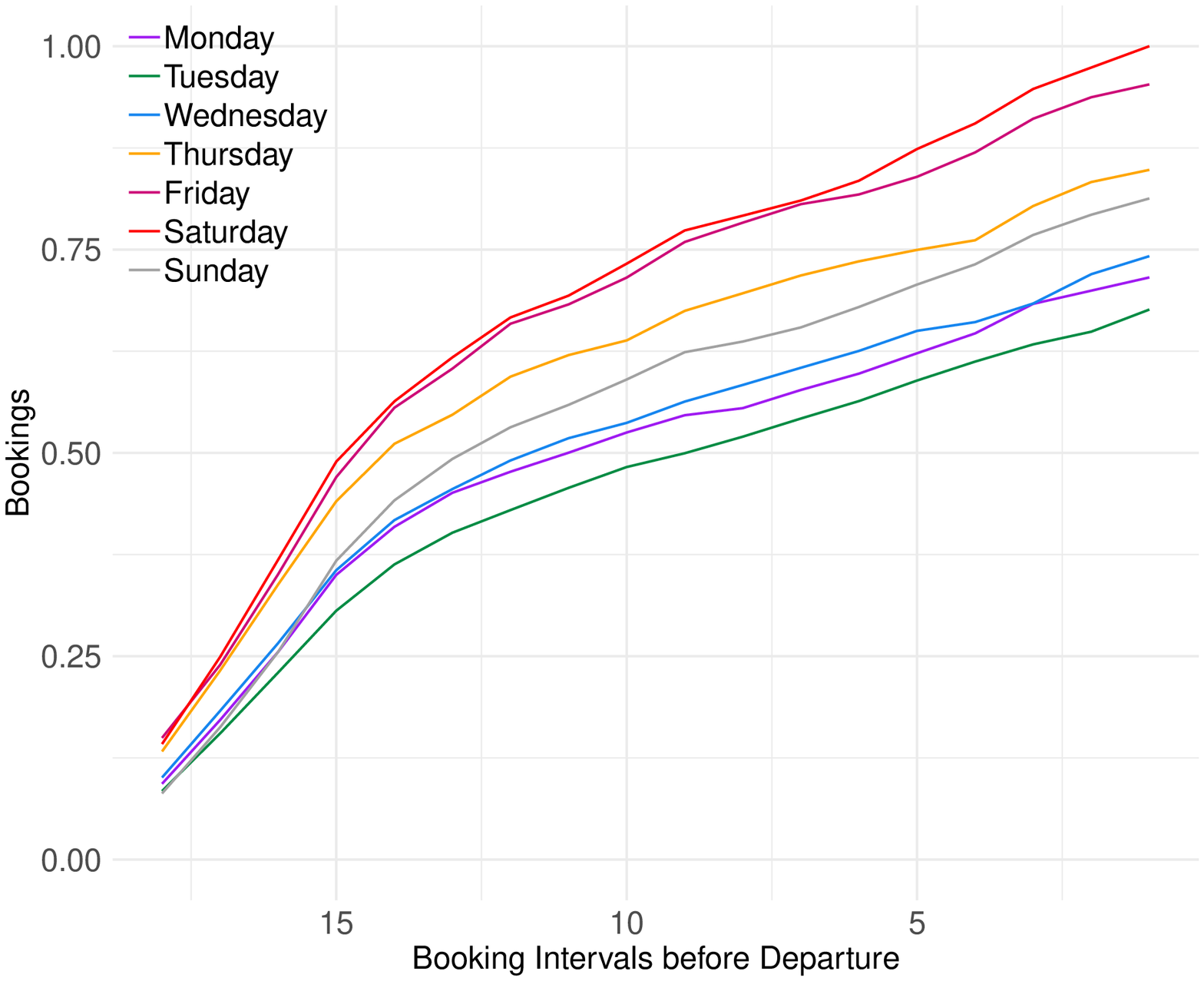}
        \caption{Functional regression curves for departures days of the week}
		\label{fig:day_reg}
    \end{subfigure}
    \quad
    \caption{Functional regression to homogenise booking patterns}
    \label{fig:func_reg}
\end{figure}

As discussed in Section 6.4, we compare booking patterns for different days of the week by applying pairwise functional ANOVA tests \citep{Cuevas2004}. We test the null hypothesis that, for two different days \(m\) and \(n\), their mean functions are equal:
\begin{equation}
    H_0: \mu_m(t) = \mu_n(t),\mbox{ vs. } H_A: \mu_m(t) \neq \mu_n(t),
\end{equation}
\begin{table}[htbp]
\centering
\begin{tabular}{r|ccccccc}
\hline \hline 
\textbf{}  & \textbf{Mon} & \textbf{Tue} & \textbf{Wed} & \textbf{Thu} & \textbf{Fri}  & \textbf{Sat} & \textbf{Sun}  \\ \hline
\textbf{Mon}  &       &   &   &   &   &   &   \\ 
\textbf{Tue}  & 0.001 &   &   &   &   &   &   \\ 
\textbf{Wed}  & \textbf{0.093} & 0.000 &   &   &   &   &   \\ 
\textbf{Thu}  & 0.000 & 0.000 & 0.000 &   &   &   &   \\ 
\textbf{Fri}  & 0.000 & 0.000 & 0.000 & 0.000 &   &   &   \\ 
\textbf{Sat}  & 0.000 & 0.000 & 0.000 & 0.000 & \textbf{0.122} &   &   \\ 
\textbf{Sun}  & 0.000 & 0.000 & 0.000 & 0.001 & 0.000 & 0.000 &  \\ \hline \hline
\end{tabular}
\caption{p-values for functional ANOVA test}
\label{tab:pvalues_anova}
\end{table}
The p-values are shown in Table \ref{tab:pvalues_anova}. The only non-significant p-values are for comparison between Monday-Wednesday and Friday-Saturday. However, the p-values are not overly convincing, especially when considering multiple testing issues, so we choose to model each departure day separately. A similar comparison can be made between booking patterns which are affected by the shortened booking horizons (see Figure \ref{fig:shorter_bh}), and those of standard length. In that test, all of the p-values were 0.

We account for both the shortened booking horizons and the effect of different departure days through fitting a functional regression model, as per Equation (12). Figure \ref{fig:day_reg} shows the regression curves for each day of the week (without shortened booking horizon effects). The functional regression model works by fitting a linear regression at each time point. That is a different value of at each booking interval. In order to make the $\beta_j(t)$ smooth functions, we penalise the integrated square error such that we seek to minimise \citep{Ramsay2009}:
\begin{equation}
\sum_{i=1}^{n} \int (y_{i}(t) - \hat{y}_{i}(t))^2 dt + \sum_{j=0}^7 \lambda_j \int [L_j \beta_j ]^2 dt,
\end{equation}
where
\begin{equation} 
\begin{split}
   \hat{y}_{i}(t) = \beta_0(t) + \beta_1(t)I_{Monday_{i}} + \beta_2(t)I_{Tuesday_{i}} + \beta_3(t)I_{Wednesday_{i}} + \\ \beta_4(t)I_{Thursday_{i}} + \beta_5(t)I_{Friday_{i}} + \beta_6(t)I_{Saturday_{i}} + 
   \beta_7(t)I_{Shorter\mbox{ }Horizon_{i}}.
\end{split}
\end{equation}
and $lambda_j$ a non-negative real number controlling the amount of smoothing, and $L_j$ is either a non-negative integer or a linear differential operator object. Due to the relatively short nature of the booking patterns (18 observations), for this data set we use a smoothing parameter of $\lambda_j = 0 \forall j$.

\end{document}